\documentclass{emulateapj}
\newcommand{\gsim}{\hspace{0.3em}\raisebox{0.4ex}{$>$}\hspace{-0.75em}\raisebox{-.7ex}{$\sim$}\hspace{0.3em}}
\newcommand{\lsim}{\hspace{0.3em}\raisebox{0.4ex}{$<$}\hspace{-0.75em}\raisebox{-.7ex}{$\sim$}\hspace{0.3em}}

\shorttitle{A Subaru/Suprime-Cam Survey of M31's Stellar Halo}
\shortauthors{Tanaka et al.}

\begin{document}


\title{STRUCTURE AND POPULATION OF THE ANDROMEDA STELLAR HALO FROM A
SUBARU/SUPRIME-CAM SURVEY\altaffilmark{1}} 


\author{Mikito~Tanaka\altaffilmark{2,3,4},
        Masashi~Chiba\altaffilmark{2},
        Yutaka~Komiyama\altaffilmark{4},
        Puragra~Guhathakurta\altaffilmark{5}, 
	Jason~S.~Kalirai\altaffilmark{6},
	and
        Masanori~Iye\altaffilmark{4}}


\altaffiltext{1}{Based on data collected at the Subaru Telescope, which
 is operated by the National Astronomical Observatory of Japan.}
\altaffiltext{2}{Astronomical Institute, Tohoku University, Aoba-ku,
 Sendai 980-8578, Japan (current address); mikito@astr.tohoku.ac.jp} 
\altaffiltext{3}{University of Tokyo, 7-3-1 Hongo, Bunkyo-ku, Tokyo
 113-0033, Japan}
\altaffiltext{4}{National Astronomical Observatory of Japan, 2-21-1
 Osawa, Mitaka, Tokyo 181-8588, Japan}
\altaffiltext{5}{University of California Observatories/Lick
 Observatory, University of California Santa Cruz, 1156 High Street,
 Santa Cruz, California 95064, USA} 
\altaffiltext{6}{Space Telescope Science Institute, Baltimore, MD 21218} 


\begin{abstract}
We present a photometric survey of the stellar halo of the nearest giant
spiral galaxy, Andromeda (M31), using Suprime-Cam on the Subaru Telescope.
A detailed analysis of $VI$ color-magnitude diagrams of the resolved stellar
population is used to measure properties such as line-of-sight distance,
surface brightness, metallicity, and age, and these are used to isolate and
characterize different components of the M31 halo:
(1) the giant southern stream (GSS), (2) several other substructures, and
(3) the smooth halo. 
First, the GSS is characterized by a broad red giant branch (RGB) and a
metal-rich/intermediate-age red clump (RC).  The $I$-magnitude of the
well-defined tip of the RGB
suggests the distance to the observed GSS field of $(m-M)_0 = 24.73 \pm 0.11$
($883 \pm 45$ kpc) at a projected radius of $R \sim 30$ kpc from M31's center.
The GSS shows a high metallicity peaked at [Fe/H]$\gsim -0.5$
with a mean (median) of $-0.7$ ($-0.6$), estimated via
comparison with theoretical isochrones. Combined with
the luminosity of the RC, we estimate the mean age of its stellar population to be
$\sim 8$ Gyr.  The mass of its progenitor galaxy is likely in the
range of $10^7$ to $10^9 M_{\sun}$.
Second, we study M31's halo substructure
along the north-west/south-east minor axis out to
$R \sim 100$ kpc and the south-west major axis region at $R \sim 60$ kpc.
We confirm two substructures in the south-east halo reported by
Ibata et al. (2007) and discover two overdense substructures
in the north-west halo. We investigate the properties of these four
substructures as well as other structures including the western shelf
and find that differences in stellar populations among these systems,
thereby suggesting each has a different origin. Our statistical analysis
implies that
the M31 halo as a whole may contain at least 16 substructures, each with
a different origin, so its outer halo has experienced at least this many
accretion events involving
dwarf satellites with mass
$10^7$--$10^9 M_{\sun}$ since a redshift of $z \sim 1$.
Third, we investigate the properties of an underlying, smooth and extended
halo component out to $R > 100$ kpc. We find that the surface density of
this smooth halo can be fitted to a Hernquist model of scale radius
$\sim 17$ kpc or a power-law profile with $\propto R^{-2.17 \pm 0.15}$.
In contrast to the relative smoothness of the halo density profile, its
metallicity distribution appears to be spatially non-uniform
with non-monotonic variations with radius, suggesting that the halo
 population has not had sufficient time to dynamically homogenize the
 accreted populations. Further implications for the formation of the M31
 halo are discussed.
\end{abstract}



%%% Section 1 %%%
\section{Introduction}\label{sec:intro}
The advent of large ground-based telescopes and the {\it Hubble Space
Telescope\/} ({\it HST\/})
has made it possible to resolve faint individual stars in
nearby galaxies, thereby enabling us to study galactic archaeology for such
external galaxies in the same way that one studies our home galaxy,
the Milky Way. In particular,
our knowledge of the stellar halo of the Andromeda galaxy (M31) has dramatically
expanded over the last two decades. Indeed, M31 provides us
with a global yet detailed view of a galaxy with a similar
morphology to the Milky Way. 

The most significant discovery in the M31 halo is probably its complex
substructures such as the giant southern stream (GSS) in the outer halo
\citep{Ibata2001,Ferguson2002}. This Andromeda GSS has
attracted particular attention, as it reveals an on-going hierarchical
formation process of M31 and thus places invaluable constraints on
galaxy formation models. The GSS contains a high
concentration of metal-rich stars \citep{Ibata2001,Guhathakurta2006}
and is located behind M31 with a significant degree of elongation along
the line of sight \citep{McConnachie2003}. \citet{Ferguson2002}
discovered significant substructures around M31 by mapping
red giant branch (RGB) stars with the Isaac Newton Telescope. 

The GSS was once thought to be tidal debris dispersed by
interactions with M32 or NGC~205 as judged from its projected trajectory on
the sky \citep{Ferguson2002,Choi2002}, but subsequent studies have revealed
that it is probably not related to these two satellites
\citep[e.g.,][]{Ibata2004, McConnachie2004,Guhathakurta2006,Kalirai2006a}. In
particular, spectroscopic observations of GSS stars with Keck/DEIMOS revealed
that they show a markedly different radial motion from M32 and
NGC~205. Recently,
$N$-body numerical simulations have been performed in view of these
kinematic results and these have showed that the GSS's progenitor had
a highly eccentric orbit and a mass of about
$10^8--10^9$ M$_\sun$
\citep{Font2006,Fardal2006,Fardal2007}. Furthermore, deep
photometric observations with {\it HST\/}/ACS
reaching down to below the main-sequence
turn-off \citep{Brown2006a,Brown2006b} detected intermediate-age stars in
a GSS field located at a projected distance of $R = 20$ kpc
from the M31 center as well as in the general halo field at $R = 11$ kpc
along the southern minor axis \citep{Brown2003}. 

In spite of these detailed studies however, it remains unclear what
the detailed properties of the GSS's progenitor galaxy are.
Besides, we do not fully understand
the connection between the GSS and the other
substructures in the M31 halo such as the NE- and W-shelves,
although the latest simulations \citep{Fardal2008,Mori2008} imply a
connection among them. To
unveil the origin of the GSS, it is of particular importance to
carry out the detailed analysis of its stellar populations by
assembling a statistically significant set of
observational data such as color, metallicity, and age distributions of
the stars, in addition to the other halo fields \citep[e.g.,][]{Bellazzini2003}. 
Stellar halos generally have such low surface brightness (SB)
that detecting them beyond the Milky Way is a major challenge. For example,
\citet{Morrison1993} estimate the SB of the Galactic halo at the solar
radius to be $\mu_V \sim 27.7$ mag arcsec$^{-2}$.
The M31 halo was
long believed to be an outward extension of its bulge, since
surprisingly metal-rich stars ($\rm\langle[M/H]\rangle\sim -0.6$) were
found in the inner halo of M31
\citep[e.g.,][]{Mould1986,Couture1995,Holland1996,Reitzel1998}. Furthermore,
the SB profile measured along the south-east minor axis from integrated
light and resolved star counts follows a de
Vaucouleurs $R^{1/4}$ profile out to $R = 20$~kpc \citep{Pritchet1994},
quite unlike the
power-law behavior deduced for the halo of the Milky Way. Both the de 
Vaucouleurs profile and high metallicity are suggestive of an active
merger history for the halo (or bulge) component of M31. 

The existence of metal-rich populations in the M31 halo has been
confirmed by several subsequent studies. \citet{Durrell2001} found that
a high-metallicity population with ${\rm [M/H]} \sim -0.5$ is dominated
in a field located at $R = 20$ kpc along the south-east minor
axis, based on a wide-field mosaic CCD camera on CFHT. They also
discovered that 30\%--40\% of the stars at this location belong to a
metal-poor population. Furthermore, in their later complementary paper 
\citep{Durrell2004}, they showed that a similar high-metallicity
population is distributed in a field at $R = 30$ kpc along
the south-east minor axis. Therefore, they concluded that the outer halo
shows little or no radial metallicity gradient. 

{\it HST\/} has made great contributions to the exploration of
the inner halo populations of M31. Using {\it HST\/}/WFPC2
\citet{Bellazzini2003} analyzed the data for 16 fields at different
distances from the M31 center, from $R \sim 4.5$ to 35 kpc, down to
limiting $V$ and $I$ magnitudes of $\sim 27$. From the metallicity
distributions in each field, obtained by comparison of the RGB stars
with Galactic globular cluster templates, they detected 
similar abundance distributions peaked at ${\rm [Fe/H]} \sim -0.6$
in all of the sampled fields. Although in some fields a very
metal-rich (${\rm [Fe/H]} \geq -0.2$) component is clearly present,
possibly attributable to disrupted disk populations, they identified
no evident metallicity gradient in their limited
region. \citet{Ferguson2005} and \citet{Richardson2008} investigated M31
inner substructures discovered by \citet{Ferguson2002}, based on the deep
{\it HST\/}/ACS survey spanning the range from $R \sim 11.5$ to 45 kpc
reaching a few magnitudes below the RC. Their systematic surveys confirmed
that the inner halo of M31 is strongly polluted by metal-rich and
intermediate-age populations disrupted from the M31 disk and the GSS
by interactions between M31 and dwarf galaxies.

Brown et al.'s studies based on ultra-deep {\it HST\/}/ACS
observations have unraveled
star-formation histories in selected locations of the M31
halo \citep{Brown2003,Brown2004,Brown2006a,Brown2006b,Brown2007,Brown2008}. 
With a minimum of 32 orbits per field, these observations have been
sufficiently deep to resolve stars below the oldest main-sequence
turn-off. \citet{Brown2003} first demonstrated that a minor axis field
projected at $R = 11$ kpc contains an indisputable population in age
(predominantly 6--10 Gyr) and abundance (${\rm [Fe/H]} >
-0.5$). \citet{Brown2006b} present results from inner halo fields they
associate with the ``tidal stream,'' ``outer disk,'' and
``spheroid''. In all cases, the fields were shown to have experienced an
extended star-formation history, in contrast, they quantified
differences between them. \citet{Brown2007} investigated a field at 21
kpc and found evidence that its population is marginally older and more
metal poor than the inner halo field. Furthermore, \citet{Brown2008},
based on an additional field at $R = 35$ kpc, found that the mean ages (the
mean [Fe/H] values) at 11, 21 and 35 kpc gradually change like 9.7
($-0.65$), 11.0 ($-0.87$) and 10.5 ($-0.98$) Gyr (dex), respectively.

The inclusion of kinematic information has been extremely useful but has
also added another dimension of complexity to the
puzzle. Spectroscopic studies of \citet{Guhathakurta2005} uncovered successfully
the presence of an extended halo component having a power-law SB profile of
$R^{-2.3}$ over the range $R = 30$--165~kpc resembling the Milky Way's
halo, while the SB of M31's spheroid at $R < 30$~kpc has been known to show
an $R^{1/4}$ law resembling a galactic bulge
\citep{Pritchet1994,Durrell2004,Irwin2005,Gilbert2006}. They also
reported that M31's halo 
may extend to $R\gtrsim165$~kpc, suggesting that halo stars of M31 and
the Milky Way occupy a substantial volume fraction of our Local Group.
However, it remains to be seen
whether this flat portion of the halo profile is affected by
substructure.

In the meantime \citet{Reitzel2002} analyzed a sample of 29 stars in a
field at $R = 19$ kpc on the minor axis and found the mean metallicity
to be in the range of ${\rm [Fe/H]} = -1.9$ to $-1.1$. Despite
still remaining uncertainties of calibration and sample selection
issues, their estimated mean metallicity is significantly lower than the
results deduced from the above photometric analyses. \citet{Kalirai2006}
and \citet{Chapman2006} recently discovered a metal-poor halo out to
$R = 165$ kpc that is similar to that of the Galaxy, both in terms of its
metallicity, radial profile, as well as the amount of substructure.

As noted above, the stellar halo of M31 as well as the Milky Way is
filled with quite complex fossil records. Our principal goal is thus
to identify each component of the halo and clarify its formation history.
To do so, we need to obtain both the global and local structures in
the stellar halo and deduce plausible origins of them. Detailed studies
based on large numbers of stars, such as with wide-field imagers, are
required to investigate the fundamental nature of the large halo regions
in M31. Indeed, our external viewpoint for this galaxy makes it possible to
study the entire area of its halo \citep{Ibata2007}; detailed stellar
maps will allow us to detect giant streams as well as a wealth of fainter
substructures at SB levels of $\sim 30$ mag arcsec$^{-2}$. In the
south-quadrant halo of M31, Ibata et al. also discovered much fainter stream-like
substructures reaching $\sim 34$ mag
arcsec$^{-2}$ as well as previously unknown surviving dwarf satellites
and peculiarly extending outer globular clusters. The properties of
these faint substructures in the outer halo, such as radial distributions
of surface brightness and metallicity, have been
reproduced by some recent numerical simulations
\citep[e.g.,][]{BJ2005,Abadi2006,Johnston2008}, which thus 
provide us with useful theoretical templates to understand the formation
history of the stellar halo.

Here we report on our wide-field, photometric observations of the M31 halo,
using Subaru/Suprime-Cam, aimed at obtaining the detailed physical properties
of the M31 halo, such as color-magnitude diagrams (CMDs), metallicity
distributions, surface brightness and age, compared with previous studies using
CFHT/MegaCam \citep[e.g.,][]{Ibata2007}. Our
Suprime-Cam survey is deeper and covers as-yet unexplored fields along the north-west
minor axis of the halo as well as some fields in the south-quadrant halo
already explored by other telescopes. A large set of our imaging data provides
an important clue to understanding the complex halo substructures and
their origin in the course of the formation history of the M31 halo.

The layout of this paper is as follows. In \S~\ref{sec:data}, we present
observations and detailed procedure for calibrating the photometric
Suprime-Cam data. \S~\ref{sec:stream} is devoted to our results
for the quantitative analysis of the GSS, including
the morphology of the obtained CMDs, the
distance estimate using the tip of the RGB stars (TRGB), their
metallicity distribution using the observable globular cluster templates
and theoretical isochrones, and their age distribution based on the
luminosity distribution of the stars. In \S~\ref{sec:halo}, we expand
our method to the entire regions of the M31 halo in order to globally
determine the fundamental properties of the M31 halo. Then, we elaborate
the technique for separating the foreground and background contaminants.
The spatial distributions of the stellar populations in the M31 halo are
also presented in the section. Furthermore, we present the CMDs of the
detected streams and other spatial substructures, and show the
radial profiles and the radial metallicity distributions of the stellar 
populations in the halo. Finally, we discuss the implications of our results
and compare with previous studies. In \S~\ref{sec:summary} we draw
conclusions and future directions. 

Throughout this work we adopt the M31 Cepheid distance
modulus of $(m-M)_0 = 24.43 \pm 0.06$ ($\sim$ 770 kpc) from
\citet{Freedman1990}, if not otherwise specified. Reddening corrections
are applied in each field based on the extinction maps of
\citet*{Schlegel1998}, and the \citet{Dean1978} reddening law
$E(V-I)=1.34E(B-V)$ and $A_I=1.31E(V-I)$. We also adopt the convention
that $R$ and $r$ denote, respectively, the projected radius and
three-dimensional radius from the M31 center.

%%%%%%%%%%%%%%%%%%%%%%%%%%%%%%%%%%%%%%%%%%%%%%%%%%%%%%%%%%%%%%%%%%%%%%%%%%%
%%\end{document}
%%%%%%%%%%%%%%%%%%%%%%%%%%%%%%%%%%%%%%%%%%%%%%%%%%%%%%%%%%%%%%%%%%%%%%%%%%%

%%% Section 2 %%%
\section{Data}\label{sec:data}

\subsection{Suprime-Cam Observations}
\begin{figure*}[htpd]
 \epsscale{1}
 \plotone{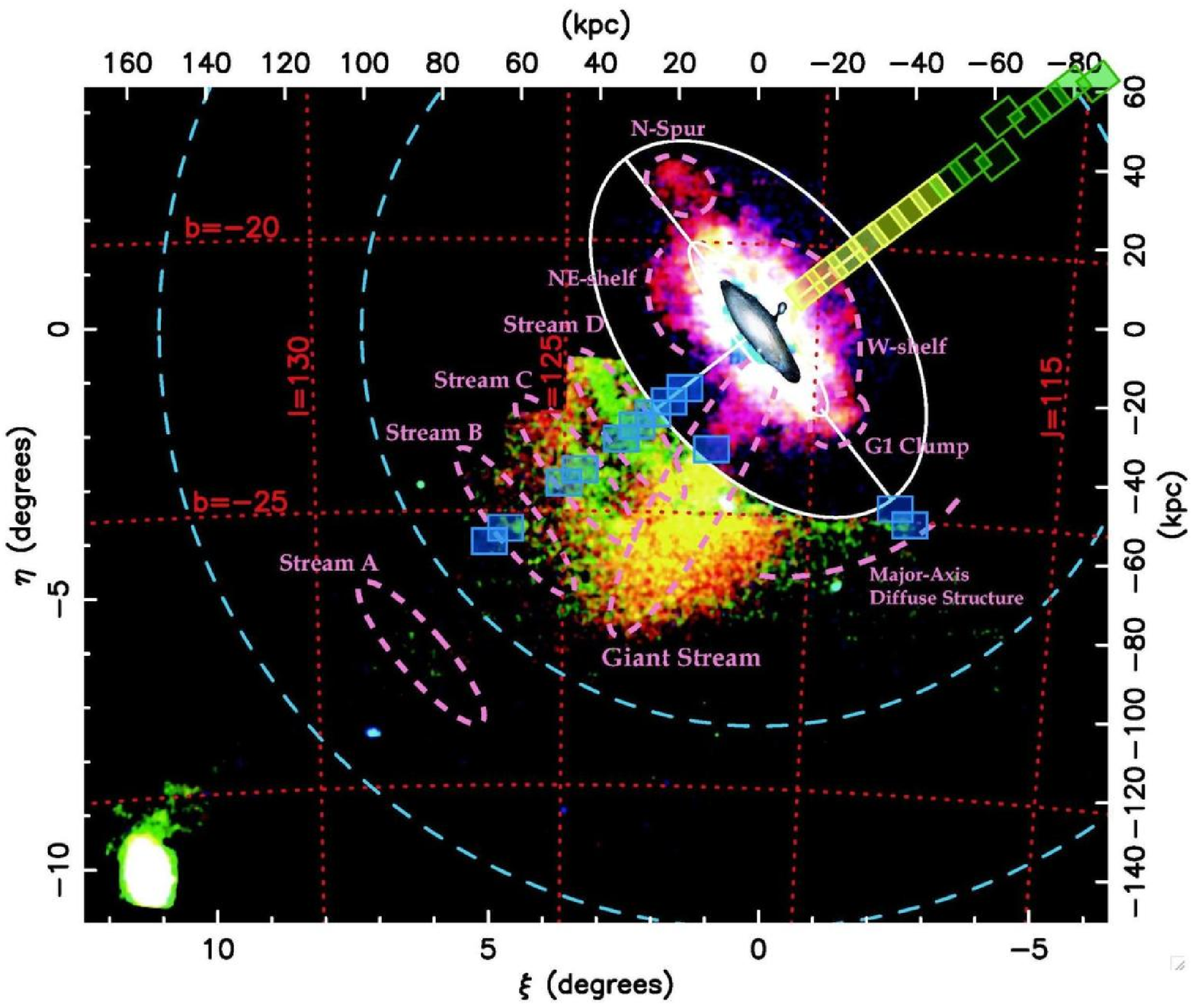}
 \caption[a]{
%
% XXX Suggested changes to the figure: Indicate the R_proj values for the
% dashed circles.  Mark locations of Streams E and F.  Add "SW" before the
% label "Major-Axis Diffuse Substructure".  Change the "Giant Stream" label
% to "Giant Southern Stream".
%
 Locations of our Subaru/Suprime-Cam fields (rectangular areas), overlaid
 on the color-composited stellar density map [taken from Figure 53 of
 the Ibata et al. (2007) paper], in which red, green, and
 blue colors show, respectively stars with $-0.7 < {\rm [Fe/H]} < 0.0$, $-1.7 <
 {\rm [Fe/H]} < -0.7$ and $-3.0 < {\rm [Fe/H]} < -1.7$. Our Suprime-Cam
 fields characterized by blue, yellow, and green rectangles were
 obtained in 2004, 2007 and 2008, respectively. Table~1 presents a log
 of these observations.}
 \label{fig:map}
\end{figure*}

In our observational studies of Andromeda's stellar halo, we use
the Suprime-Cam imager \citep{Miyazaki2002} on the 8.2-m Subaru Telescope
\citep{Iye2004} on Mauna Kea in Hawaii. Suprime-Cam consists of
ten $2045\times4070$ CCDs with a scale of $0\farcs202$ per pixel and
covers a total field-of-view $34\arcmin \times 27\arcmin$. Using
this wide-field imager, we have carried out a systematic imaging survey
of two series of fields far from the M31 center: fields
along the minor and major axis of the spheroid (hereafter referred to
as ``halo'' fields) and fields along the Andromeda GSS (``stream'' fields).
Our targeted fields of the halo and stream fields are, respectively,
located at between 22 and 91~kpc (Figure~\ref{fig:map}) and
at $\sim$ 32~kpc ($2\fdg4$) from the M31 center. 
Tables~\ref{tab:infoOBS} summarizes the names of these targeted fields
and their observations.

During four nights in August 2004, we obtained the Suprime-Cam image of
the one stream field (hereafter referred to as GSS1) and also made
imaging surveys for eleven halo fields in the south part of M31
(SE1--9 and SW1--2). In addition to these targeted fields, we obtained the
imaging data at the control field (hereafter referred to as CompL103Bm21)
to estimate the number of foreground dwarf stars of the Milky Way and
unresolved background galaxies in each M31's field. The weather conditions
in August 2004 were unfortunately slightly poor, with seeing varying from
$0\farcs7-1\farcs5$. Therefore, we carried out additional imaging of
GSS1, SE1, SE6, SE9 and SW1 on a photometric night in August 2005 in
order to calibrate the data. In August 2007 and August 2008, the halo
fields in the north part of M31 (NW1--15) were observed in photometric
conditions, along with the control fields (CompL103Bm25 and
CompL103Bm16\_5). 

The observations were made with Johnson
$V$-band and Cousins $I$-band filters. Typical exposure times of each object
field observed in 2004--2005 and 2007 are 1800 sec and 720 sec in $V$ and $I$ 
band, respectively, and those in 2008 are 1440 and 2160 sec in $V$ and $I$ band, 
respectively, so that the images reach down to magnitude of the RC of M31. 
Each field was observed with adequate dithering pattern to cover gaps between 
adjacent CCD chips. %
Some of these fields were observed in short
exposures of 5 and 10\,s (12 and 15\,s) in $V$ ($I_c$) bands.
They are used to improve photometric accuracy and to
investigate space variation of foreground dwarf stars along the galactic
latitude at the same galactic longitude. In addition, all the targeted
fields in the M31 halo are overlapping with neighboring fields in order to 
calibrate the adjacent fields whose photometric zero points are unknown 
in poor condition, such as SE2, SE3 and so on. 

The position angle of each field observed in 2004--2005 is 0\arcdeg,
i.e., North is up and East is to the left of each image. 
In the 2007--2008 runs,
the position angle of each field is 38\arcdeg, i.e., the longer side 
of each image is aligned to the minor axis of M31
in order to obtain many pointings along the far outer part of M31's halo. 
For more details, see Table~\ref{tab:infoOBS}.

\begin{deluxetable*}{lccccccc}
 \tablewidth{0pt}
 \tabletypesize{\footnotesize}
 \tablecaption{Observational status for target fields\label{tab:infoOBS}}
 \tablehead{
 \colhead{Field} & \colhead{Coordinates} & \colhead{Date} &
 \colhead{Filter} & \colhead{Exposure} & \colhead{Airmass} &
 \colhead{Seeing} & \colhead{50\% completeness}\\
 \colhead{$R_{\rm proj}$} & \colhead{$(\alpha_{2000}/\delta_{2000})$} &
 \colhead{mm/yyyy} & & \colhead{(sec)} & & \colhead{(mag)} &
 }
%
% XXX Add a column giving the position angle (PA) of the fields.  It looks
% like all the 2004 fields had a certain PA (0 degrees?) while all the 2007
% and 2008 fields have a different PA (parallel to M31's disk major axis -
% PA ~ -53 degrees?)
%
 \startdata
 GSS1                 & $00^{\rm h}47^{\rm m}38\fs11$        & 08/2004 &
 $V$ & $600 \times 5$ & 1.7 & $1\farcs07$ & 25.95 \\
 ($\approx 32$ kpc) & $+39{\arcdeg}05{\arcmin}57{\farcs}4$ & 08/2004 &
 $I$ & $300 \times 6$ & 1.2 & $0\farcs92$ & 25.00 \\
%
% XXX Add a footnote here and in the main body of the text explaining why
% the GSS2 field is never used or referred to in the paper.
%
 \hline
 SE1                  & $00^{\rm h}50^{\rm m}09\fs16$        & 08/2004 &
 $V$ & $600 \times 3$ & 1.3 & $1\farcs00$ & 26.40 \\
 ($\approx 24$ kpc) & $+40{\arcdeg}08{\arcmin}36{\farcs}4$ & 08/2004 &
 $I$ & $300 \times 6$ & 1.1 & $0\farcs71$ & 25.66 \\
 SE2                  & $00^{\rm h}51^{\rm m}38\fs06$        & 08/2004 &
 $V$ & $600 \times 5$ & 1.2 & $1\farcs13$ & 26.34 \\
 ($\approx 29$ kpc) & $+39{\arcdeg}55{\arcmin}03{\farcs}4$ & 08/2004 &
 $I$ & $300 \times 6$ & 1.1 & $0\farcs93$ & 25.24 \\
 SE3                  & $00^{\rm h}53^{\rm m}06\fs96$        & 08/2004 &
 $V$ & $600 \times 3$ & 1.4 & $1\farcs03$ & 26.30 \\
 ($\approx 34$ kpc) & $+39{\arcdeg}41{\arcmin}30{\farcs}4$ & 08/2004 &
 $I$ & $300 \times 6$ & 1.1 & $0\farcs82$ & 25.40 \\
 SE4                  & $00^{\rm h}54^{\rm m}35\fs86$        & 08/2004 &
 $V$ & $600 \times 5$ & 1.3 & $1\farcs13$ & 26.34 \\
 ($\approx 39$ kpc) & $+39{\arcdeg}27{\arcmin}57{\farcs}4$ & 08/2004 &
 $I$ & $300 \times 6$ & 1.1 & $0\farcs87$ & 25.27 \\
 SE5                  & $00^{\rm h}56^{\rm m}04\fs76$        & 08/2004 &
 $V$ & $600 \times 5$ & 1.5 & $1\farcs21$ & 26.24 \\
 ($\approx 44$ kpc) & $+39{\arcdeg}14{\arcmin}24{\farcs}4$ & 08/2004 &
 $I$ & $300 \times 6$ & 1.1 & $0\farcs99$ & 25.03 \\
 SE6                  & $01^{\rm h}00^{\rm m}04\fs68$        & 08/2004 &
 $V$ & $600 \times 2$ & 1.6 & $1\farcs05$ & 26.12 \\
 ($\approx 55$ kpc) & $+38{\arcdeg}54{\arcmin}58{\farcs}5$ & 08/2004 &
 $I$ & $300 \times 6$ & 1.1 & $1\farcs00$ & 25.37 \\
 SE7                  & $01^{\rm h}01^{\rm m}31\fs73$        & 08/2004 &
 $V$ & $600 \times 2$ & 1.4 & $1\farcs01$ & 26.12 \\
 ($\approx 60$ kpc) & $+38{\arcdeg}41{\arcmin}25{\farcs}5$ & 08/2004 &
 $I$ & $300 \times 5$ & 1.1 & $0\farcs89$ & 25.09 \\
 SE8                  & $01^{\rm h}09^{\rm m}04\fs58$        & 08/2004 &
 $V$ & $600 \times 3$ & 1.4 & $0\farcs94$ & 26.48 \\
 ($\approx 84$ kpc) & $+37{\arcdeg}37{\arcmin}39{\farcs}5$ & 08/2004 &
 $I$ & $300 \times 6$ & 1.1 & $0\farcs74$ & 25.77 \\
 SE9                  & $01^{\rm h}10^{\rm m}30\fs09$        & 08/2004 &
 $V$ & $600 \times 3$ & 1.2 & $1\farcs00$ & 26.47 \\
 ($\approx 89$ kpc) & $+37{\arcdeg}24{\arcmin}06{\farcs}5$ & 08/2004 &
 $I$ & $300 \times 6$ & 1.1 & $0\farcs71$ & 25.77 \\
 \hline
 SW1                  & $00^{\rm h}29^{\rm m}16\fs28$        & 08/2004 &
 $V$ & $600 \times 3$ & 1.6 & $1\farcs08$ & 26.31 \\
 ($\approx 58$ kpc) & $+37{\arcdeg}49{\arcmin}40{\farcs}4$ & 08/2004 &
 $I$ & $300 \times 6$ & 1.1 & $0\farcs75$ & 25.58 \\
 SW2                  & $00^{\rm h}28^{\rm m}12\fs40$        & 08/2004 &
 $V$ & $600 \times 3$ & 1.6 & $0\farcs98$ & 26.42 \\
 ($\approx 63$ kpc) & $+37{\arcdeg}33{\arcmin}23{\farcs}4$ & 12/2005 &
 $I$ & $300 \times 7$ & 1.1 & $0\farcs98$ & 25.23 \\
 \hline
 NW1                  & $00^{\rm h}38^{\rm m}07\fs50$        & 08/2007 &
 $V$ & $240 \times 3$ & 1.4 & $0\farcs65$ & 25.99 \\
 ($\approx 15$ kpc) & $+41{\arcdeg}56{\arcmin}47{\farcs}0$ & 08/2007 &
 $I$ & $180 \times 4$ & 1.3 & $0\farcs67$ & 24.95 \\
 NW2                  & $00^{\rm h}36^{\rm m}12\fs20$        & 08/2007 &
 $V$ & $240 \times 3$ & 1.1 & $0\farcs69$ & 26.18 \\
 ($\approx 21$ kpc) & $+42{\arcdeg}13{\arcmin}31{\farcs}8$ & 08/2007 &
 $I$ & $180 \times 4$ & 1.1 & $0\farcs65$ & 25.35 \\
 NW3                  & $00^{\rm h}34^{\rm m}16\fs90$        & 08/2007 &
 $V$ & $240 \times 3$ & 1.1 & $0\farcs69$ & 26.34 \\
 ($\approx 27$ kpc) & $+42{\arcdeg}30{\arcmin}16{\farcs}6$ & 08/2007 &
 $I$ & $180 \times 4$ & 1.2 & $0\farcs59$ & 25.87 \\
 NW4                  & $00^{\rm h}32^{\rm m}21\fs60$        & 08/2007 &
 $V$ & $240 \times 3$ & 1.1 & $0\farcs63$ & 26.74 \\
 ($\approx 33$ kpc) & $+42{\arcdeg}47{\arcmin}01{\farcs}4$ & 08/2007 &
 $I$ & $180 \times 4$ & 1.1 & $0\farcs61$ & 25.74 \\
 NW5                  & $00^{\rm h}30^{\rm m}26\fs30$        & 08/2007 &
 $V$ & $240 \times 3$ & 1.1 & $0\farcs61$ & 26.72 \\
 ($\approx 39$ kpc) & $+43{\arcdeg}03{\arcmin}46{\farcs}2$ & 08/2007 &
 $I$ & $180 \times 4$ & 1.1 & $0\farcs61$ & 25.74 \\
 NW6                  & $00^{\rm h}28^{\rm m}31\fs00$        & 08/2007 &
 $V$ & $240 \times 3$ & 1.1 & $0\farcs59$ & 26.63 \\
 ($\approx 45$ kpc) & $+43{\arcdeg}20{\arcmin}31{\farcs}0$ & 08/2007 &
 $I$ & $180 \times 4$ & 1.1 & $0\farcs65$ & 25.60 \\
 NW7                  & $00^{\rm h}26^{\rm m}35\fs70$        & 08/2007 &
 $V$ & $240 \times 3$ & 1.3 & $0\farcs65$ & 26.46 \\
 ($\approx 51$ kpc) & $+43{\arcdeg}37{\arcmin}15{\farcs}8$ & 08/2007 &
 $I$ & $180 \times 4$ & 1.2 & $0\farcs57$ & 25.93 \\
 \hline
 NW8                  & $00^{\rm h}24^{\rm m}25\fs10$        & 08/2008 &
 $V$ & $240 \times 5$ & 1.1 & $1\farcs07$ & 25.70 \\
 ($\approx 57$ kpc) & $+43{\arcdeg}52{\arcmin}56{\farcs}5$ & 08/2008 &
 $I$ & $360 \times 6$ & 1.1 & $0\farcs87$ & 25.16 \\
 NW9                  & $00^{\rm h}22^{\rm m}26\fs10$        & 08/2008 &
 $V$ & $240 \times 5$ & 1.1 & $1\farcs07$ & 25.72 \\
 ($\approx 63$ kpc) & $+44{\arcdeg}09{\arcmin}41{\farcs}3$ & 08/2008 &
 $I$ & $360 \times 6$ & 1.1 & $0\farcs89$ & 25.02 \\
 NW10                 & $00^{\rm h}19^{\rm m}58\fs60$        & 08/2008 &
 $V$ & $240 \times 8$ & 1.1 & $1\farcs05$ & 25.94 \\
 ($\approx 70$ kpc) & $+44{\arcdeg}20{\arcmin}03{\farcs}1$ & 08/2008 &
 $I$ & $360 \times 6$ & 1.1 & $0\farcs71$ & 25.47 \\
 NW11                 & $00^{\rm h}18^{\rm m}23\fs20$        & 08/2008 &
 $V$ & $240 \times 7$ & 1.1 & $0\farcs99$ & 25.98 \\
 ($\approx 78$ kpc) & $+44{\arcdeg}59{\arcmin}02{\farcs}1$ & 08/2008 &
 $I$ & $360 \times 6$ & 1.1 & $0\farcs71$ & 25.62 \\
 NW12                 & $00^{\rm h}16^{\rm m}30\fs30$        & 08/2008 &
 $V$ & $240 \times 29$ & 1.2 & $0\farcs71$ & 27.06 \\
 ($\approx 82$ kpc) & $+45{\arcdeg}00{\arcmin}59{\farcs}4$ & 08/2008 &
 $I$ & $360 \times 18$ & 1.1 & $0\farcs67$ & 25.94 \\
 NW13                 & $00^{\rm h}14^{\rm m}29\fs10$        & 08/2008 &
 $V$ & $240 \times 6$ & 1.2 & $0\farcs63$ & 26.88 \\
 ($\approx 88$ kpc) & $+45{\arcdeg}17{\arcmin}44{\farcs}1$ & 08/2008 &
 $I$ & $360 \times 6$ & 1.1 & $0\farcs67$ & 25.60 \\
 NW14                 & $00^{\rm h}12^{\rm m}27\fs20$        & 08/2008 &
 $V$ & $240 \times 6$ & 1.2 & $0\farcs67$ & 26.60 \\
 ($\approx 94$ kpc) & $+45{\arcdeg}34{\arcmin}28{\farcs}9$ & 08/2008 &
 $I$ & $360 \times 6$ & 1.1 & $0\farcs75$ & 25.53 \\
 NW15                 & $00^{\rm h}09^{\rm m}46\fs50$        & 08/2008 &
 $V$ & $240 \times 6$ & 1.1 & $0\farcs75$ & 26.80 \\
 ($\approx 100$ kpc) & $+45{\arcdeg}42{\arcmin}43{\farcs}1$ & 08/2008 &
 $I$ & $360 \times 6$ & 1.1 & $0\farcs75$ & 25.57 \\
 \hline
 CompL103Bm21                  & $23^{\rm h}17^{\rm m}00\fs00$
 & 08/2007 &
 $V$ & $240 \times 3$ & 1.1 & $0\farcs69$ & 26.43 \\
 $[l=103\arcdeg,b=-21\arcdeg]$ & $+38{\arcdeg}32{\arcmin}00{\farcs}0$
 & 08/2007 &
 $I$ & $180 \times 4$ & 1.1 & $0\farcs53$ & 26.13 \\
 CompL103Bm25                  & $23^{\rm h}25^{\rm m}30\fs00$
 & 08/2008 &
 $V$ & $240 \times 8$ & 1.1 & $0\farcs71$ & 26.53 \\
 $[l=103\arcdeg,b=-25\arcdeg]$ & $+34{\arcdeg}29{\arcmin}50{\farcs}0$
 & 08/2008 &
 $I$ & $360 \times 6$ & 1.0 & $0\farcs67$ & 25.76 \\
 CompL103Bm16\_5                  & $23^{\rm h}06^{\rm m}30\fs00$
  & 08/2008 &
 $V$ & $240 \times 22$ & 1.2 & $0\farcs71$ & 27.15 \\
 $[l=103\arcdeg,b=-16.5\arcdeg]$ & $+42{\arcdeg}13{\arcmin}00{\farcs}0$
 & 08/2008 &
 $I$ & $360 \times 27$ & 1.1 & $0\farcs71$ & 26.13 \\
  \enddata
\end{deluxetable*}

\subsection{Reduction and Photometry}\label{sec:reduction}
The raw data were reduced in the standard procedures, with the software
package SDFRED, a useful pipeline developed to optimally deal
with Suprime-Cam images \citep{Yagi2002,Ouchi2004}. 
Using SDFRED, we obtained a mean-stacked image for each field, 
which covers $34\arcmin \times 27\arcmin$ field-of-view. 

We then conducted PSF-fitting photometry using the IRAF version of the
DAOPHOT-I\hspace{-.1em}I software \citep{Stetson1987}. We adopted a
3~$\sigma$ detection threshold for the initial object
detection/photometry pass and repeated the PSF-fitting photometry twice
with 5~$\sigma$ and 7~$\sigma$ detection thresholds, respectively, in
order to account for blended stars. A stellar PSF template was constructed 
from 10--100 bright, not saturated, isolated stellar objects per
image. Finally, we merged two independent $V$-band and $I$-band catalogues
into a combined catalogue using 2-pixel matching radius. Note that we
cannot detect objects near the edge of a field in case of adopting a
1-pixel matching radius. 

When conducting photometry, we divide a Suprime-Cam field into a few
subfields and photomery is carried out for each subfield separately. 
For example, for fields with PA$=0\arcdeg$ we divide them to
four subfields (labeled by ``a'', ``b'', ``c'' and ``d''), while for
fields with PA$=38\arcdeg$, we divide them to two subfields (one is not
labeled and the other is labeled by ``d''). 
If we perform detection and photometry using DAOPHOT without dividing 
to subfields, stellar-like objects near the edge of field-of-view 
($34\arcmin \times 27\arcmin$) are not properly detected because the
actual PSF is largely varying depending on the location. Even if we
consider the space variation of PSF by means of making a model PSF function
to detect objects ($PSF$ task), it results in inducing erratic values
of $sharpness$ and $chi$ in some DAOPHOT parameters. Actually, when we
have detected objects without dividing one field-of-view, we have 
failed to detect approximately 20\% of stellar samples of M31 red giants and
Milky Way dwarf stars identified by Keck/DEIMOS spectroscopic
observations (see \S\ref{sec:keck}).

\subsection{The Artificial Star Experiments}
\subsubsection{Completeness}
To evaluate incompleteness due to low signal-to-noise ratio and crowding, 
We have conducted the artificial star experiments as detailed below. 
We have added a sufficient number of artificial stars to all the
original images, where each star has the PSF as determined from many
bright stars in each frame in \S~\ref{sec:reduction}, and the range of the
magnitudes is $20 \leq V, I \leq 28$ with a binning step of 0.2~mag. Note
that for a portion of the large images the binning is 0.5~mag in the brighter
parts with $V = 23$~mag. To prevent these stars from interfering with one another, 
we have divided the Suprime-Cam frames into grids of cells of $200$ pixels
width and have randomly added one artificial star to each cell for each run. 
In addition, we constrain each artificial star to have 20 pixels
from the edges of the cell. 
We note that a similar procedure to ours has recently been
adopted by \citet{PZ1999} and \citet{Bellazzini2002}. In this way we can
control the minimum distance between adjacent artificial stars. At each
run the absolute position of the grid is randomly changed in a way that,
after a large number of experiments, the stars are uniformly distributed
in coordinates \citep[see also][]{Tosi2001}.

The stars, which have a Penny2 function expressed as the analytic component
of the PSF model computed by the DAOPHOT-I\hspace{-.1em}I PSF task and are
quadratically variable over the image, were added on the original frame
including Poisson photon noise. Each star has been added to the $V$ and $I$ 
coadded frames. For the artificial frames we have
performed detection and photometry following the method described in
\S~\ref{sec:reduction} and extracted stellar-like objects by applying
our established selection criteria based on DAOPHOT parameters (see
\S~\ref{sec:daopars}). 

\begin{figure}[htpd]
 \begin{center}
  \epsscale{1.0}
  \plotone{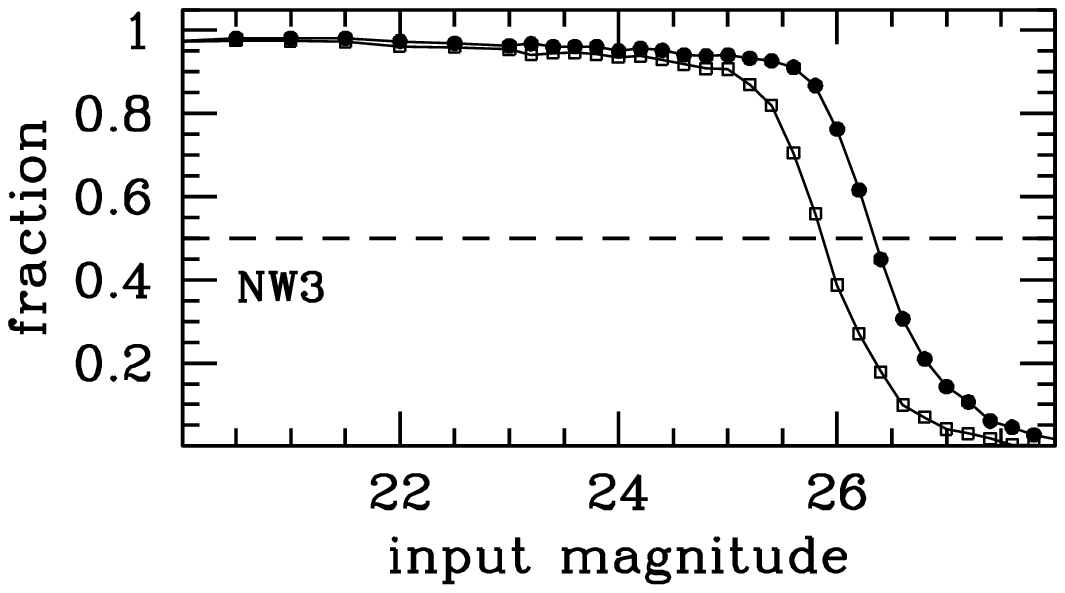}
  \caption[a]{Completeness fraction ($=N_{\rm output}/N_{\rm input}$) as
  a function of input $V$ (filled circles) and $I$ (open rectangles)
  magnitudes in the NW3 field as estimated by the current artificial
  star experiments.}
  \label{fig:completeNW3}
 \end{center}
\end{figure}

About 500 to 4300 stars,  were configured for each run, and the total number
is about 18,000 to 140,000 in each band image, where the number of added
artificial stars depends on the subfield size.  After the above-mentioned
detection/photometry and selection, the final recovered objects were
regarded as stars, if they reside within 2 pixel of the position of the
added star and their magnitudes are within 0.75 mag of the assigned
values. In this task, the completeness is defined as a number fraction
of recovered stars in added stars. In 
Figure~\ref{fig:completeNW3}, we plot the
completeness fractions ($=N_{\rm output}/N_{\rm input}$) as a function
of input $V$ (filled circles) and $I$ (open rectangles)
magnitudes in the NW3 field as estimated by the current artificial star
experiments. Note that the continuous line is not a fit to the data but
is simply the line connecting them. Alphabetical letters labeled in each
field of the south quadrant area of the M31 halo, such as ``a'', ``b'',
``c'' and ``d'' represent the subfields which are obtained by dividing a single
Suprime-Cam field into four subfields. In these subfields, the 25\%
overlapping regions (labeled by ``c'' for SE and SW fields and by ``d''
for GSS fields) indicate slightly deep completeness limits. In contrast,
the regions which are overlapping the nearby fields along the
North-West minor axis of the M31 halo are labeled by ``d'', 
meaning ``deep''.  The rightmost column of Table~\ref{tab:infoOBS} shows
the 50\% completeness limits of V- and I-band in each subfield. In all the
fields the completeness is larger than 80\% at $I \leq 24.5$, and our
imaging data are much deeper than the previous studies using ground-based
telescopes, e.g. \citet{McConnachie2003},
\citet{Durrell2001,Durrell2004} and \citet{Ibata2007}. The 50\%
completeness is reached at $V \geq 25.70$ and $I \geq 24.85$ in all the fields. 

\subsubsection{Selection of Stellar-Like Objects Based on DAOPHOT Parameters}\label{sec:daopars}
In extracting stellar-like objects from the catalogs containing all
the detected objects, we apply selection criteria based on DAOPHOT
parameters such as the $sharpness$, $chi$, and  the photometric
error, which are simulated in the above artificial star experiments. 
The $sharpness$ parameter estimates the intrinsic angular size of the
measured objects. It is roughly defined as the difference between the
square of the width of the object and the square of the width of PSF and
has values close to zero for single stars, large positive values for
blended doubles and partially resolved galaxies, and large negative
values for cosmic rays and blemishes. The $niter$ parameter is the
number of iterations the solution required to achieve convergence. The
$chi$ parameter assesses the estimated goodness-of-fit. It is the ratio
of the observed pixel-to-pixel mean absolute deviation from the profile
fit, to the value expected on the basis of the noise as determined from
Poisson statistics and the readout noise. 

\begin{figure*}[htpd]
 \begin{center}
  \epsscale{1}
  \plottwo{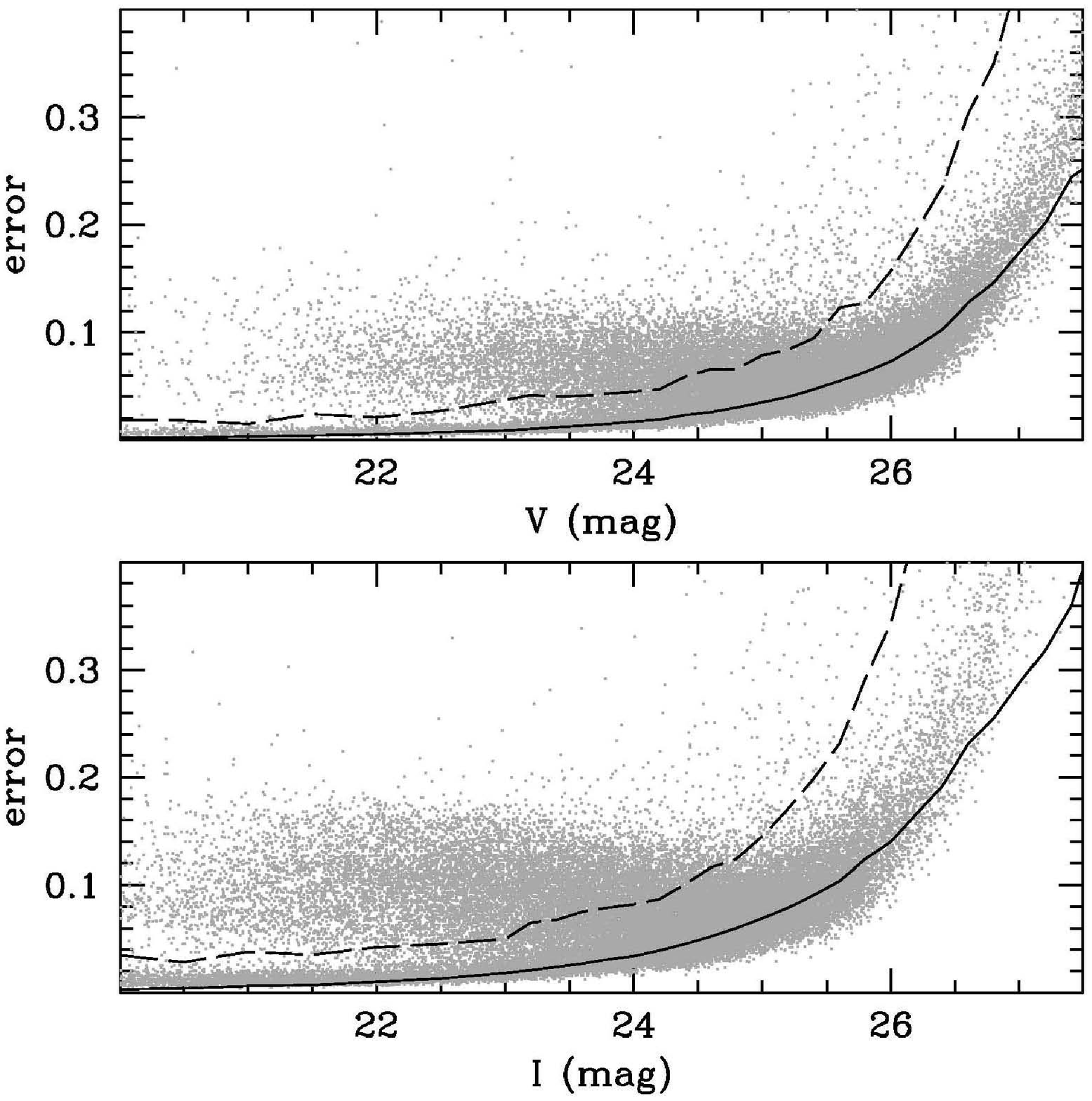}{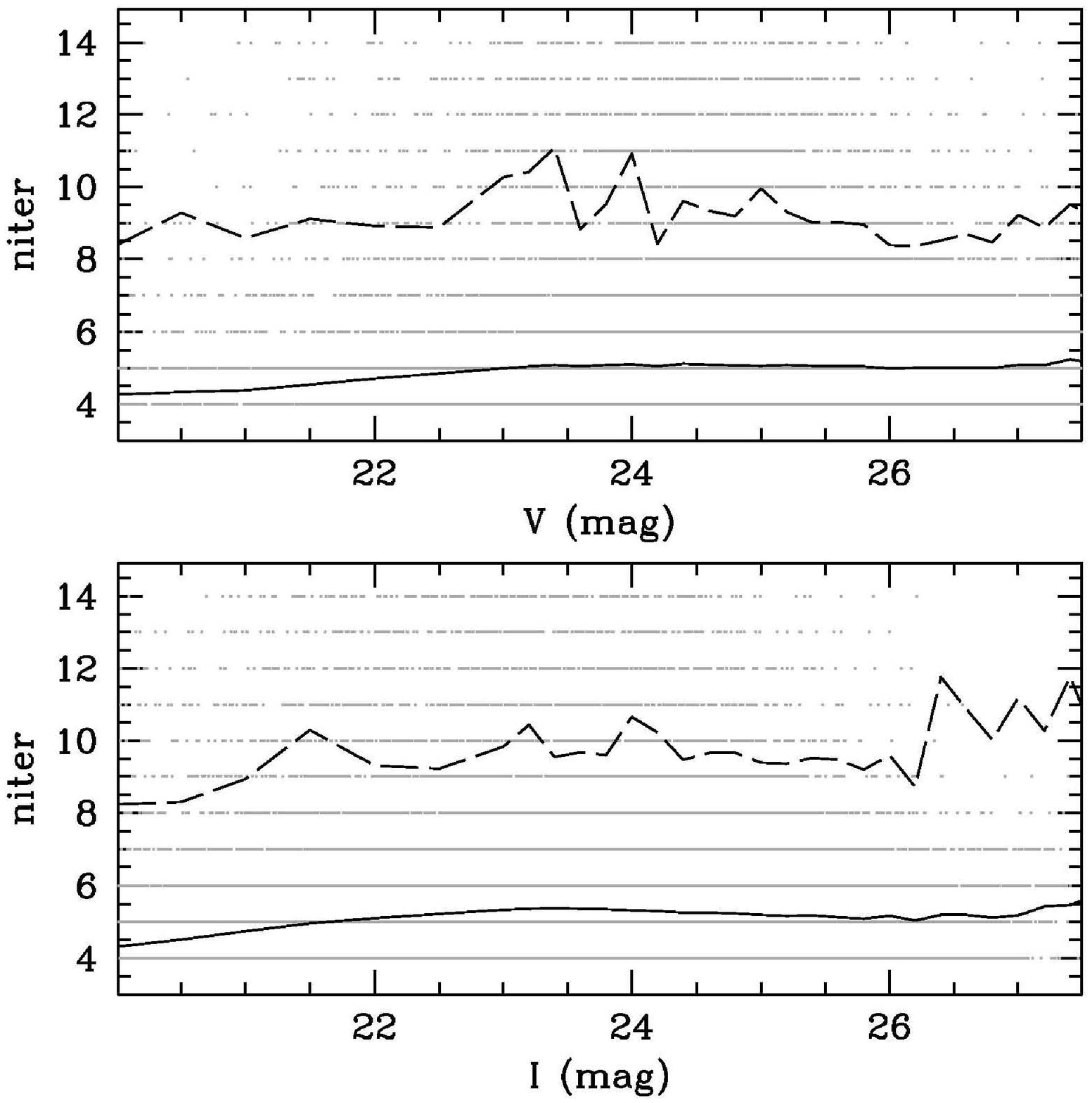}
  \plottwo{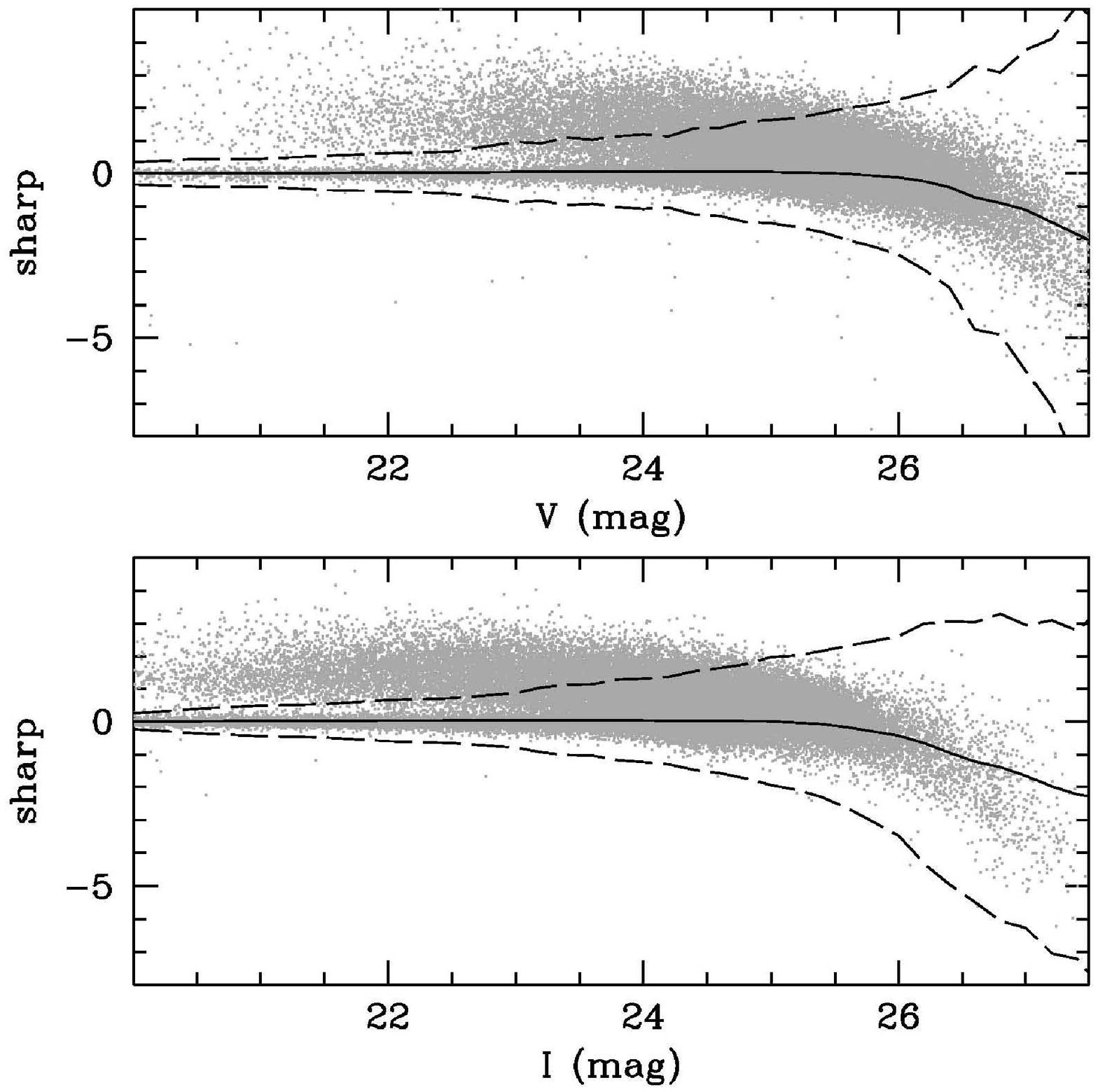}{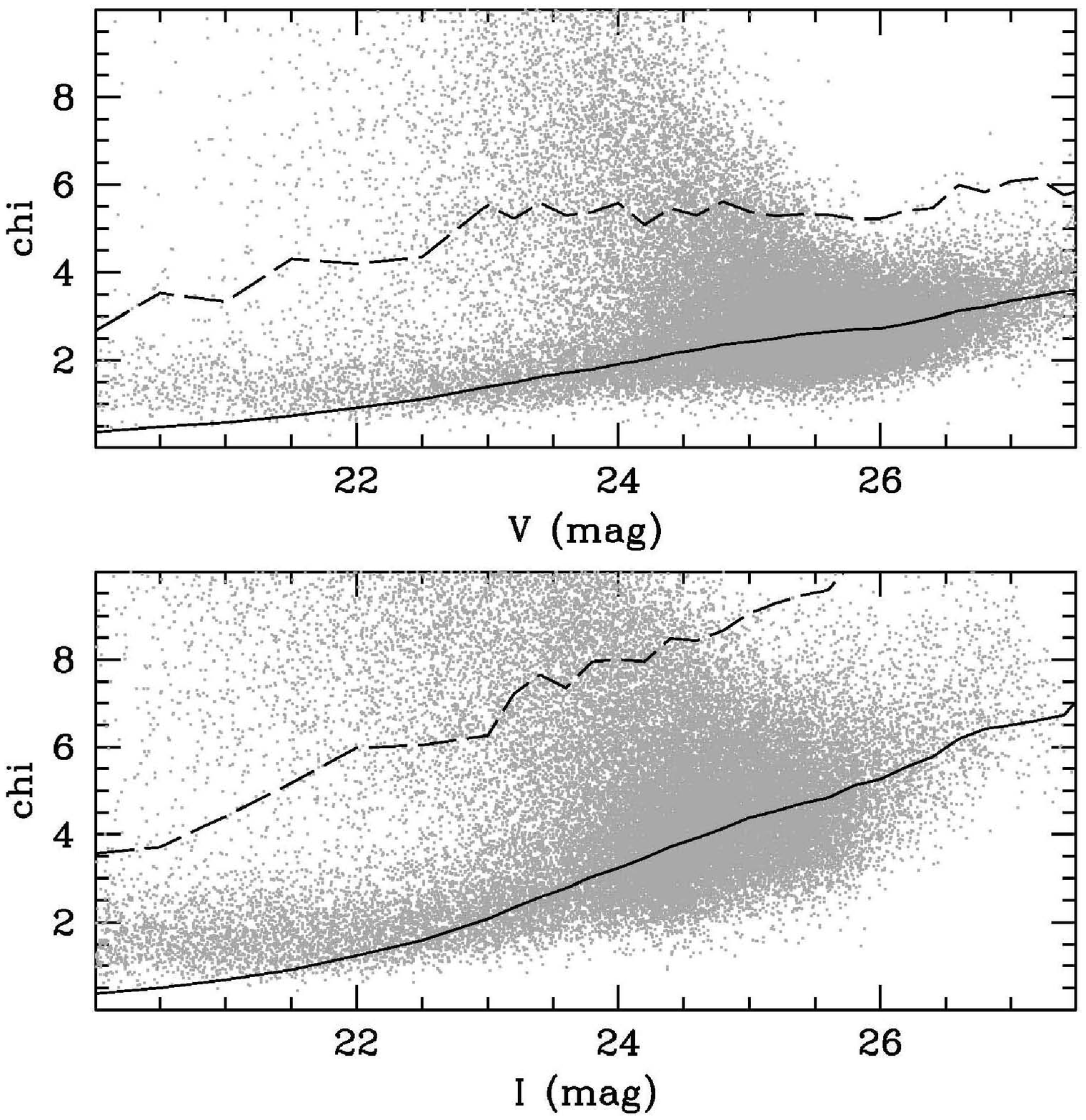}
  \caption[a]{Photometry selection criteria based on the photometric error,
  niter, sharpness and chi parameters from DAOPHOT programs are plotted
  as functions of $V$ and I magnitudes. All of the detected objects after
  matching up to separate $V$ and I catalogs derived from the
  observation of NW3 field are illustrated as light gray dots. The
  solid lines show returned mean values of each simulated parameter,
  and the dashed lines denote the  4 sigma deviations. We select
  final objects within the 4 sigma deviations.}
  \label{fig:art_daopars}
 \end{center}
\end{figure*}

The simulation results of the four parameters to select stellar-like
objects are illustrated in Figure~\ref{fig:art_daopars}. The solid
and dashed lines show mean value and 4 $\sigma$ deviations
of each parameter measured for artificial stars, 
respectively. We choose as ``good'' stars those
sources lying within two hyperbolic envelopes around the sharpness value
of zero (Fig.~\ref{fig:art_daopars} bottom left panel), that is, within
the 4 sigma deviation drawn by the dashed lines, below the 4 sigma
lines of the niter value (top right panel) and the chi value (bottom
right panel), and to the right of the maximum allowed $\sigma$ error at a
given magnitude (top left panel). The remaining cosmic-ray blemishes and
resolved background galaxies were then rejected. However, this sigma
criterion slightly changes with the sky condition because this simulation
is prone to misclassify nearly point-like background galaxies as
stellar objects under the worse seeing condition. We have tested the
seeing effect by comparing the overlapping region between the NW7 field
with good seeing and the NW8 field with bad seeing. To adjust the surface
brightness of the NW8 subfield to that of the NW7 subfield, we have
found a 1 sigma shift of our adopted selection criteria. Therefore, we
apply the 3 sigma criteria for the poor data with the larger seeing size
than 0\farcs8. 

However, this morphological segregation method cannot completely
distinguish M31 halo stars from other point sources such as
Galactic foreground dwarf stars, and compact extragalactic
objects. 
Thus, we have to remove these contaminations from the targeted fields 
statistically. To estimate the contaminations, 
we adopted a control field located at the Galactic
latitude comparable to that of our targeted fields, on the assumption
that it has almost the same abundance of foreground and background
objects as that in the M31 spheroid fields. Actually, we found uniform
galaxy distribution in our control fields. As described in
\S~\ref{sec:comp}, to further refine our control field, we will consider
the finite spatial gradient in the number of foreground disk dwarf stars
of the Milky Way along the Galactic latitude.

\subsubsection{Photometric Errors}
\begin{figure*}[htpd]
 \begin{center}
  \epsscale{1}
  \plotone{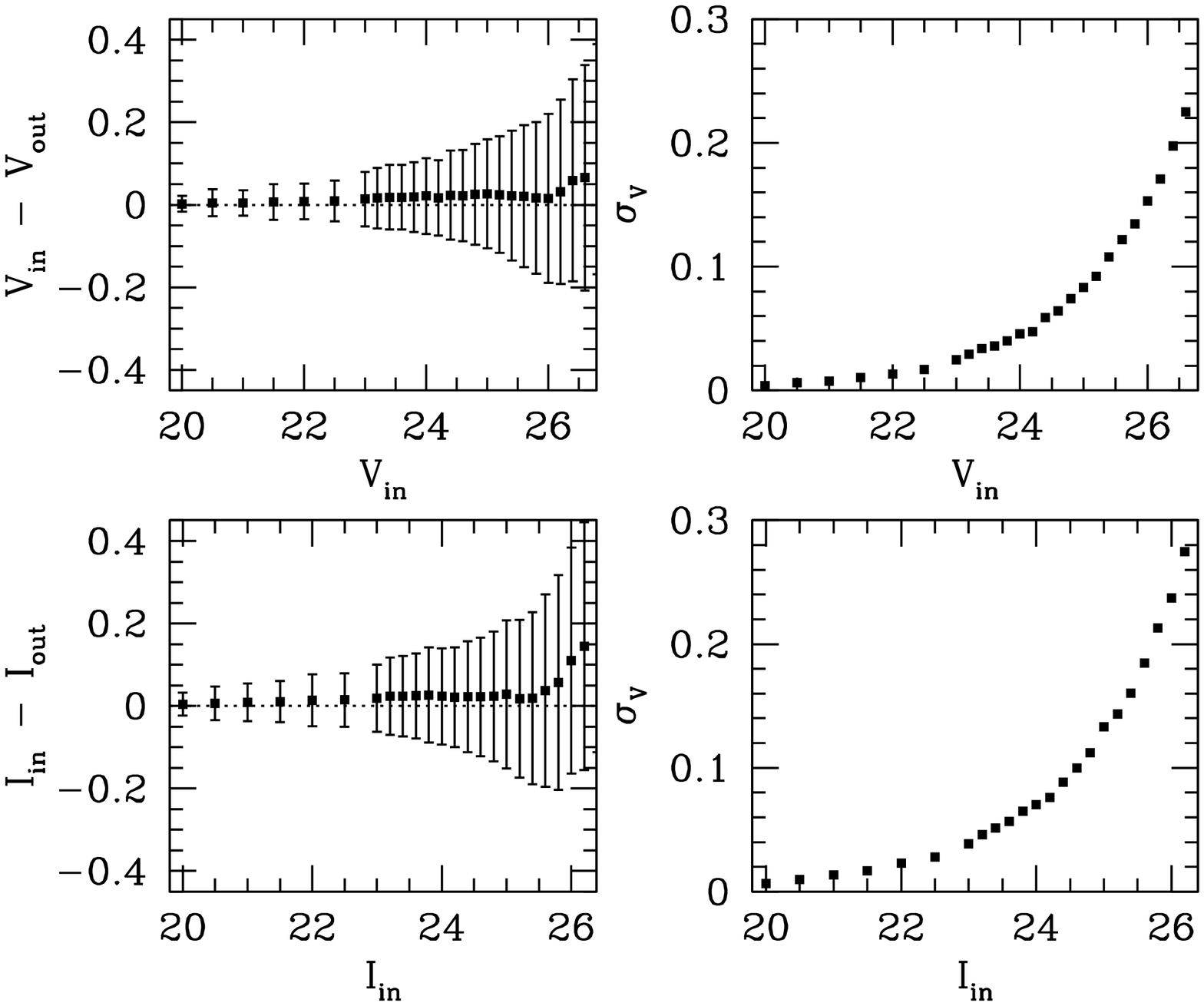}
%
% XXX Italicize V and I in the axis labels.  Check other figures too (e.g.
% Fig 7).
%
  \caption[a]{Typical mean error as a function of magnitude for NW3
  field. On the left we plot difference between the measured and input
  magnitude as a function of input magnitude for all the stars from our
  completeness simulations, while the panels on the right show mean
  error as a function of magnitude as measured from these
  simulations. Only those simulated stars that were recovered within 2
  pixels of the input position and that had magnitudes within 0.75 of
  the input magnitude are considered to be recovered. There is no
  systematic shift of the mean magnitude difference, and the typical 1
  $\sigma$ errors at the 50\% completeness magnitude limits are of the
  order of 0.3 mag.}
  \label{fig:perror}
 \end{center}
\end{figure*}

These artificial star experiments are implemented to ascertain not only
the completeness but also more accurate photometric errors in our data. 
These are quantified as the average of differences between input and
output magnitudes of an artificial star at a specific magnitude, shown
in Figure~\ref{fig:perror}. The mean magnitude difference $\Delta {\rm mag}
={\rm mag}_{\rm input}-{\rm mag}_{\rm output}$ is consistent with 0 for
magnitudes brighter than a 80\% completeness limit ($V \sim 26$ and $I
\sim 25.5$ in case of the NW3 field). The shift toward
systematically brighter measured magnitudes, which signals the
occurrence of blending of the stellar images, is smaller than 0.09 mag at
the magnitude bin corresponding to 50\% completeness level. The right
panels of Figure~\ref{fig:perror} show mean error as a function of
magnitude as measured from these simulations.

\subsection{Comparison with Published Data}
\subsubsection{Detection featuring Keck/DEIMOS Spectroscopic Samples}\label{sec:keck}
In SE3 field, there are some M31's RGB stars identified by
\citet{Gilbert2006}. Using the DEIMOS instrument on the
Keck\,I\hspace{-.1em}I 10m telescope, their RGB samples have been
confirmed based on five criteria: radial velocity, photometry in the
intermediate-width DDO51 band to measure the strength of the MgH/Mg$b$
absorption features, strength of the Na${\rm I}~\lambda$8190 absorption line
doublet, location within an ($V-I, I$) color-magnitude diagram, and
comparison of photometric (color-magnitude diagram based) versus
spectroscopic (Ca${\rm I\hspace{-.1em}I}~\lambda$8500 triplet based)
metallicity estimates. Therefore, to test the validity of our detection
method, we match up our photometric catalogs with their spectroscopic
catalogs using a 10-pixel matching radius. 

\begin{figure}[htpd]
 \begin{center}
  \epsscale{1}
  \plotone{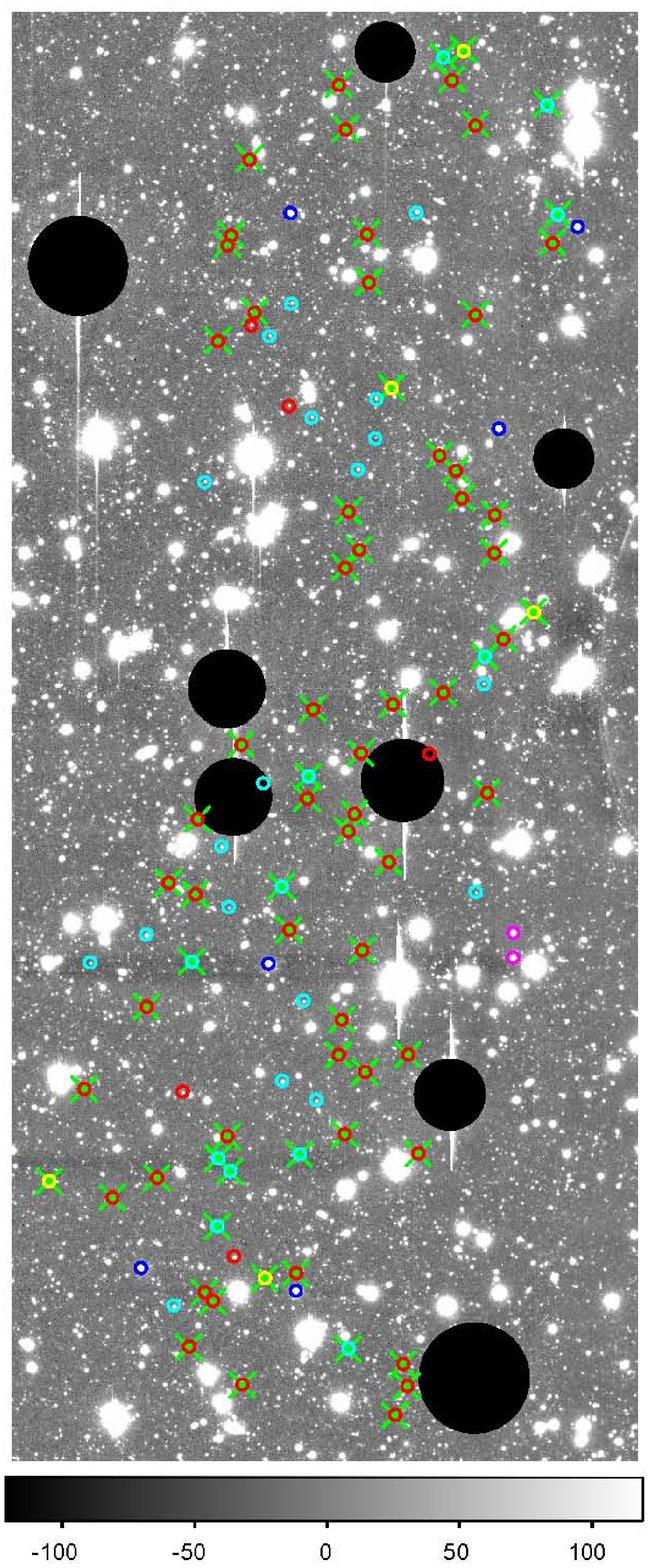}
  \caption[a]{A small section of the SE3 field. This is a V-band coadded image 
  with the total exposure time of 30 min and covers a field-of-view 
  $7.6\arcmin \times 17.5\arcmin$.
  }
  \label{fig:spec_check}
 \end{center}
\end{figure}

Figure~\ref{fig:spec_check} shows a small section of the SE3 field ($2250
\times 5200$ pixels, the dotted rectangle in
Figure~\ref{fig:map_KPNO}). This field is located close to the center of
the DEIMOS spectroscopic field, at $\alpha_{2000} = 00^{\rm h}54^{\rm
m}08\fs34, \delta_{2000} = +39\arcdeg41\arcmin51\farcs7$. The colored
circles denote those objects whose spectra are measured by Keck: M31
red giants and Milky Way dwarf stars (red, 61 objects), background
galaxies/QSOs (cyan, 31 objects), alignment stars (blue, 6 objects) used
to carry out fine astrometric alignment of the DEIMOS multislit mask,
spectroscopic failures (yellow, 5 objects) and guide stars (magenta, 2
objects) only used for coarse astrometric alignment - no spectra
are obtained for them. In contrast, green X points show our Subaru
objects corresponding to all the Keck samples in our final star catalog. 

The number of stellar-like objects detected by our Subaru
observation is 56 objects, and the number of missing objects is less than
7\% of the total. Note that one out of 61 spectroscopic stellar-like objects
is in the masked region, and therefore the total number is
60 objects. Figure~\ref{fig:diff_phot} shows the
magnitude difference between these different systems. The 4 missing objects are
removed from the final catalog because they have somewhat large $sharpness$
and $chi$ values in their magnitude. 

We have also detected 12 objects classified as background galaxies/QSOs
by the Keck spectroscopic observation. This is 40\% of the total number,
and the contaminations attributable to these background objects are
moderately removed. This suggests that all the background contaminations
are never rejected using only DAOPHOT morphological parameters.
Therefore, the remaining background contaminations are removed in comparison 
to the data of Control Fields located outside M31's halo (for more details, see
\S~\ref{sec:daopars}). 

\subsubsection{Photometry featuring KPNO 4m Telescope Samples}\label{sec:KPNO}
In this section, we compare our photometric data with other
photometric data taken with the Mosaic camera on the Kitt Peak National
Observatory (KPNO) 4m telescope in the Washington System $M$ and
$T_2$. Photometric transformation relations from \citet{Majewski2000}
were used to derive Johnson-Cousins $V$ and $I$ magnitudes from the $M$
and $T_2$ magnitudes. The color transformation follows:
\begin{equation}
 (V-I) = (V - T_{\rm 2}) = -0.006 + 0.008(M-T_{\rm 2}).
\end{equation}
With the assumption $I = T_{\rm 2}$ and the above relation, we 
converted to a $V$ magnitude
\begin{equation}
 V = M - 0.006 - 0.200(M-T_{\rm 2}).
\end{equation}

\begin{figure}[htpd]
 \begin{center}
  \epsscale{1}
  \plotone{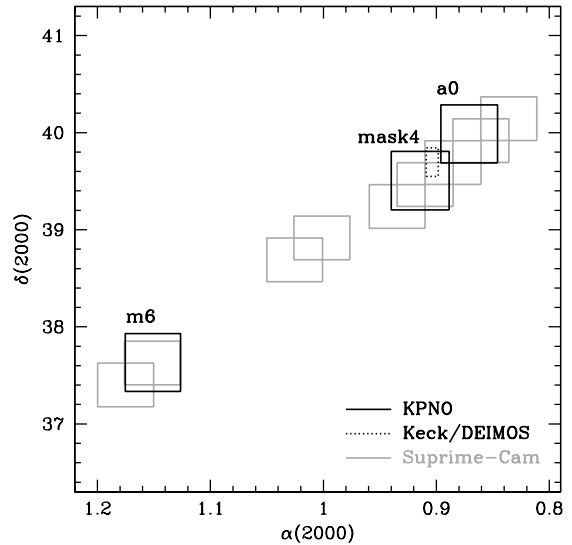}
  \caption[a]{Sky positions observed by three different telescopes,
  which they are used to evaluate efficiency of our
  Detection/Photometry. The gray rectangles
  represents the position and area of the Subaru/Suprime-Cam image. The
  solid squares indicate the approximate positions of the KPNO fields. The
  dotted vertically long rectangle approximates the size and position
  angle of the DEIMOS spectroscopic slitmask.}
  \label{fig:map_KPNO}
 \end{center}
\end{figure}

Figure~\ref{fig:map_KPNO} shows sky positions observed by three different
telescopes, which are used to evaluate the efficiency of our detection/photometry.
%%The black rectangles represents the position and
%5area of the Subaru/Suprime-Cam image. The red rectangles indicate the
%%approximate positions of the KPNO fields. The blue vertically long rectangle
%%approximates the size and position angle of the DEIMOS spectroscopic slitmask.
The ``a0'' field is associated with our SE1, SE2 and SE3
fields, and the ``mask4'' field covers our SE3, SE4 and SE5 fields, and
the ``m6'' field overlaps our SE8 and SE9 fields. 

\begin{figure*}[htpd]
 \begin{center}
  \epsscale{1}
  \plotone{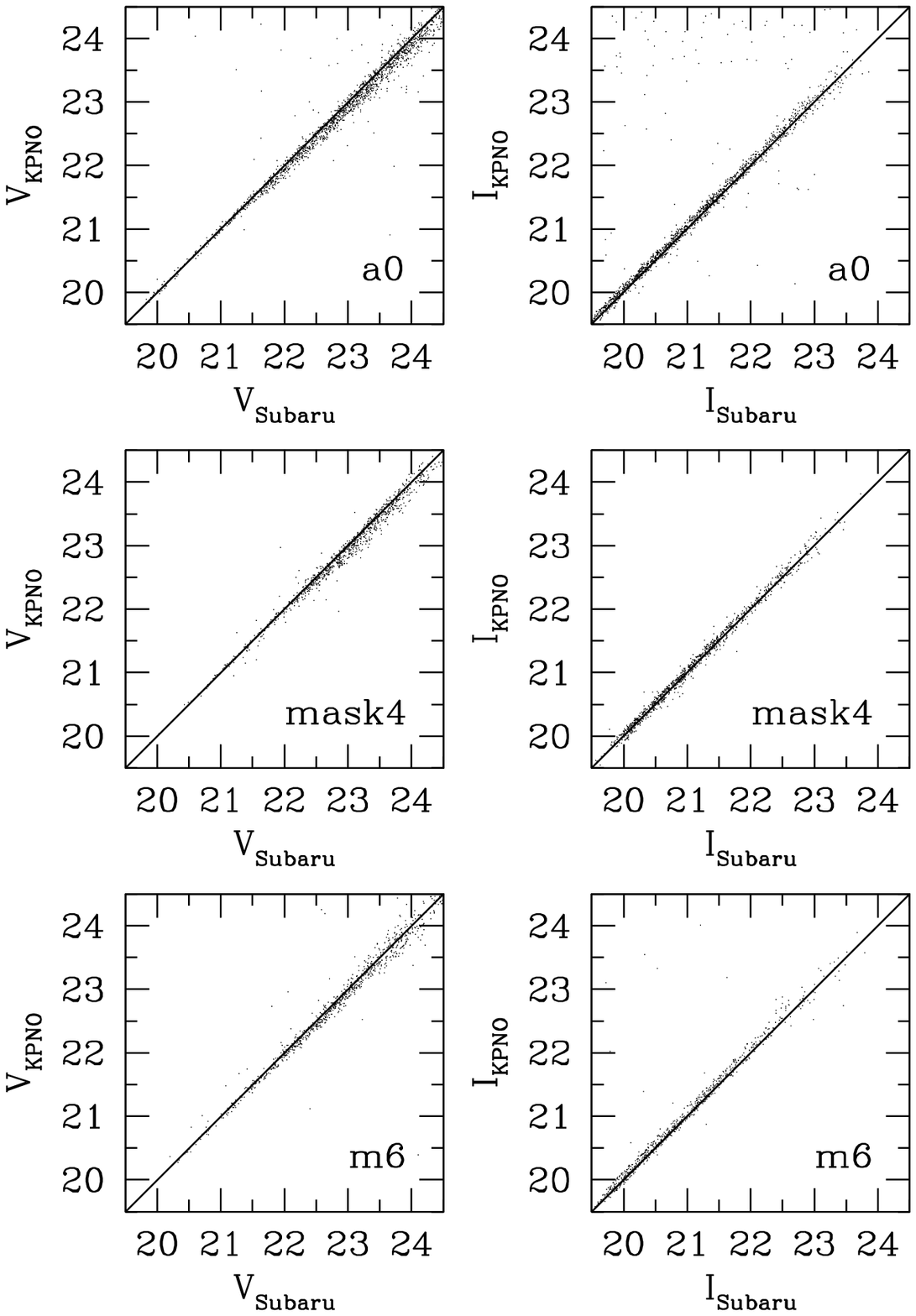}
%
% XXX Might be nice to add a second subpanel to each panel showing \Delta V
% and \Delta I (i.e. Subaru minus KPNO mag) vs Subaru mag.  That will make
% it much easier to see the finer details of the comparison relative to a
% \Delta mag = zero horizontal line.
%
  \caption[a]{Comparison of Subaru versus KPNO magnitudes of stellar-like
  objects in the three overlapping regions. For reference, a 1:1
  relation is shown as a solid line.}
  \label{fig:diff_phot}
 \end{center}
\end{figure*}

We compare the magnitudes based on the Subaru calibration with the
magnitudes based on the KPNO calibrations after matching up two
different catalogs using a 3-pixel matching radius. We extract 
stellar-like objects based on the DAOPHOT parameters 
($merr < 0.02$, $-0.3 < sharp < 0.3$ and $chi < 4$).
In Figure~\ref{fig:diff_phot}, we present a comparison of
the two independently determined $V$- and $I$- band flux measurements
for the three KPNO fields. The fact that the number of faint stars with
$V \gtrapprox 23.5$ and $I \gtrapprox 22.5$ decreases with decreasing
flux is due to low signal-to-noise in fainter part of the KPNO survey. 
The individual $V_{\rm Subaru}$ ($I_{\rm Subaru}$) and $V_{\rm KPNO}$
($I_{\rm KPNO}$) are found to be in good agreement with one another over
most of the magnitude range, indicating that
there are no systematic variations in our magnitude measurements. 

%%% Section 3 %%%

\section{The Andromeda Giant Stream}\label{sec:stream}
\subsection{Morphology of the Color-Magnitude Diagram}\label{sec:GScmd}
\begin{figure*}[htpd]
 \begin{center}
  \epsscale{1}
  \plotone{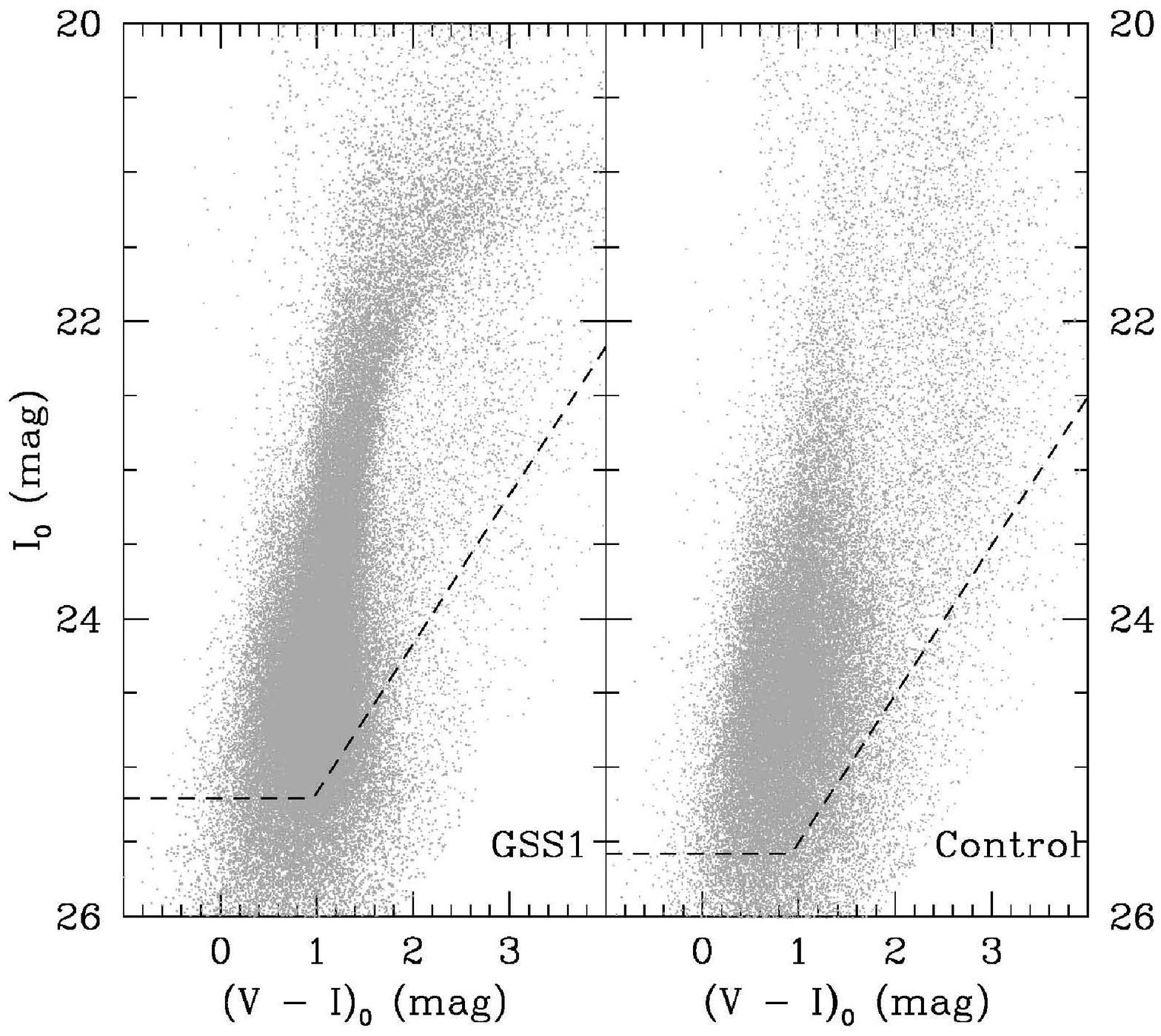}
  \caption[a]{Color-magnitude diagrams in the $I_0$ vs. $(V-I)_0$ plane for GSS1
  and the Control Field (SE3).
  The dashed lines denote the full ranges of the 50\% completeness levels for the mosaic images of the Suprime-Cam. 
  Extended background galaxies have been removed via image classification. 
  }
  \label{fig:cmdGS1}
 \end{center}  
\end{figure*}

Figure~\ref{fig:cmdGS1} shows the CMDs for the GSS (left) and for the
control field (right). In this section, we use the SE3 field as the
control field. The clearest feature in the stream field is a broad RGB
characterized by low-mass, hydrogen-shell-burning stars. It also follows
that the GSS has a broad metallicity distribution and to some extent
metal-rich RGB attributable to a larger stellar envelope opacity
\citep[e.g.,][]{Salaris2005}, in good agreement with the previous
studies \citep[e.g.,][]{McConnachie2003,Ferguson2005}. The CMD of the
control field also shows an apparently broad RGB feature, although it is
less dense than that of the GSS. It is worth noting that our control
field is located slightly more remote ($\sim 5$ kpc) from M31's center
than the field analyzed by \citet{Durrell2004} on the minor axis,
thereby suggesting that metal-rich halo stars are distributed even in
the outer parts of the M31 halo \citep[e.g.,][]{Irwin2005}. 

\begin{figure}[htpd]
 \begin{center}
  \epsscale{1}
  \plotone{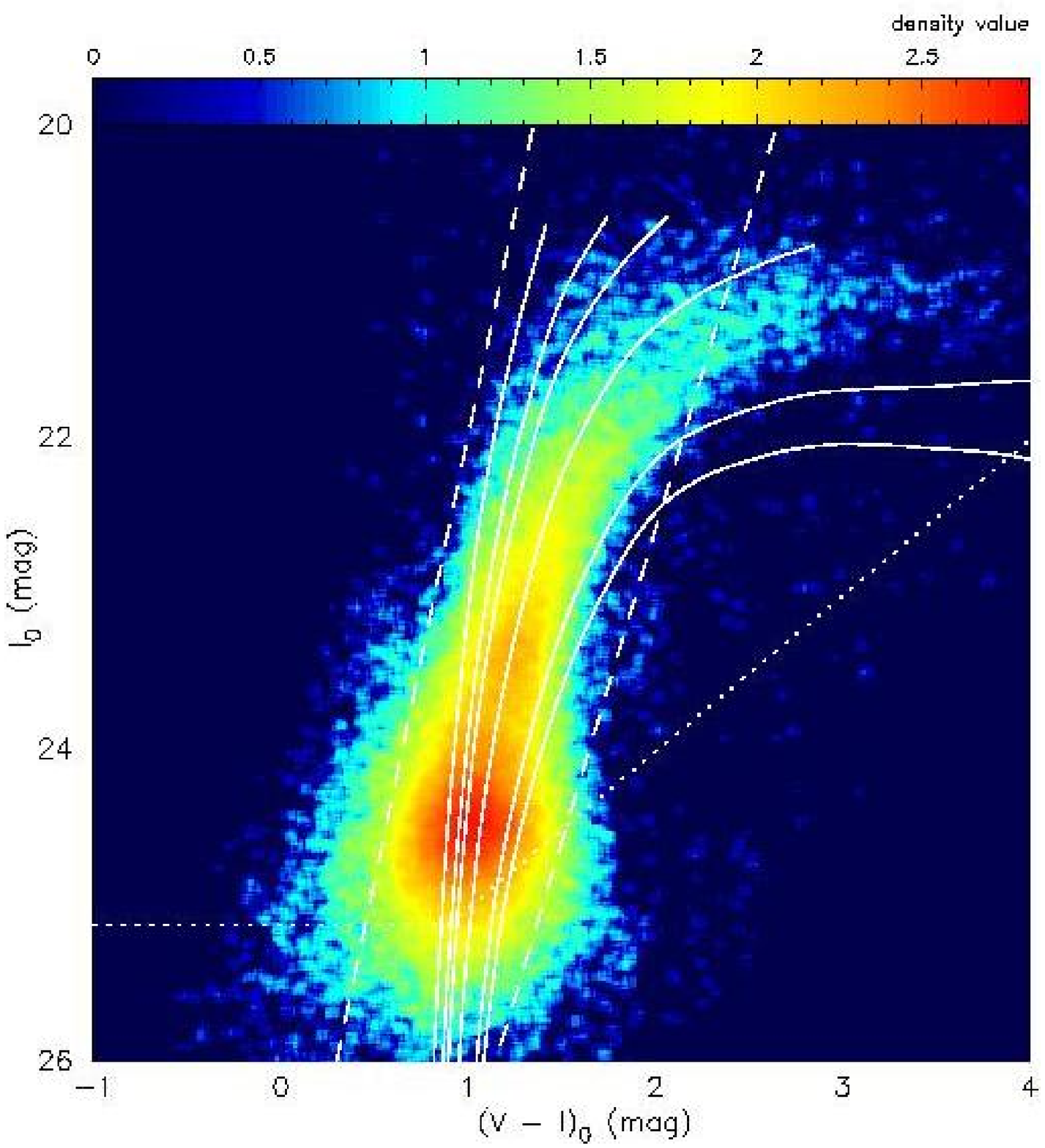}
  \caption[a]{A log-scaled CMD for GSS1. 
  The peak number densities of the AGB bump and the RC amount to
  224 and 657 stars per 0.07 $\times$ 0.07 mag box, respectively. Stellar
  density increases with redder color. The main features are
  distinguished by an edge of stellar density: the TRGB at $I_0 \sim
  20.5$ and by the reddest peaks of high stellar density: the AGB bump at
  $I_0 \sim 23.5$ and the RC at $I_0 \sim 24.5$. The solid lines are
  theoretical isochrones \citep{VandenBerg2006} of age 12 Gyr spanning
  the metallicity range [Fe/H]$ = -2.14$, $-1.41$, $-1.14$, $-0.71$,
  $-0.20$ and $0.00$. The dotted line denotes the full ranges of the
  50\% completeness level. Stars in the region between the two dashed
  lines are used to determine the edge of the TRGB and the peaks of the AGB bump
  and the RC in the following section. 
}
  \label{fig:cmdGS1cont}
 \end{center}  
\end{figure}

To extract the detailed distribution of stellar populations in the GSS
alone, the field stars other than the stream stars must be
removed from the CMDs. For this purpose, we have made statistical
subtraction of the field stars taken from the CMD of the control field,
following the method by \citet{Durrell2004}. In
Figure~\ref{fig:cmdGS1cont}, we show a colored contour for the
log-scaled CMD of GSS1 field after this procedure, illustrating detailed
characteristic features of the GSS stellar populations
\citep[cf.][]{Bellazzini2003}. It follows from this figure that in
addition to the RGB feature, we identify a Red Clump (RC) at $I \sim
24.5$, structure in the CMD consisting of a number of helium
core burning stars. More accurate estimates of the RC magnitude will be
made below, by correcting for the incompleteness of magnitudes
(Sec.~\ref{sec:GSLF}). The detection of a clear RC thus indicates the
presence of metal-rich and/or young populations in the GSS. 

In Figure~\ref{fig:cmdGS1cont}, there is also a second peak at $I \sim 23.5$,
which is apparently brighter than the RC. This feature can be regarded as
an Asymptotic Giant Branch (AGB) bump, not a RGB bump. This is for the reason
that while most stars in the GSS appear to be old and/or metal-rich
populations, a RGB bump with such old and/or metal-rich populations
lies at a fainter magnitude than a RC magnitude in the CMD,
following various theoretical models as calibrated by \citet{Alves1999}.
There is also a possibility that a clumpy feature being slightly brighter
than a RC is just a portion of the red horizontal branch (RHB) clump
as shown in \citet{Bellazzini2003}. Their G76 field having such a feature
is located at an edge of a luminous thin disk of M31, so only young,
metal-rich populations like disk stars give rise to this kind of a feature
slightly brighter than a RC magnitude.

At $I \sim 20.5$, we also recognize the tip of the RGB (TRGB). 
The presence of a few stars brighter
than the TRGB magnitude level presumably suggests the presence of
long-period variable stars in metal-rich and/or young population like
bright AGB stars,  as is the case in the halo of NGC~5128
\citep{Rejkuba2003}.

\subsection{Distance to the Andromeda Giant Stream}\label{sec:GStrgb}

As mentioned above, the TRGB is a useful indicator to estimate the distance 
to resolved galaxies \citep[e.g.,][]{Lee1993,Madore1995,Ferrarese2000,Salaris2005}. 
If there are a sufficient number of old and
metal-poor RGB stars in a targeted field, the TRGB is easily detected as
a sharp cut-off of the luminosity function (LF) with the application of 
an edge detection algorithm, Sobel filter with a kernel $[-1,-2,0,+2,+1]$
\citep{Madore1995,Salaris2005}.
In this manner, we have identified the magnitudes of the TRGB from $I$-band
LFs and determined the distance to the GSS.
The TRGB is safely detectable, since it is bright enough not to
be affected by incompleteness (i.e., more than 95\% completeness for both $V$
and $I$-band magnitudes near the TRGB) and since it is fainter
than a saturation level ($I \sim 19.5$) of the CCDs. 

\begin{figure}[htpd]
 \begin{center}
  \epsscale{1}
  \plotone{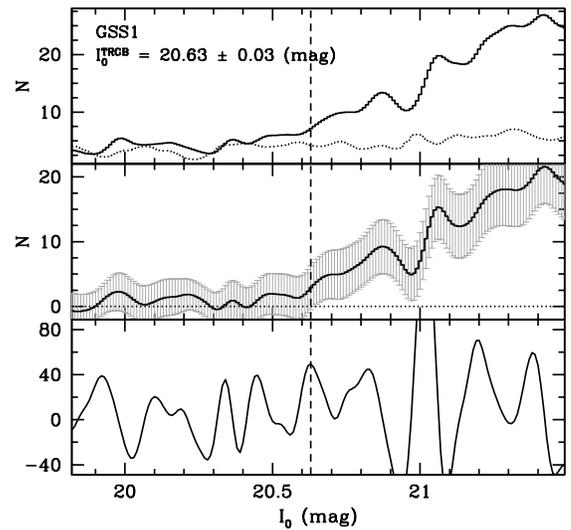}
%
% XXX Remove the dotted horizontal line from the middle panel.  The y axis
% label N should be italicized.
%
% XXX It is not clear to me why this particular peak in the Sobel filter
% response was picked as the correct one.
%
  \caption[a]{TRGB detection in the GSS. 
  The top panel presents the smoothed LFs as a function of the
  extinction-corrected $I$-band magnitude zoomed in around
  the TRGB magnitude. The solid line shows the original
  GSS1 LF before subtracting the background and the
  dotted line shows the background
  LF (which is the SE3 field). The middle panel shows the backgrond-subtracted
  GSS1 LF with Poisson error bars.
  The bottom panel indicates the Sobel
  filter response to the background-subtracted LF. The TRGB magnitude
  (the vertical dashed line) including the smoothing error is also shown.
%
% XXX What does "smoothing error" mean?
%
  }
  \label{fig:LFGS1trgb}
 \end{center}
\end{figure}

Detection of the TRGBs for the GSS1 field is shown in Figure~\ref{fig:LFGS1trgb}
(see also Fig.~\ref{fig:LFGS1}). These smoothed LFs are constructed
from the raw catalog by treating each star as a unit Gaussian with width
$\sigma(I)$ and summing all the Gaussians \citep{Durrell2001}. 
Taking into consideration that the $I$ magnitude of the TRGB changes only
by less than 0.1 mag for a metallicity range of
$-2.2 <$ [Fe/H] $< -0.7$ dex \citep{Lee1993}, which corresponds to
the color range defined by the RGB of the metal-poor globular cluster NGC~6341
and metal-rich one 47~Tuc, we select the
stars in this color range (between two dashed lines plotted in
Figure~\ref{fig:cmdGS1cont}) to estimate TRGB magnitudes. We then calibrate a residual LF,
for which contaminations are removed by comparison with the LF
of the control field. As deduced from Figure~\ref{fig:LFGS1trgb},
the final selected LFs of the GSS's stars show the leading edge of
the LFs, indicating that the number density of RGB stars significantly exceeds the Poisson
noise. In fact, the location of the TRGB on the LF is at the point that 
the contamination-subtracted LF with the vertical Poisson noise drawn in the middle panel
of Fig.~\ref{fig:LFGS1trgb} is the most steeply rising up from the noise. Also,
as shown later in Figure~\ref{fig:LFGS1}, the global shape of the LFs allows
us readily to identify the TRGB. 

Table~\ref{tab:GStrgb} shows the distance to our two stream fields, on the
assumption that the absolute $I$-band magnitude of the TRGB is
$M_I^{\rm TRGB} = -4.1\pm0.1$ for metal-poor TRGB stars. This $M_I^{\rm TRGB}$
adopted here corresponds to the $I$-band brightness of the TRGB
at [Fe/H] $= -1.3$ derived from various empirical and semi-empirical calibrations
\citep{Ferrarese2000,Ferraro2000,Lee1993,Salaris1998} by \citet{Salaris2005}. 
The TRGB absolute magnitude suffers from statistical uncertainty of the order
of $\sim$ 0.1 mag in each data. In estimating the extinction-corrected apparent
magnitude of TRGB, $I_{\rm 0,TRGB}$, systematic errors also arise, related to
zero point uncertainties, calibration through relative photometry,
aperture corrections, photometric errors, smoothing of the LFs and
extinction law; the resulting total error is evaluated by a root-mean-square
of these errors.

Our estimated distance to the GSS in the GSS1 field is $883\pm45$ kpc
(see Tab.~\ref{tab:GStrgb}). 
It suggests that the GSS is located behind M31 (which is at
$\sim 770$ kpc by \citet{Freedman1990}) 
and extended along the line of sight, which is in good agreement with 
the distance to Field~5 in \citet{McConnachie2003} within errors
($840\pm20$ kpc), the closest field to GSS1.

\begin{deluxetable}{lccc}
 \tablewidth{0pt}
 \tablecaption{Distance to the GSS by analysis of the TRGB\label{tab:GStrgb}}
 \tablehead{
 \colhead{Name} & \colhead{$I_{\rm 0,TRGB}$} & \colhead{Distance Modulus} & \colhead{Distance (kpc)}
 }
 \startdata
%
% XXX Good place to explain why GSS2 is not used.
%
 GSS1   & $20.63\pm0.05$ & $24.73\pm0.11$ & $883\pm45$ \\ 
 \enddata
\end{deluxetable}

\subsection{Metallicity Distributions}\label{sec:GSmdf}

We derive the metallicity distribution (MD) of the RGB stars in the GSS,
based on the comparison with the RGB templates defined by
Galactic globular clusters \citep{Bellazzini2003}.
In this procedure, the assembly of a sufficient number of the RGB stars
with small photometric errors is important for deriving an accurate MD,
otherwise large photometric errors lead to the contamination of
red or metal-rich stars; our Suprime-Cam data combined with the careful
correction for contaminations shown in \S~\ref{sec:daopars} are advantageous
in this respect. It is also remarked that recent studies by
\citet{Brown2006a,Brown2006b} suggest an intermediate age, i.e., younger
than 10 Gyr, for about half the population of the GSS stars,
in contrast to Galactic globular clusters.
However, for low-mass RGB stars with age of 6-14 Gyr, the redward shift of
an RGB sequence with increasing age is minor, compared with its sensitive
redward shift with increasing metallicity \citep{Girardi2002}.
This is a favorable behavior of the RGB, lending support to our procedure of
deriving an MD in comparison with old globular clusters, but the photometric
results should be followed up by spectroscopic observations.

\begin{figure*}[htpd]
 \begin{center}
  \epsscale{1}
  \plotone{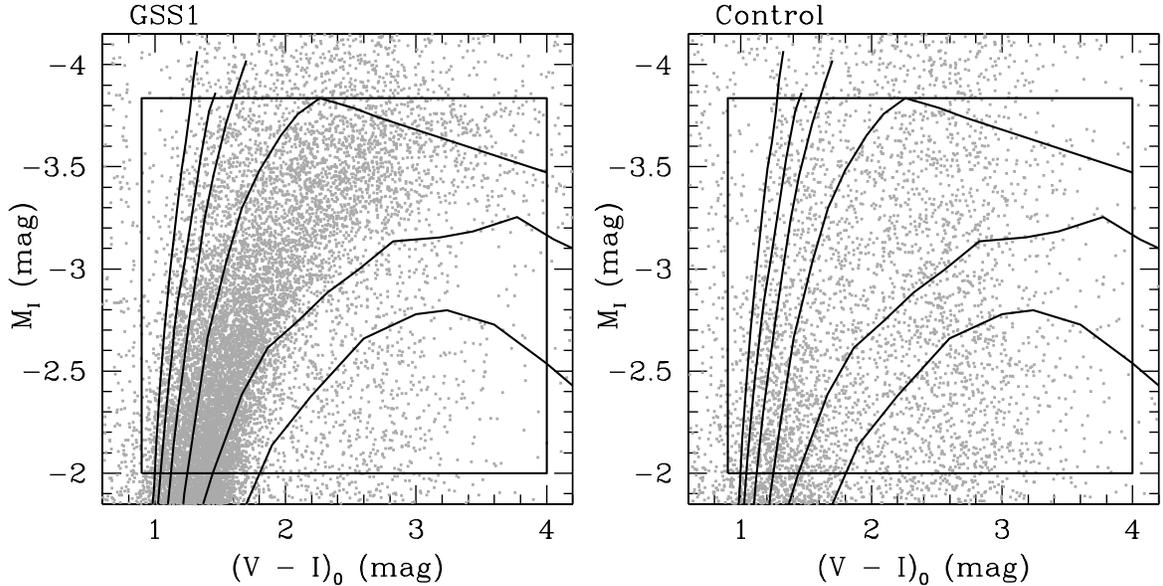}
  \caption[a]{CMDs of the upper RGB part for two fields. The ridge lines of template 
  Galactic globular clusters are superimposed to each plot as solid lines, which are
  from left to right,
  NGC~6341 ([Fe/H]$ = -2.16$), NGC~6205 ([Fe/H]$ = -1.39$), NGC~5904 ([Fe/H]$ = -1.11$) 
  and NGC~104 ([Fe/H]$ = -0.70$), NGC~6553 ([Fe/H]$ = -0.16$) and NGC~6528 ([Fe/H]$ = -0.07$). 
  The inner square frame encloses the stars whose metallicities are determined using
  the interpolating scheme \citep{Bellazzini2003}.}
  \label{fig:cmdGS1metal}
 \end{center}  
\end{figure*}

As RGB templates we adopt the six ridge lines of Galactic globular clusters
as in \citet{Bellazzini2003}.
The metallicities of these clusters are calibrated in the \citet{CG97}
abundance scale, hereafter denoted as $\rm [Fe/H]_{CG}$.
We take four metal-poor clusters in the Galactic
halo from \citet{Saviane2000}:
NGC 6341 ($\rm [Fe/H]_{CG}=-2.16$), NGC 6205 ($\rm [Fe/H]_{CG}=-1.39$), 
NGC 5904 ($\rm [Fe/H]_{CG}=-1.11$) and 47 Tuc ($\rm [Fe/H]_{CG}=-0.70$).
In addition, we take two metal-rich clusters in the Galactic bulge,
NGC 6553 ($\rm [Fe/H]_{CG}=-0.16$) from \citet{Sagar1999} and 
NGC 6528 ($\rm [Fe/H]_{CG}=0.07$, \citealp{Carretta2001})
from \citet{Ortolani1995} (see Table~\ref{tab:gc}).
For secure determinations of MDs using these templates, we select
the targeted RGB stars having
$-3.8 < M_I < -2.0$ and $0.9 < (V-I)_0 < 4.0$ as shown in Figure~\ref{fig:cmdGS1metal}. 
These selection criteria also allow us to remove a number of contaminations,
such as AGB stars, AGB bump stars, young stars, and foreground and/or
background objects.
We perform the interpolation procedure to obtain the metallicities of
the stars both in the stream and control fields, and subtract the derived
distribution of metallicities in the control field from that in the stream
field in order to remove the effects of the M31 field halo stars.

\begin{deluxetable}{llcc}
 \tablewidth{0pt}
 \tablecaption{Galactic globular cluster data used in the present study\label{tab:gc}}
 \tablehead{
 \colhead{} & \colhead{[Fe/H]$_{\rm CG}$} & \colhead{$E(B-V)$} & \colhead{$(m-M)_{0}$}
 }
 \startdata
 NGC~6341\tablenotemark{a}~~M92 &$  -2.16$&$0.02$&$14.74$\\
 NGC~6205\tablenotemark{a}~~M13 &$  -1.39$&$0.02$&$14.38$\\
 NGC~5904\tablenotemark{a}~~M5  &$  -1.11$&$0.03$&$14.31$\\
 NGC~104\tablenotemark{a}~~47Tuc&$  -0.70$&$0.04$&$13.29$\\
 NGC~6553\tablenotemark{b}      &$  -0.16$&$0.84$&$13.44$\\
 NGC~6528\tablenotemark{c}      &$~~~0.07$&$0.62$&$14.35$\\
 \enddata
 \tablecomments{The metallicities of these clusters are calibrated in the \citet{CG97} abundance scale.}
 \tablenotetext{a}{\citet{Saviane2000}}
 \tablenotetext{b}{\citet{Sagar1999}}
 \tablenotetext{c}{Locus read off from Fig.~13 of \citet{Bellazzini2003}}
\end{deluxetable}

\begin{figure}[htpd]
 \begin{center}
  \epsscale{1}
  \plotone{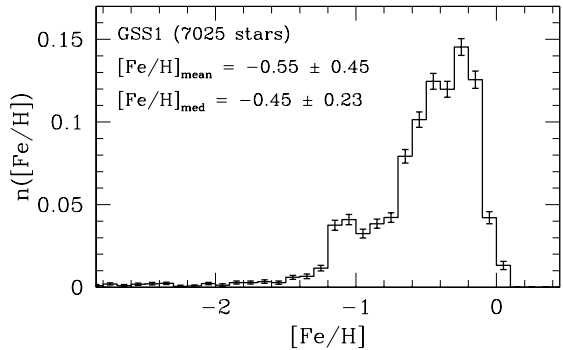}
  \caption[a]{Metallicity distribution for the GSS1 field. The number of
  stars used to derive it are shown in the upper left corner of the
  panel. The average metallicity with the standard deviation and the
  median metallicity with the quartile deviation of the GSS1 field are
  shown in the panel. 
  }
  \label{fig:mdfGS1}
 \end{center}  
\end{figure}

Figure~\ref{fig:mdfGS1} shows the MDs in the GSS1 field of the GSS.
The vertical error bars denote a nominal uncertainty
in each metallicity bin, derived from the Poisson errors equal to
$\pm \sqrt{N({\rm stream} + {\rm control})}$. 
It is worth noting that these errors are significantly small because of
a large number of the RGB stars (7051 stars in the GSS1 field) available
from our Suprime-Cam data. It follows that the MDs
have a broad distribution ranging from [Fe/H] $\sim -3$ to the near solar
metallicity and there is a clear high-metallicity peak at
[Fe/H]$_{\rm peak} \sim -0.3$. In the GSS1 field, the average metallicity
is [Fe/H]$_{\rm mean} = -0.55 \pm 0.45$ and the median metallicity is
[Fe/H]$_{\rm med} = -0.45$ with a quartile deviation of 0.23 dex. 
The average metallicities derived here are in good agreement with those
of kinematically selected RGB stars in the GSS
\citep{Guhathakurta2006,Kalirai2006}. 

While the MD is dominated by metal-rich stars, the presence of metal-poor
stars reaching [Fe/H]$ \leq -2$ is suggested from a long metal-poor tail
in the MD, based on the existence of an extended blue horizontal branch
in the CMDs derived by HST/ACS
\citep{Ferguson2005,Brown2006a,Brown2006b}. Our found metal-poor tail
feature is consistent with \citet{Bellazzini2003}. However, as the also
discussed, we cannot strongly claim really and scattered RGB stars with
[Fe/H]$ \geq -2$.

\subsubsection{Effects of Adopted Assumptions on Metallicities}\label{sec:metal_err}
\begin{figure*}[htpd]
 \epsscale{1}
 \plottwo{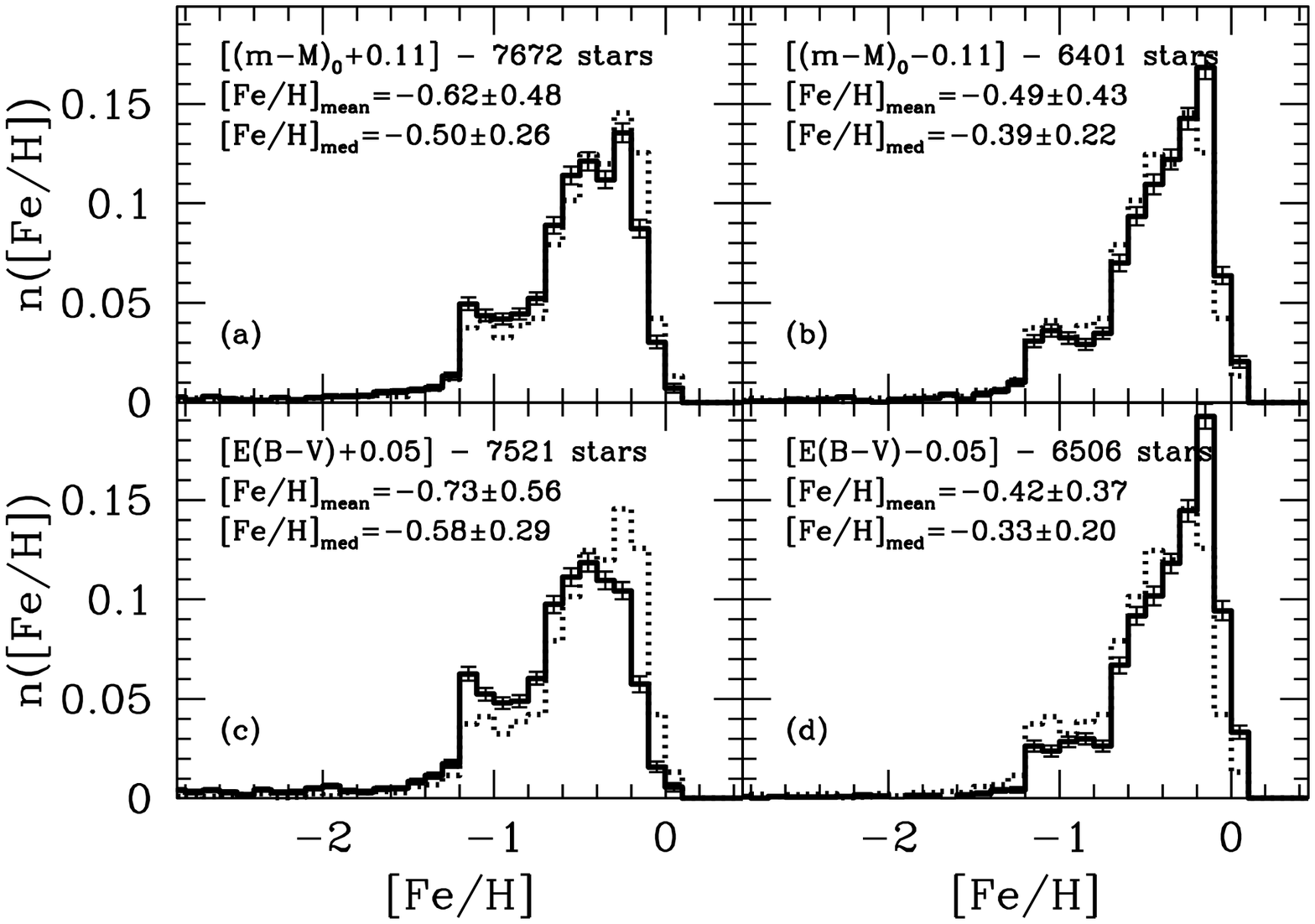}{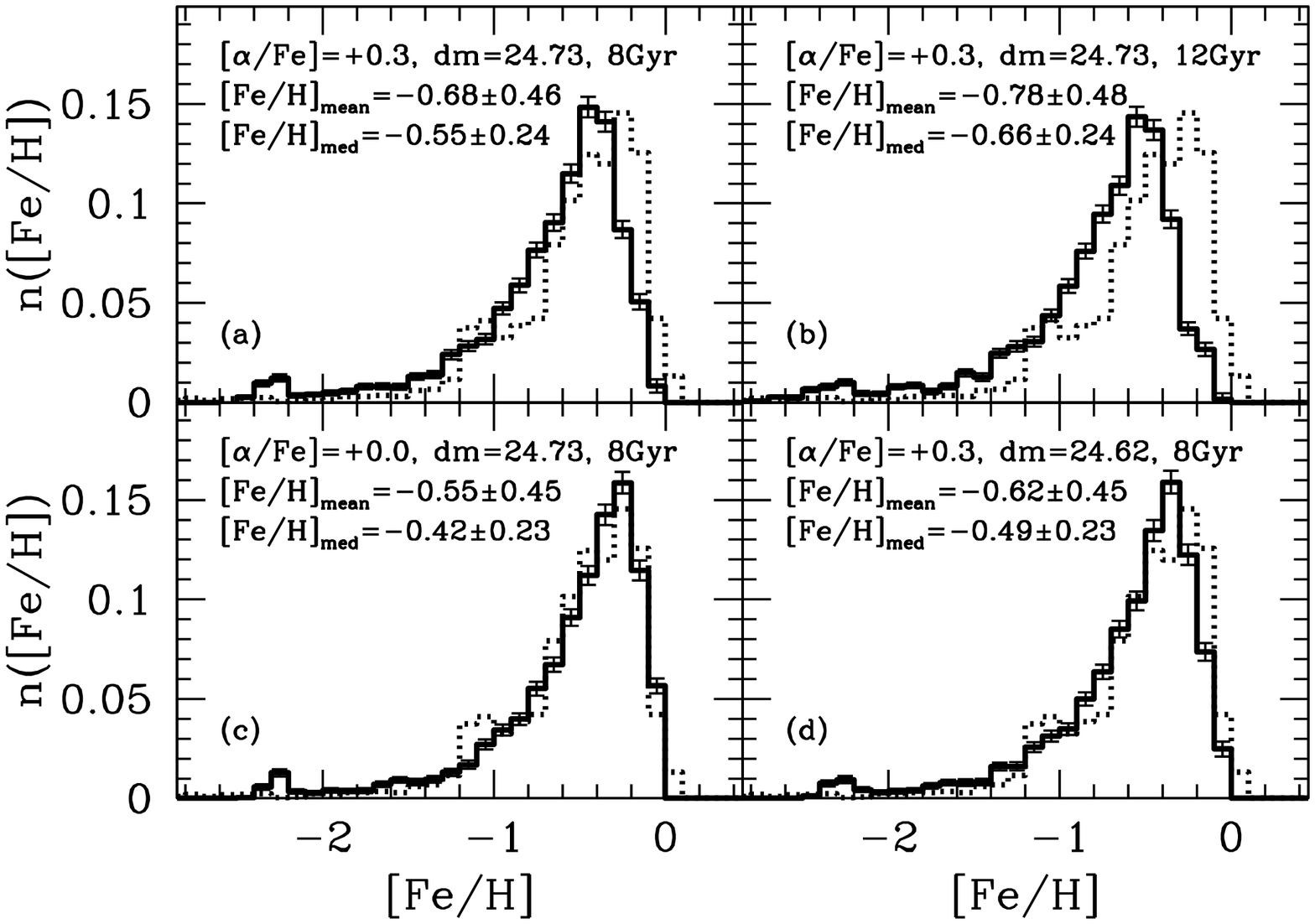}
 \caption[a]{Effects of varying assumptions on the form of the MDs.
 The dotted histogram: our standard case with $(m-M)_0 = 24.43$ and
 $E(B-V)$ from the \citet*{Schlegel1998} map, which is derived based on
 the Galactic globular cluster templates. {\it Left\/}: Panel (a):
 $(m-M)_0$ increased by 0.15 mag, (b): $(m-M)_0$ decreased by 0.15 mag,
 (c): $E(B-V)$ increased by 0.05 mag, and (d): $E(B-V)$ decreased by
 0.05 mag. {\it Right\/}: Re-constructed MDs based on the Victoria-Regina
 theoretical isochrones from \citet{VandenBerg2006} with the change of
 [$\alpha$/Fe], $(m-M)_0$ and age. 
 }
 \label{fig:metal_err}
\end{figure*}

The detailed form of the MDs is affected by several assumptions in the
analysis, such as adopted distance modulus $(m-M)_0$ and reddening
correction $E(B-V)$. To see these effects in the MDs, we examine the GSS1
field as an example and the results are shown in
Figure~\ref{fig:metal_err}. While Fig.~\ref{fig:mdfGS1} shows the case
based on our standard assumptions, the left four panels in
Figure~\ref{fig:metal_err} show 
the MDs associated with the change of our assumptions. For reference, we
also plot the MD derived based on the standard assumptions as the
dotted histogram. Panels (a) and (b) are devoted to the effect of
changing $\pm0.11$ mag in distance modulus, resulting in the change of
mean/median metallicity at only $\pm0.07$ dex level. Somewhat large
change of the MDs is seen in panels (c) and (d), where we test $\pm0.05$
mag variation in $E(B-V)$. This is inevitable because our method for
deriving MDs is based on the comparison between the color distributions
of RGB stars and that of adopted templates, thereby a change of $E(B-V)$
affects this comparison in a sensitive manner, especially in a
metal-poor range below [Fe/H] $=-1$, where the dependence on color is
large. The resultant change in metallicities is however no more than
$\pm0.18$ dex in response to a $\pm0.05$ mag change in
$E(B-V)$. 

Next, we examine the effect of adopting different RGB templates, i.e., other
than Galactic globular clusters, to estimate the metallicity of
each star. To do so, we re-construct the MDs using the Victoria-Regina
theoretical isochrones from \citet{VandenBerg2006} (whose isochrones are
partly displayed in Fig.~\ref{fig:cmdGS1cont}), in place of the
globular cluster templates of the Milky Way. In accordance with the
interpolation and extrapolation scheme of \citet{Kalirai2006}, we
calculate the metallicity for each star in the same segment of the CMD
(see also their Fig.~6). The resultant MDs with different
$\alpha$-enhancement, distance modulus and age are shown in the right
four panels of Figure~\ref{fig:metal_err}. Panels (a) and (b) are
devoted to the effect of changing 4 Gyr in age, resulting in the change of
mean/median metallicity at approximately $0.1$ dex level. Somewhat large
change of the MDs is seen in panels (a) and (c) with the two different
[$\alpha$/Fe]. The MD with $\alpha$ no-enhancement is very similar to
the one derived from the Galactic globular cluster templates, while the
$\alpha$-enhancement MD has a more metal-poor peak. Finally, in panels (a)
and (d), we test the effect of varying distance modulus. The difference
of mean metallicity between the two is 0.06 dex. The four MDs derived
from the model have gradual increase to the peak at [Fe/H] $\sim-0.3$
without sharp edge at [Fe/H]$\sim -1.1$ and $-0.7$ as in MDs constructed
based on the Galactic globular cluster templates, although all the MDs
derived from different metallicity indicators have the same global shape
such as the high metallicity peak at [Fe/H] $\sim -0.3$, the existence
of stars with near
solar metal and the long metal-poor tail. Furthermore, the GSS's mean
metallicity shown in panel (a) of the right of Figure~\ref{fig:metal_err},
where the basic assumptions behind this
is almost the same as those of \citet{Brown2006b}, is in good agreement with
the mean metallicity of $\langle$[Fe/H]$\rangle = -0.7$ derived from
their best-fit model (based
on the distance modulus of 780 kpc and the Victoria-Regina
isochrones). It is noteworthy that our mean metallicity estimated from
only RGB stars is consistent with their result by main-sequence
fitting to much deeper CMD.

Based on these experiments, we conclude that the basic features of the
MDs derived here are robust, including the existence of stars with
near solar metallicities, high metallicity peak at [Fe/H]$\sim -0.3$,
and long metal-poor tail. The mean and median metallicities remain
basically unchanged within a typical error of at most 0.1 to 0.2 dex in
the current method. 

\subsection{Stellar Populations Inferred from the AGB Bump and Red Clump}
\label{sec:GSLF}

\begin{figure}[htpd]
 \begin{center}
  \epsscale{1}
  \plotone{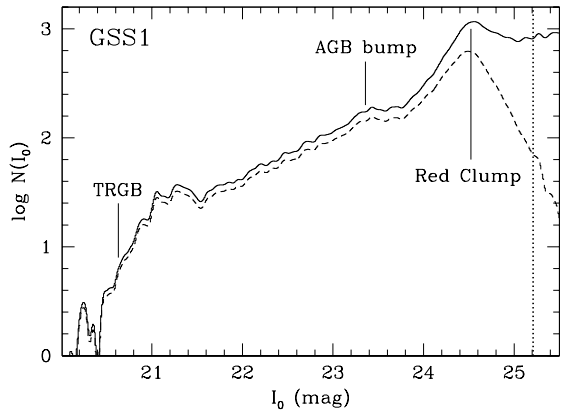}
  \caption[a]{$I$-band luminosity function corrected for extinction and
  incompleteness at magnitude brighter than horizontal branch of GSS1
  field, is plotted with solid line, while the dashed curve shows the
  curve with extinction-correction only. The vertical dotted lines indicate
  50\% completeness magnitude of $I_0 = 25.21$ for GSS1 field. The main
  features are indicated by vertical solid lines: the TRGB at $I_0 \sim
  20.63$, the AGB bump at $I_0 \sim 23.37$, and the RC at $I_0 \sim 24.52$.
  }
  \label{fig:LFGS1}
 \end{center}
\end{figure}

In addition to the broad RGB feature of the GSS, being attributable
to a composite population of metal-rich and metal-poor stars,
the CMD shows two noticeable populations (see Fig.~\ref{fig:cmdGS1cont}),
the AGB bump and RC, which are related to more advanced stellar
evolutionary phases. To highlight these features, we show, in
Figure~\ref{fig:LFGS1}, the $I$-band LFs of the GSS1 field,
which are corrected for extinction. The dashed and solid lines denote
LFs before and after correction for incompleteness, respectively. The
vertical dotted line indicates a 50~\% completeness magnitude of $I_0 =
25.22$ for the GSS1 field. The main features are marked by vertical solid
lines: the TRGB, AGB bump, and RC. More details on the latter two
features are discussed below. 

%%% 3.4.1 %%%
\subsubsection{AGB Bump}
The AGB bump is a stellar evolutionary phase of low-mass stars,
and is caused by the clustering of stars in a CMD, where AGB stars
at the beginning of helium shell-burning evolution slowly re-ascend
along the Hayashi line in a thermal nonequilibrium state.
However, due to the very short lifetime of the AGB phase,
the AGB bump feature can be detected only when dealing with a large sample
of stars \citep{Rejkuba2005}. In fact, \citet{Gallart1998} first detected
an AGB bump from the data of \citet{Holland1996} in the dense inner part of
M31's halo (at 7.6 kpc from the center approximately along the SE minor axis).
As already reported in \S~\ref{sec:GScmd}, we have also found an AGB
bump feature in the stream field at $I_0 \sim 23.5$ and $(V-I)_0 \sim 1.2$
(Fig.~\ref{fig:cmdGS1cont}). 

To more quantitatively identify the mean magnitude of the AGB bump,
we perform a Gaussian-weighted fit to the $I$-band LF
of GSS stars after correction for extinction and incompleteness. 
The upper left and right panels of Figure~\ref{fig:LFGS1agbrc} shows the
zooming CMD and the LF with Poisson errors for the GSS1 field near the AGB
bump magnitude, respectively. In this procedure, the surrounding stars 
in the shaded zones (left-hand panel) are avoided to reduce unwanted
contaminations, since the AGB bump feature is rather weak. 
At the magnitude level of the AGB bump, our photometric completeness is
more than 80~\% for $I$-band imaging. 
Although, in the usual CMD of the left panel, it is somewhat difficult for us to
detect the AGB bump feature by eye, we can easily confirm the existence
of the stellar population specified by the AGB bump through the LF of
the right panel. A red solid curve drawn in the LF is the best-fit model with the sum of 
a Gaussian function and a straight line in the range of $23.1 \la I_0 \la 23.8$. 
The best-fit function has been derived by minimizing reduced chi-squares
in the fitting procedure. The mean magnitude of the AGB bump is then
derived from an average value from these fittings: for the GSS1 field this is
$I_0 = 23.37 \pm 0.16$, corresponding to $M_I^{\rm AGBb} = -1.37$ with
a standard deviation of 0.13 mag. The estimation of this error is
based on those in the photometry, smoothing and fitting procedures as
well as the distance error using the TRGB as shown in
$\S$~\ref{sec:GStrgb}. We note that the fitting error
in the GSS1 field is negligibly small (less than $0.01$ mag). 

%%% 3.4.2 %%%
\subsubsection{Red Clump}
The RC is a clustering feature of RHB stars being metal-rich and/or
intermediate-age. {\it HST\/}/ACS observations \citep{Brown2006a,Brown2006b}
of the stream field, which is located closer to the galactic center than ours,
suggest the dominance of RHB stars in the HB, i.e., the presence of RC.
We have also found this feature in our stream fields, as already reported
in \S~\ref{sec:GScmd}, at $I \sim 24.5$ and $(V-I) \sim 1.0$
(Fig.~\ref{fig:cmdGS1cont}).

\begin{figure}[htpd]
 \begin{center}
  \epsscale{1}
  \plotone{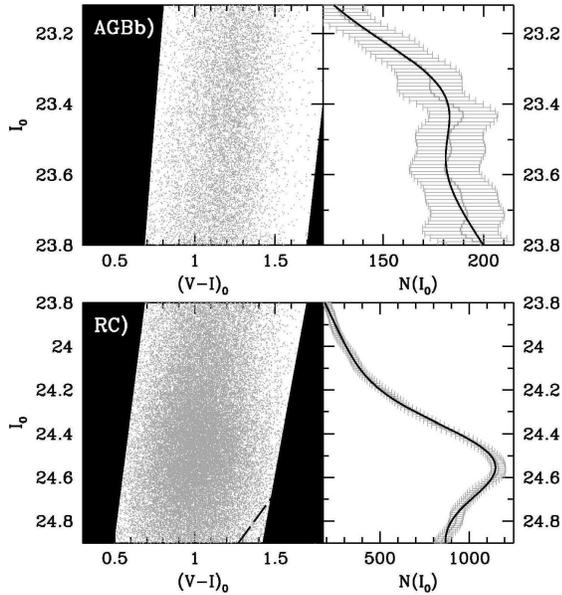}
  \caption[a]{Zoom up display of the CMD of the GSS1 field
  around the AGB bump and the RC features referring to
  \citet{Rejkuba2005}. We do not use the stars in shaded areas to
  evaluate the luminosity functions (see text). Shown on the right
  panels are the $I_0$-band completeness-corrected LFs around the AGB
  bump and the RC, 
  best-fitted with a simple model comprising of a Gaussian plus a
  straight line. The peaks of the AGB bump and the RC for the GSS1 field
  are at $I_0 = 23.37 \pm 0.16$ with $\sigma = 0.13$ and $I_0 = 24.52
  \pm 0.22$ with 0.17, respectively.
  }
  \label{fig:LFGS1agbrc}
 \end{center}  
\end{figure}

As in the case of the AGB bump, we set tighter limits on the mean magnitude
of the RC based on a Gaussian-weighted fit to the suitably corrected
$I$-band LF. The bottom left and right panels of
Figure~\ref{fig:LFGS1agbrc} shows the zooming CMD and the LF near the RC
magnitude, as in the upper panel of Figure~\ref{fig:LFGS1agbrc} for the
AGB bump. We note that particular attention must be paid to the
magnitude completeness since the RC is just beyond the detection limit:
the completeness at RC is about 60~\% for $I$-band image. 
We thus reconstruct the
completeness-corrected LF, using the completeness curves for the stars 
in the unshaded zone of the
CMD (the bottom left-hand panel of Figure~\ref{fig:LFGS1agbrc}). A red
solid curve drawn in the LF is the best-fit model with the sum of  a
Gaussian and a straight line in the range of $23.8 \la I_0 \la
24.9$. Using the curve we determine the mean magnitude of the RC: for
the GSS1 field this is $I_0 = 24.52 \pm 0.22$, corresponding to $M_I^{\rm
RC} = -0.21$ with the standard deviation of 0.17 mag. This error is
derived in the same manner as for the AGB bump; in this case the
photometric uncertainty is larger. 

\subsubsection{Age Estimation}

We estimate the mean age of the GSS's stars using the relation between
age and metallicity for the AGB bump and RC features. For this purpose,
we adopt the calibration of this relation by \citet{Rejkuba2005} using
the theoretical stellar evolutionary tracks of
\citet{Pietrinferni2004}. Applying to their observations of NGC~5128's
halo, they showed that the mean age of the halo stars in NGC~5128 is
estimated as $\sim8^{+3}_{-3.5}$ Gyr. 

\begin{figure}[htpd]
 \begin{center}
  \epsscale{1}
  \plotone{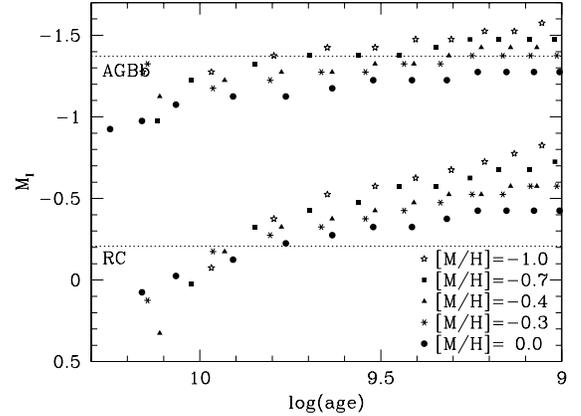}
  \caption[a]{Absolute $I$ magnitudes of the AGB bump and RC features predicted by
  the theoretical models, plotted as a function of age for five metallicities
  (kindly provided by Dr. Marina Rejkuba).
  These metallicities cover most of the metallicity range of the observed MD
  in the halo of the GSS.
  For comparison, the measured AGB bump and RC magnitudes 
  are indicated with horizontal dotted lines.}
  \label{fig:rejkuba}
 \end{center}  
\end{figure}

Figure~\ref{fig:rejkuba} shows the absolute $I$-band magnitudes of the
AGB bump and RC features predicted by the theoretical models, plotted as
a function of age for five different metallicities. The horizontal
dotted lines show the measured absolute magnitudes of the AGB bump and
RC, respectively. If we assume that the alpha elements of the GSS's
stars are enhanced such that [$\alpha$/Fe] $= +0.3$ as seen in the local
halo stars of the Milky Way, where the conversion, [M/H]$ \sim $[Fe/H]$
+ 0.3$, holds for most metallicity ranges, the average metallicity of
the stream fields is estimated as [M/H] $= -0.25$, which suggests (as
seen from asterisks in the Figure) the average age of the GSS stars of
$\sim 8$ Gyr.

\begin{figure}[htpd]
 \begin{center}
  \epsscale{1}
  \plotone{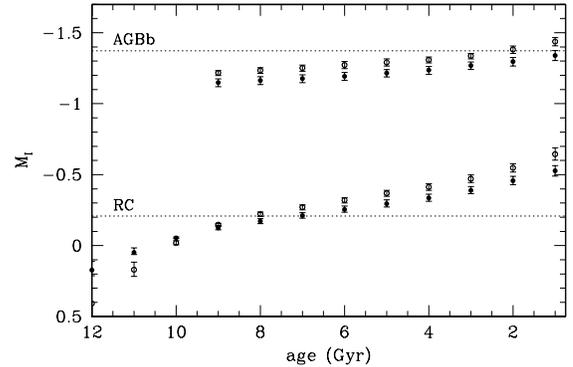}
  \caption[a]{Absolute $I$ magnitudes of the AGB bump and RC features vs. age predicted by
  the theoretical models, as obtained by convolving the metallicity dependency
  with measured metallicity distributions of
  the RGB stars \citep[see][]{Rejkuba2005}. 
  Filled circles present the relations computed with [$\alpha$/Fe] $= +0.3$, 
  while open circles indicate those computed with no $\alpha$-enhancement. 
  The measured AGB bump and RC magnitudes are shown as dotted horizontal lines.
  The error bars on the model points are calculated from assumed $\pm0.1$ dex
  uncertainty in the measured metallicity distributions.}
  \label{fig:agbrcmag}
 \end{center}
\end{figure}

To be more quantitative, we compute the average magnitude of the AGB bump and RC
as a function of age, as expected for the observed MDs using equation (9) of
\citet{Rejkuba2005}. Figure~\ref{fig:agbrcmag} shows the corresponding average
$I$ magnitudes of AGB bump and RC features plotted as a linear function of age,
where the error bars are evaluated by employing $\pm 0.1$ dex shift in the
observed MDs. The dotted horizontal lines indicate the measured mean magnitudes
of the AGB bump and RC. These magnitude measurements have a negligible statistical
error due to a huge number of available stars, but they contain a systematic
error associated with reddening and distance uncertainties,
which amounts to about $\pm 0.22$ mag. It is worth noting that
for metal-poor and old populations with [M/H] $<-1$ and age of $> 9$ Gyr,
the AGB bump feature is invisible in the theoretical luminosity functions,
and only a very weak feature is present for more metal-rich stars
\citep{Rejkuba2005}, so the model predictions for the AGB bump with
older ages are not shown.

The mean magnitude of the RC in GSS1 field indicates that the average
age of stars is 7.1 Gyr, whereas the AGB bump magnitude in the field 
suggests younger age. However, as seen in Figure~\ref{fig:agbrcmag},
the dependence of the AGB bump magnitude on age is much weaker than that
of the RC, so that small systematic errors in the measured AGB bump
magnitude can lead to a large difference in the estimated age.
The abundance of $\alpha$ elements somewhat affects the AGB bump-based
age estimation, in such a manner that lower [$\alpha$/Fe] ratios yield
older ages (see open circles for [$\alpha$/Fe]$ = 0$ in
Fig.~\ref{fig:agbrcmag}), whereas the RC-based age estimation is rather
insensitive to this effect. Much larger systematics may be caused by the
followings: (1) The AGBb feature appears only when [M/H]$ > -1$ and age
of $< 9$ Gyr, so the mean age of the stars is dominated by relatively
metal-rich and thus bright populations. (2) If the GSS contains
relatively young populations associated with secondary star formation
activities, the mean luminosity of the AGBb reflects the relative
fraction of each population, especially being biased in favor of
younger, brighter populations. This effect would be more significant for
AGBb than RC, because the number of the stars in the AGBb feature is
much small, so the Poisson error in the number of the populations,
especially for the small number of bright stars among the AGBb stars is
relatively large, thereby having a larger probability of yielding a
brighter AGB magnitude and younger ages. 

Thus, while taking into account that AGB bump magnitudes yields younger
ages, we adopt the RC-based age estimation, giving upper limits on the
age of GSS stars. Based on the available $I$-band magnitude of the RC
(because its $V$-band magnitude is unavailable in the current study), we
arrive at $7.1_{-4.7}^{+3.6}$ Gyr for the mean age of GSS stars, where
the error includes 0.22 mag systematic uncertainties in the RC magnitude
arisen from reddening, distance modulus and so on, while the
uncertainties in stellar evolutionary models are not taken into
account. \citet{Rejkuba2005} compared Teramo models
\citep{Pietrinferni2004} with Padova models \citep{Girardi2000} in their
similar study for the stellar population of NGC~5128 and found an
additional 1~Gyr uncertainty due to the age difference between the two
model predictions. 

We note that the mean age of GSS stars derived from the RC magnitude
is slightly younger compared to that derived by \citet{Brown2006b}. They
estimated the GSS's mean age to be 8.7~Gyr and the mean metallicity to be
$\langle$[Fe/H]$\rangle = -0.7$ based on the main-sequence fitting with
the distance modulus of 780 kpc and the Victoria-Regina isochrones,
whereas our re-computed mean age based on the MD constructed by the
Victoria-Regina theoretical isochrones (where the MD is shown in the panel (a)
of the right of Fig.~\ref{fig:metal_err}) is 8.1 Gyr. This result
is consistent with the GSS's mean age found by \citet{Brown2006b}. In
addition, the mean age and metallicity estimations based on different
assumptions as discussed in \S~\ref{sec:metal_err} are listed in
Table~\ref{tab:GSage}. In particular, all mean ages derived from the AGB bump
magnitude are much younger than those derived from the RC magnitudes. The
theoretical models provide robust mean ages without assumptions such as 
$\alpha$-enhancement and age to estimate the metallicity, while the ages are
somewhat sensitive to the assumption of distance as well as in the case
of using the Galactic globular cluster templates. Therefore, we note
that our technique applied to calculation of mean age based on the RC
magnitudes produces several Gyr uncertainty in accordance with
assumptions such as distance, extinction, $\alpha$-enhancement and
stellar evolutionary models.

The use of the AGB bump magnitude suggests systematically younger ages,
although this age estimation is rather sensitive to the accurate
determination of the AGB bump magnitude. We note that the presence of
the AGB bump feature implies the dominance of young and metal-rich
populations in the GSS; \citet{Rejkuba2005} reported that in the models
with ages $\gsim 10$ Gyr there exist very few AGB bump stars compared to
the RC or horizontal-branch stars and that in metal-poor ([M/H]$ < -1$
dex) and old (age $> 9$ Gyr) models the AGB bump is not present at
all. Thus, the presence of the AGB bump itself yields the younger age
estimate. It is also noted that the current implication for young and
metal-rich populations in the GSS may be associated with the result of
\citet{Brown2006b} (as depicted in their Figure 17b), suggesting that
star formation activities in the GSS had continued until some Gyrs ago;
it is inferred that the progenitor of the GSS may have had three main
star formation activities: (1) 30\% of the stars were provided in the
first star formation (e.g., by the collapse of the initial gas) before 8
Gyrs ago, (2) 40\% of the stars were provided in the second star
formation (e.g., supernova feedback or galactic interactions) about 8
Gyrs ago, and (3) remaining 30\% of the stars were provided in the third
star formation (e.g., the tidal interaction in the first pericentric
encounter with M31) after the second star formation.

\subsection{Mass of the Progenitor Galaxy}

The total mass of the stellar system expected for the progenitor galaxy
of the GSS can be estimated from the mean metallicity of stars in the
GSS ([Fe/H]$_{\rm mean} = -0.7$), combined with the
metallicity-luminosity relation for Local Group dwarfs
\citep{Cote2000}. The relation shows, on average, more metal-rich
populations in more luminous galaxies, and we obtain $M_{\rm stars} \sim
10^7$ M$_{\odot}$ for the progenitor galaxy if the mass-to-light ratio
is $M/L \sim 2$. In contrast, assuming the stellar mass-metallicity
relation observed for the Local Group dwarf galaxies of
\citet{Dekel2003}, we find $M \sim 4 \times 10^9 M_{\sun}$ as the
progenitor mass, which is consistent with the work of \citet{Mori2008}
who restricted the progenitor mass to be $M \leq 5 \times 10^9
M_{\sun}$, taking into account the effect of disk heating by dynamical
friction. Such a dwarf galaxy may have been accreted to the M31 halo at
recent epochs, possibly within some billion years, leaving yet a
conspicuous giant stream like the Sgr stream in the Milky Way.

\begin{deluxetable*}{lccc}
 \tablewidth{0pt}
 \tablecaption{Mean age and metallicity based on various assumptions\label{tab:GSage}}
 \tablehead{
 \colhead{Various Assumption} & \colhead{$\langle$[Fe/H]$\rangle$} &
 \colhead{$\langle$age$\rangle_{\rm RC}$} &
 \colhead{$\langle$age$\rangle_{\rm AGBb}$}
 }
 \startdata
 \cutinhead{Based on the Galactic globular cluster templates}
 Standard & $-0.55$ & $7.1$ & $0.7$ \\ 
 Standard with no $\alpha$-enhancement & $-0.55$ & $8.2$ & $2.3$ \\ 
 $(m-M)_0+0.11$ (24.84) & $-0.62$ & $4.8$ & $1.0$ \\
 $(m-M)_0-0.11$ (24.62) & $-0.49$ & $9.4$ & $2.8$ \\
 $E(B-V)+0.05$ (0.10)   & $-0.73$ & $5.6$ & $0.4$ \\
 $E(B-V)-0.05$ (0.00)   & $-0.42$ & $8.9$ & $1.4$ \\
 \cutinhead{Based on the Victoria-Regina theoretical isochrones}
 $[\alpha$/Fe$]=+0.3$, dm$=24.73$,  8 Gyr & $-0.68$ & $8.0$ & $1.9$ \\
 $[\alpha$/Fe$]=+0.3$, dm$=24.73$, 12 Gyr & $-0.79$ & $8.1$ & $2.3$ \\
 $[\alpha$/Fe$]=+0.0$, dm$=24.73$,  8 Gyr & $-0.55$ & $8.2$ & $2.5$ \\
 $[\alpha$/Fe$]=+0.3$, dm$=24.62$,  8 Gyr & $-0.62$ & $9.4$ & $4.9$ \\
 \enddata
\end{deluxetable*}

%%% Section 4 %%%
\section{Panoramic Views of the Andromeda Stellar Halo}\label{sec:halo}

In this section, we report on the results for the entire halo regions of M31
observed by the current survey, i.e., including those along its minor axis
both in the north and south parts of the galaxy and a south-west major-axis
region.

%%% 4.1 %%%
\subsection{The Color-Magnitude Diagram}\label{sec:cmdall}

\begin{figure*}[htpd]
 \begin{center}
  \epsscale{1}
  \plottwo{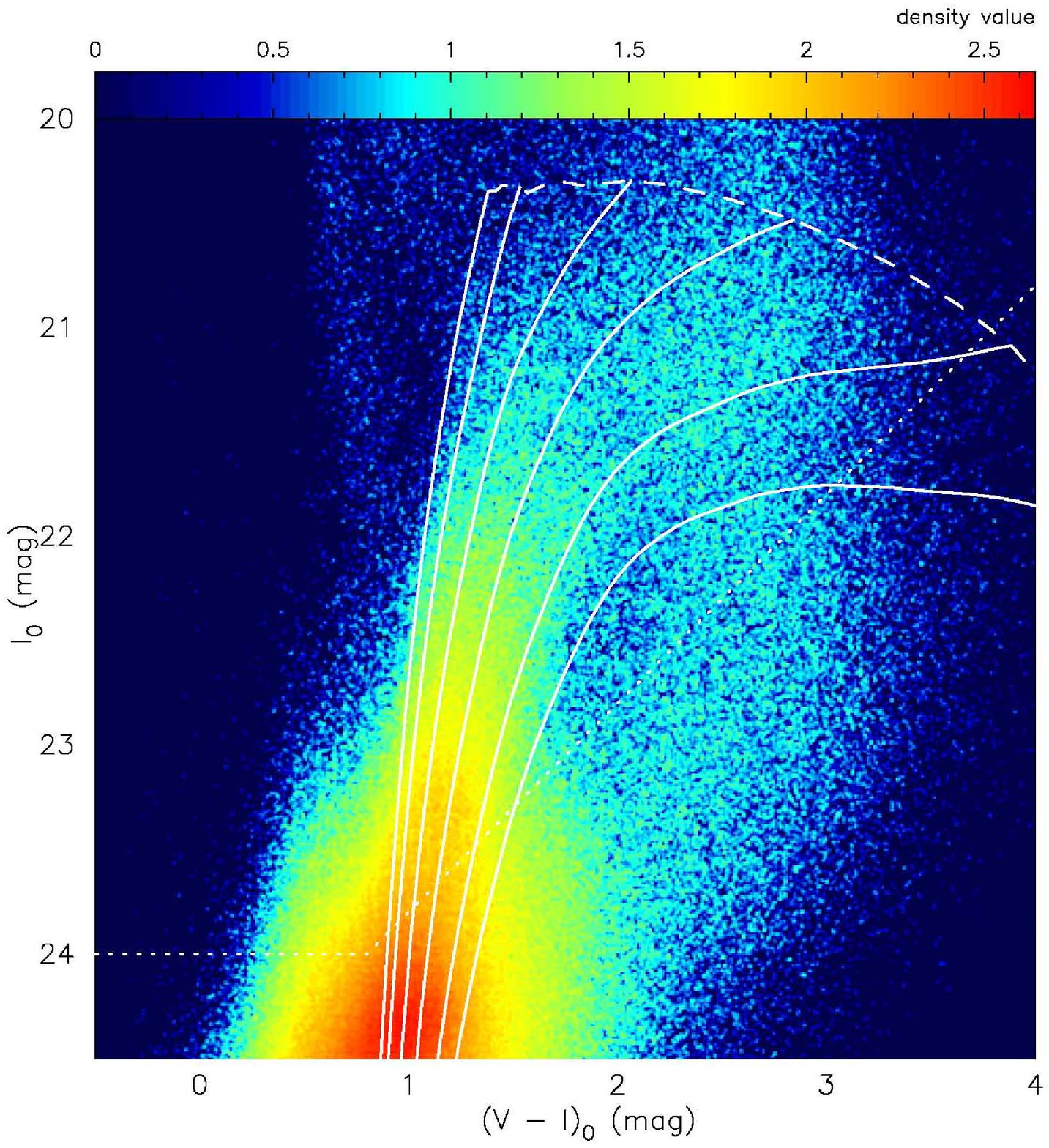}{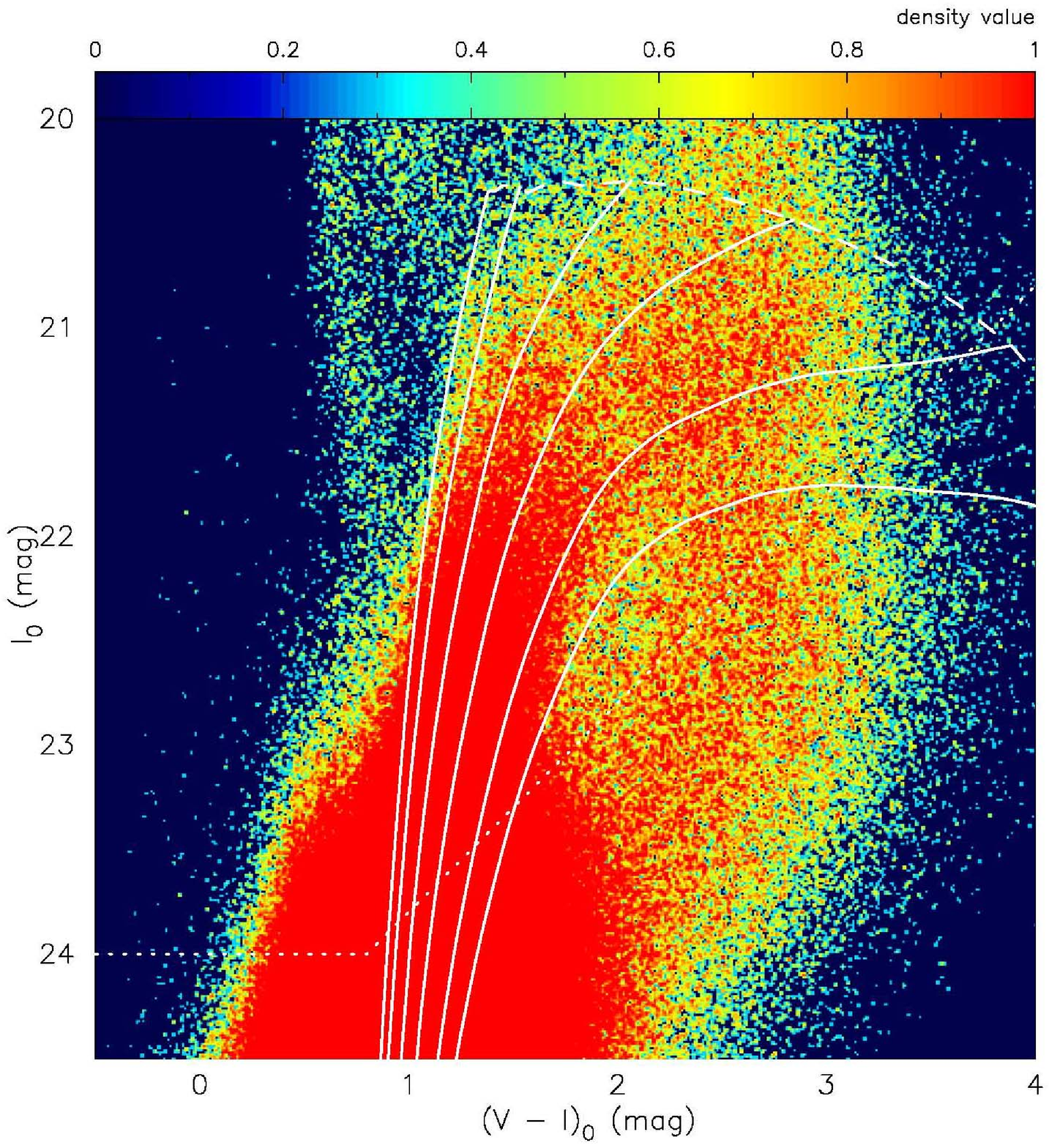}
  \caption[a]{A composite CMD in the $I_0$ vs. $(V-I)_0$ plane for stellar
  sources detected in all the observed fields (including the GSS
  field). The dotted lines denote the full range of the 90\% completeness
  levels for the mosaic images of the Suprime-Cam in the NW8 field. The
  superposed solid lines in the CMD show theoretical RGB tracks from
  \citet{VandenBerg2006} for an age of 12 Gyr, [$\alpha$/Fe]$=+0.3$, and
  metallicities ({\it left to right\/}) of [Fe/H]$ = -2.31$, $-1.71$,
  $-1.14$, $-0.71$, $-0.30$ and $0.00$. Extended background galaxies
  have been removed via image classification. The right panel is the
  same CMD as the left panel, but with the higher maximum level of 
  representation in the stellar densities. 
  }
  \label{fig:cmdall}
 \end{center}
\end{figure*}

Figure~\ref{fig:cmdall} shows a co-added CMD from all of the observed fields
(including the GSS field as shown in the previous section), after
removing extended sources such as background galaxies and cosmic rays
based on DAOPHOT parameters (see \S~\ref{sec:daopars}). The
thin solid lines in the CMD show theoretical RGB tracks from
\citet{VandenBerg2006} for an age of 12 Gyr, [$\alpha$/Fe]$=+0.3$, and
metallicities ({\it left to right\/}) of [Fe/H]$ = -2.31$, $-1.71$,
$-1.14$, $-0.71$, $-0.30$ and $0.00$. The dotted line denotes
about 90~\% completeness limit as determined by artificial star experiments,
where the limit is estimated by making use of the NW8 field with worse
seeing and shorter exposures. Although redundant stars and galaxies are
not completely removed from the CMD, we can clearly identify a broadly
distributed RGB feature attributable to overdense regions with a number
of metal-rich stars such as the GSS. At fainter than
$I_0 \sim 23$ (mag), the RGB feature appears to be much broader because of
the presence of background galaxies which passed through our selection
of separate stellar objects. Although it is contaminated by these objects,
a population extending toward fainter and bluer region than the most
metal-poor theoretical RGB isochrone at $I_0 \ga 24$ (mag) corresponds
to the horizontal branch stars in M31's halo. 

In the right panel of Fig.~\ref{fig:cmdall} which is the same CMD
as the left one but with the higher maximum level of representation in
the stellar densities, there exists
a sequence of Galactic disk dwarfs at $(V-I)_0\ga2$ having a
broad RGB; the vertical sequence is the result of low-mass stars
accumulating in a narrow color range, yet being seen over a large range
in distance along the line of sight. In addition, halo stars in the
Milky Way can be seen as vertically-distributed stars on the blue side
of this diagram, $(V-I)_0\la0.8$ and $I_0<23$. Usually this appears as
a smooth vertical structure in a CMD, which corresponds to the stars
at or close to the
main-sequence turnoff at increasing distance through the Galactic halo
\citep{Martin2007,Ibata2007}. 

\subsubsection{Calibration of the Control Field}\label{sec:comp}

\begin{figure*}[htpd]
 \epsscale{0.35}
 \plotone{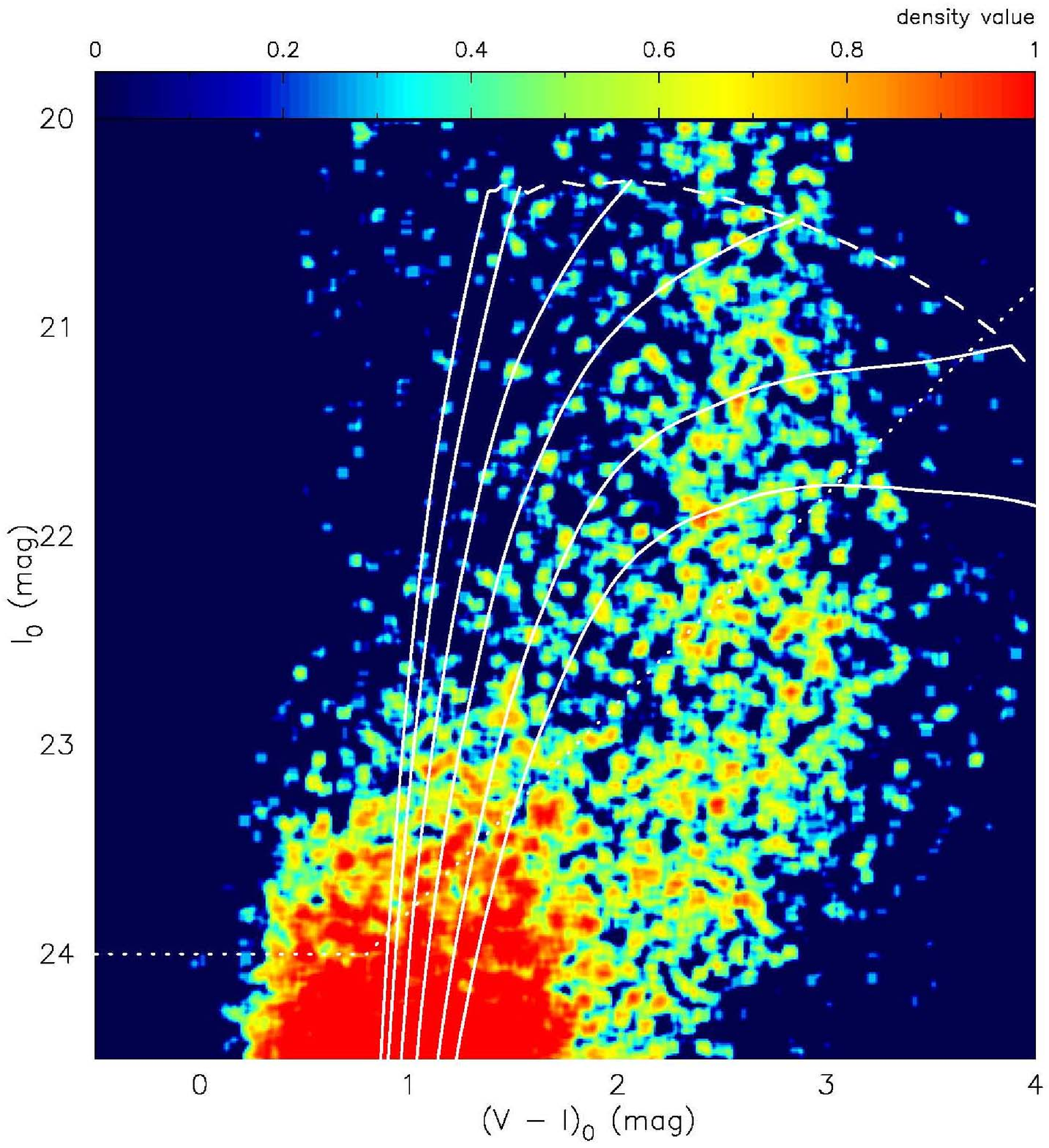}
 \plotone{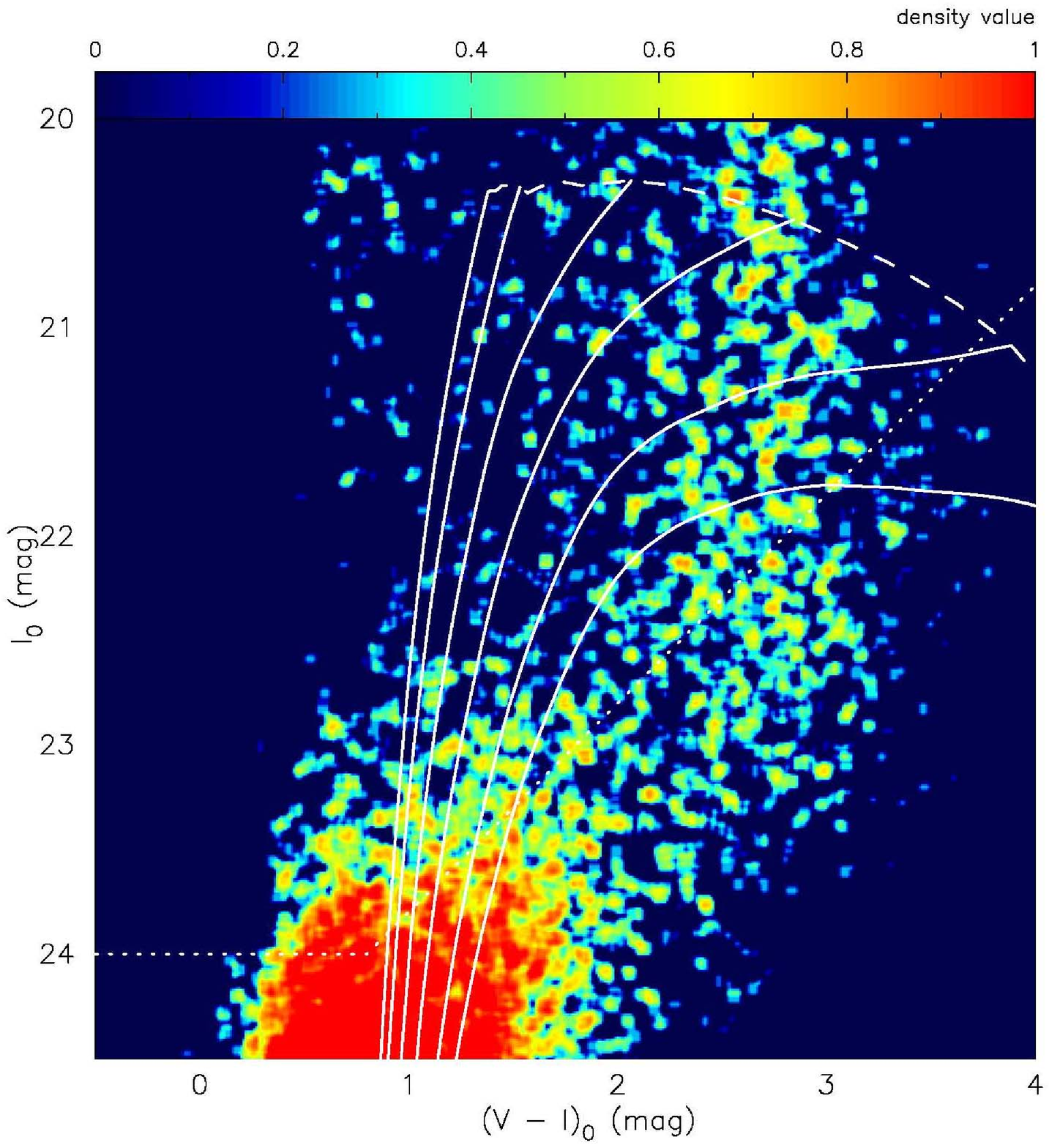}
 \plotone{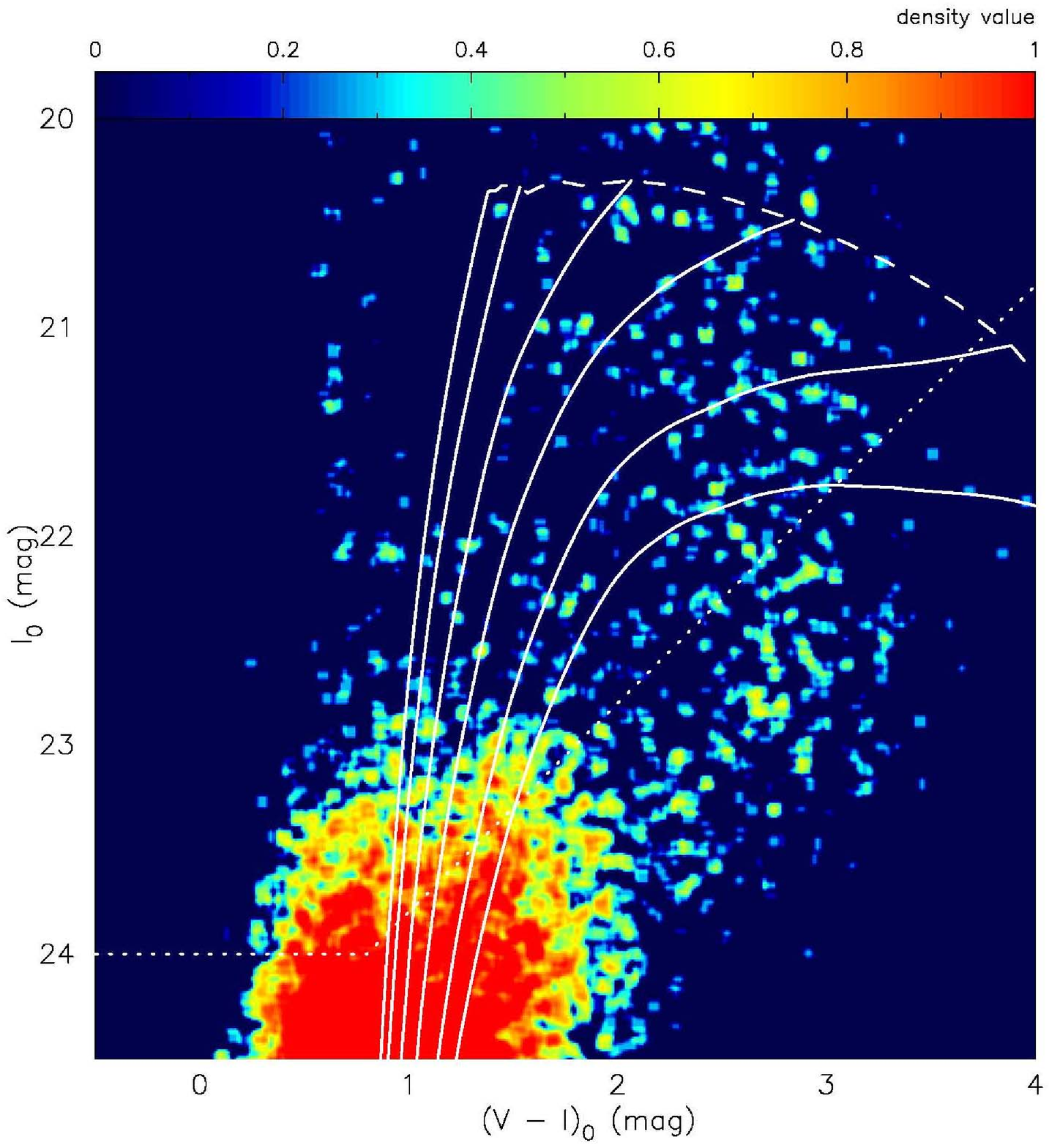}
 \caption[a]{CMDs of the three control fields in the $I_0$ vs. $(V-I)_0$
 plane. For reference, the white solid/dotted lines as shown in
 Fig.~\ref{fig:cmdall} are also overlaid. {\it left\/}: $(l, b) = (103\fdg0, 
 -16\fdg5)$, {\it middle\/}: $(l, b) = (103\fdg4, -20\fdg7)$, {\it right\/}:
 $(l, b) = (103\fdg4, -25\fdg1)$. 
 }
 \label{fig:cmdcomp}
\end{figure*}

Our three control fields, to be utilized for removing contaminations
such as Galactic dwarf stars which cannot be separated morphologically,
are located at the same galactic longitude $l = 103\degr$ but at
different galactic latitudes with $b = -16\fdg5$, $-20\fdg7$, and
$-25\fdg1$, i.e., far from the M31 direction ($l = 121.2\degr, b = -21.6\degr$). 
The CMDs of the three control fields are shown in
Figure~\ref{fig:cmdcomp}; from left to right, galactic latitude is
higher. For reference, the solid/dotted white lines as shown in 
Fig.\ref{fig:cmdall} are also plotted in these CMDs. A prominent feature is that
the number of Galactic disk dwarf stars is gradually decreasing with increasing
galactic latitudes. In contrast, there is little change in the
distribution of remaining faint background galaxies
between these three CMDs. Finally, it is found that the number variation of the
Galactic halo stars in these CMDs is also small. 

These properties of the control fields indicate that the contaminants
in our target fields of M31's halo are dominated by the faint background
galaxies and the Galactic disk dwarf stars, while the effect of the
Galactic halo stars can easily be corrected because such stars distribute
outside the M31 RGB sequence in the CMD. Regarding the unresolved
background galaxies, one can assume that such galaxies are uniformly
distributed in a Suprime-Cam field-of-view, whereby these background
contaminants per field are simply regarded as having a uniform density,
regardless of the galactic latitude. Thus, the remaining task
is to understand the effects of the foreground
Galactic disk dwarf stars on our target fields.

\begin{figure}[htpd]
 \epsscale{1}
 \plotone{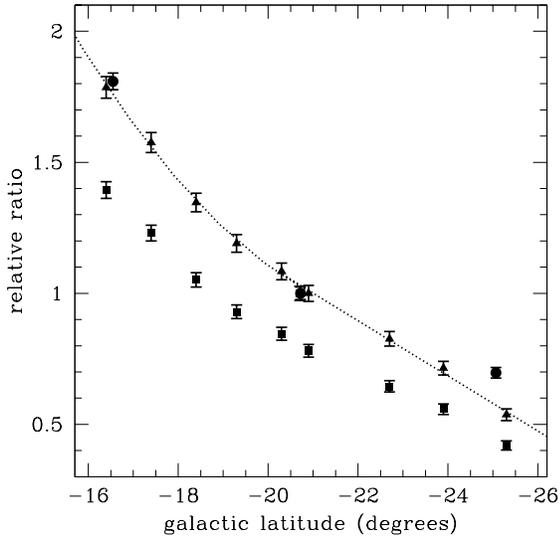}
 \caption[a]{The variation of the number density of the Galactic disk stars
 as normalized by that in the control field
 with $b=-20\fdg7$ as a function of the galactic latitude. The
 disk stars are taken from a box defined as $20.5<I_0<22.5$ and
 $1.2<(V-I)_0<3.0$ in the CMD. The three big solid circles
 denote our observational results,
 while the small solid circles show the model prediction by \citet{Robin2003}
 with the same galactic coordinates as our M31 target fields and the
 solid triangles are the scaled plots. The solid triangles are fitted by
 exponential (at the lower latitude than that of the NW1 field)
 and linear profiles (at the higher latitude). 
 }
 \label{fig:diskdwarfgrad}
\end{figure}

In order to correct the effect
of these dwarf stars in each target field of M31's halo, we estimate the fraction
of such stars in each target field based on the knowledge of the number of
the contamination stars in the CMDs of the control fields, by combining the
observed CMDs and those predicted by the models of \citet{Robin2003}. 
Firstly, we use two CMDs of the short exposed NW1 field and one of the
three control fields with almost the same galactic latitude of
$b\sim-21\degr$. We have obtained 0.78 as the scale factor to adjust the number
difference arising from the galactic longitudinal difference between the 
M31 with $l\approx121\degr$ and our control fields with $l\approx103\degr$,
based on the comparison of the star counts of clean Galactic dwarfs in a box
defined as $17.0<I_0<18.0$ and $1.5<(V-I)_0<2.2$. Note that these selection
criteria enable us to avoid the effects of bright AGB and RGB stars of M31. This scale
factor derived from our observation is consistent with the
model prediction from \citet{Robin2003} including Poisson noise. In addition,
the scale factor calculated from the star count of the disk stars in a
box defined as $20.5<I_0<22.5$ and $1.2<(V-I)_0<3.0$ of the
corresponding model CMDs without the M31 population is also in good
agreement with the scale factor from the comparison of the bright disk
stars. Secondly, we investigate the variation of the number density of 
the disk dwarf stars along galactic
latitude. Figure~\ref{fig:diskdwarfgrad} shows the relative number
variation normalized by the number at the control field with $b=-20\fdg7$ as a
function of the galactic latitude, where the disk stars reside in a
box defined as $20.5<I_0<22.5$ and $1.2<(V-I)_0<3.0$. The three big solid
circles denote our observed results, while the small solid circles show the model
prediction by \citet{Robin2003} with the same galactic coordinates as
our M31 target fields and the solid triangles are the scaled plots by the
above-mentioned scale factor of 0.78. It is evident that the spatial variations
of both the observational result and the model prediction are reasonably
consistent with each other. Based on the results of these experiments,
we thus adopt the theoretical galactic model of \citet{Robin2003}
to estimate the spatial variation of the number of the disk dwarf stars.
This spatial variation of the number density is well represented by
an exponential profile at the lower galactic latitude than that of the NW1 field
and is represented by a linear profile at the higher galactic latitude,
as judged by a (dotted) fitting line in the figure.
For this correction of the contaminations, we adopt the blank
field with the lowest galactic latitude of $b=-16\fdg5$ as the control
field for the north part of the M31 halo, whereas the blank field with
the galactic latitude of $b=-20\fdg7$ is adopted as the
control field for the south part of the M31 halo. 

\begin{figure*}
 \begin{center}
  \epsscale{1}
  \plottwo{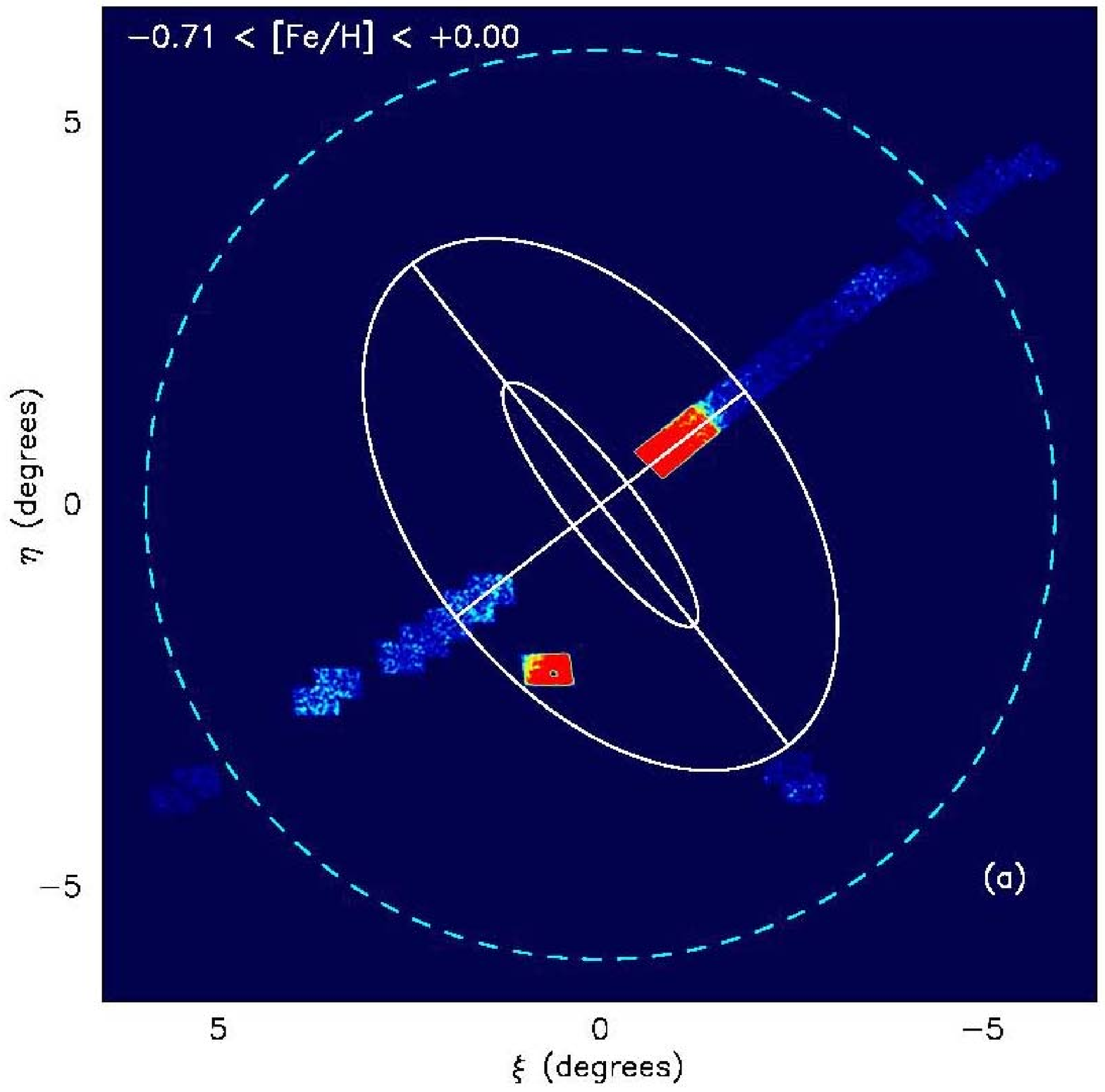}{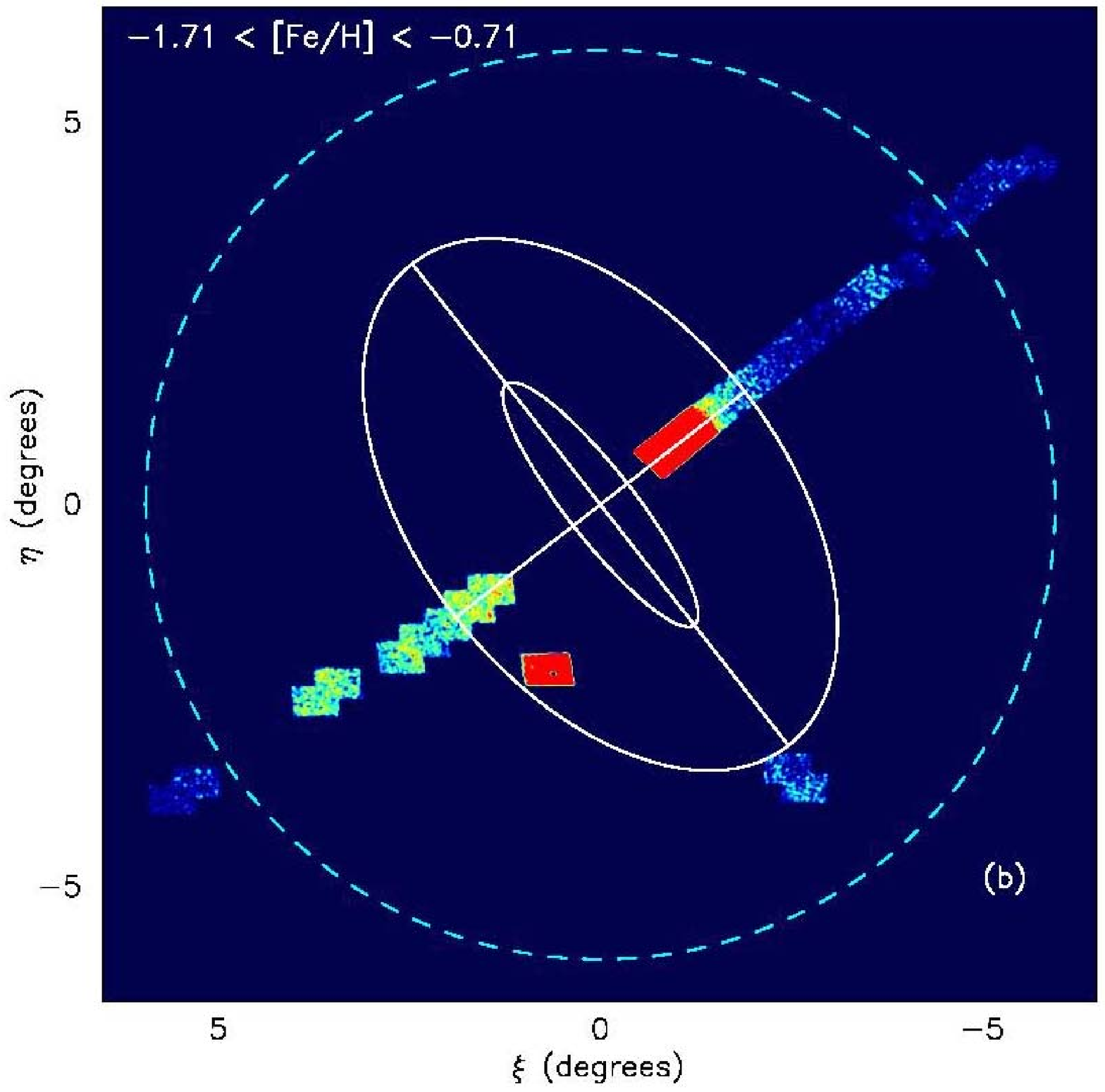}
  \plottwo{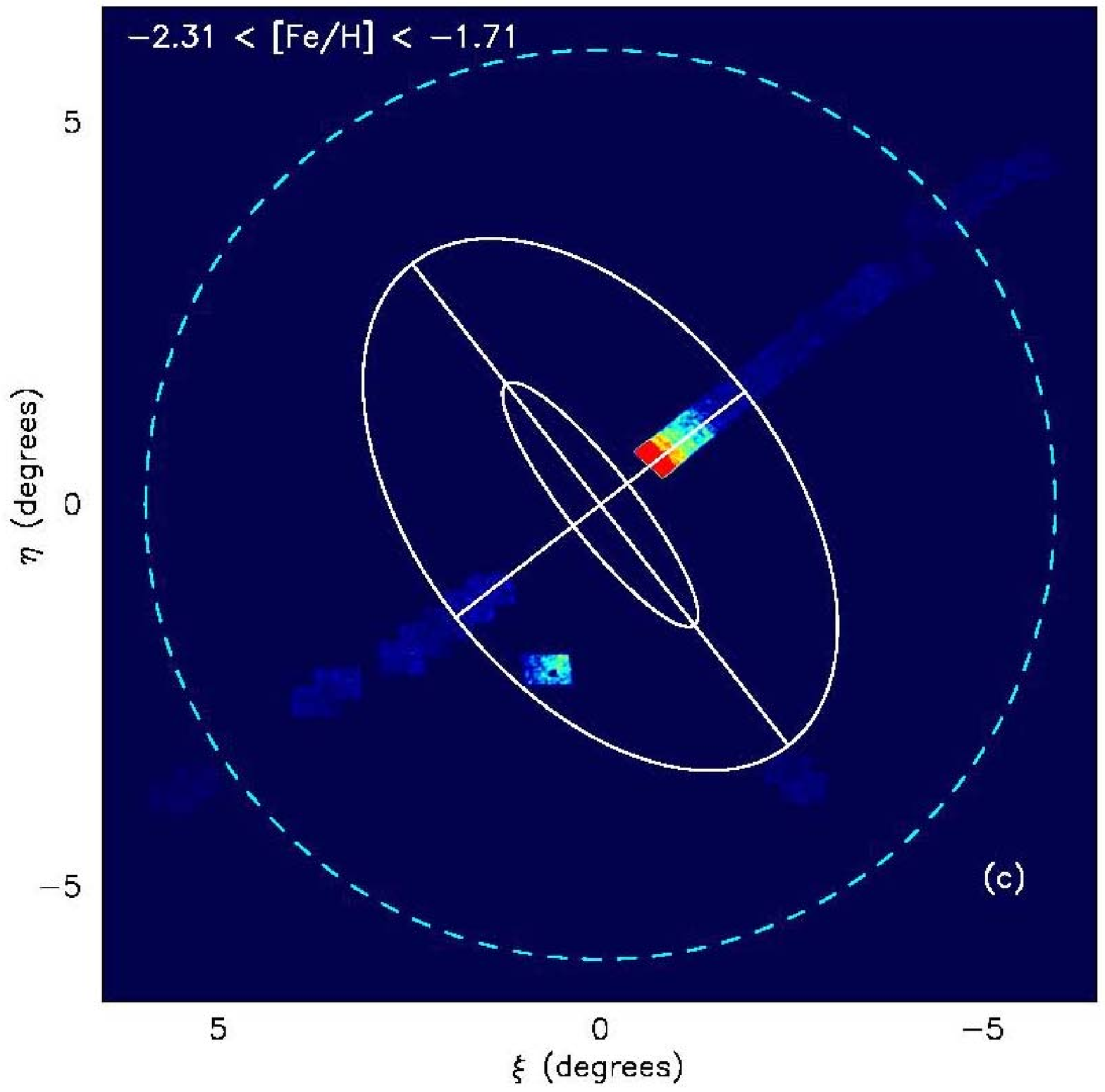}{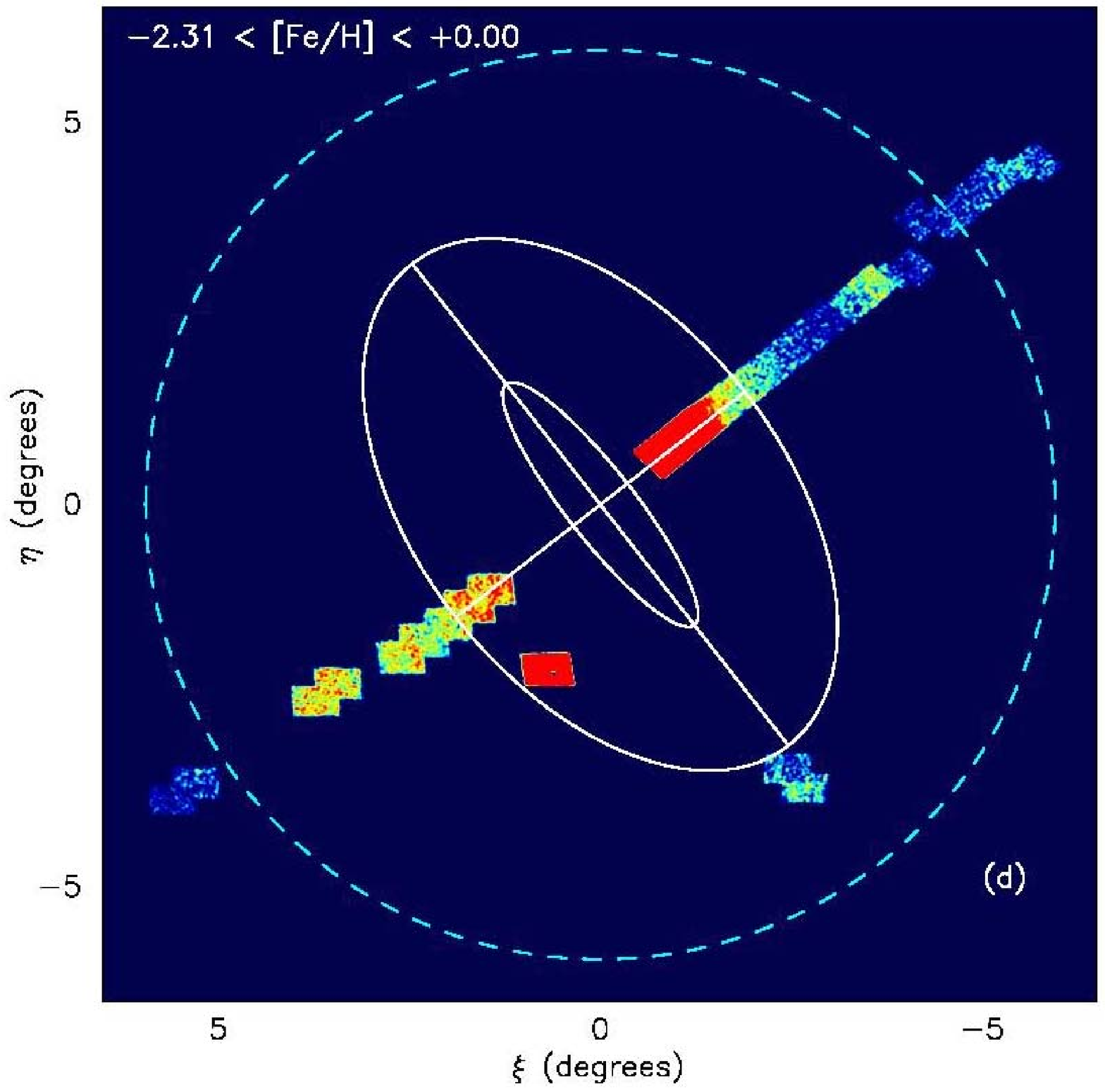}
  \caption[a]{Matched filter maps to a limiting magnitude of $I_0 = 24.0$ and
  $V_0 = 24.8$. Log-scaled images are shown on the left, while
  linear-scale versions are presented on the right. The resolution of
  this map is $0.018\degr \times 0.018\degr$ pixels, smoothed with a
  Gaussian kernel over 3 pixels. 
  }
  \label{fig:m31map}
 \end{center}
\end{figure*}

\subsection{Stellar Population Maps}\label{sec:M31map}

In this subsection, we investigate how the RGB populations of the M31 halo,
as identified in Fig.~\ref{fig:cmdall}, are distributed spatially out to about
100 kpc from the M31 center. To do so, we extract these stellar populations
of M31 alone from our imaging data. However, it is noted that the outer halo of
M31 is known to have a very diffuse stellar density as deduced from previous studies
\citep[e.g.,][]{Gilbert2006,Irwin2005}, so a clean signal of these M31
populations is buried under heavy contamination from foreground
disk dwarf stars and the background galaxies. In particular, the amount
of the disk dwarfs in the northern part of the M31 halo are about twice as
much as in the southern region, as suggested in Fig.~\ref{fig:diskdwarfgrad}. 
In order to reveal the signal from these heavily contaminated data,
we adopt the so-called matched filter method as has been utilized
in \citet{Ibata2007}. This method is
an optimal search strategy if one has a precise idea of the properties
of both the signal and the contamination. 

\begin{figure}[htpd]
 \epsscale{1}
 \plotone{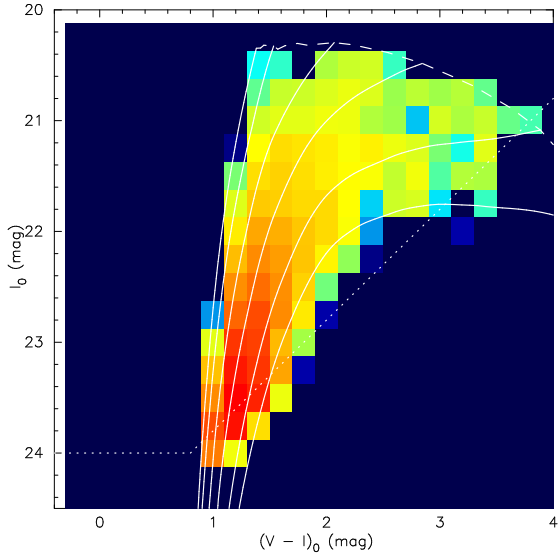}
 \caption[a]{Matched filter weight map on a log scale, trimmed to the
 color-magnitude region encompassing stars of metallicity $-2.31<{\rm
 [Fe/H]}<+0.00$. The resolution of this map is 0.2 (mag) $\times$ 0.25 (mag).
 }
 \label{fig:matchedcmd}
\end{figure}

We first divide the $I_0$ vs. $(V-I)_0$ plane into uniformly-spaced rectangular
grids. Then, based on the simple weighting of each CMD in a box defined by
these grids, in terms of the ratio of signal to contamination as expected
for the concerned CMD, we can optimally
boost the signal and suppress the contamination. Since we have the objective CMDs
(i.e., containing both the signal and contamination) as well as the CMDs containing
only the contamination, it is possible to obtain the so-called weight matrix,
which is given by subtracting the CMD for the contamination from the objective CMD
(using the correction method for the contamination as discussed in the previous
subsection), and finally we sum up these corrected CMDs in each field of M31's halo.
The weight matrix obtained by this procedure is shown in
Figure~\ref{fig:matchedcmd}. This is reasonably similar to the one derived by
\citet{Ibata2007}. Here, to construct the weight matrix, we have
selected the stars filed with 90\% completeness limit, the fainter stars
than an upper dashed line of Fig.~\ref{fig:cmdall} and the stars in the
possible metallicity range of $-2.31<{\rm [Fe/H]}<+0.00$, as an
example. Note that, as expected, the greatest weight arises at faint
magnitudes in the color range of $0.9<(V-I)_0<1.5$, so of course, stars
with this photometric property will contribute most strongly to the
following matched filter maps. 

Based on this method, we have analyzed all of the CMDs and obtained the matched filter
maps. In Figure~\ref{fig:m31map}, we present the final matched filter maps
over the entire survey region for four different ranges in metallicity. The maps are
smoothed with a Gaussian kernel over 0.054$\degr$. The limiting
magnitudes are chosen to be $I_0=24.0$ and $V_0=24.8$; these magnitudes
keep the detection efficiency over 90\% completeness in the entire
region. The images with the resolution of $0.018\degr \times 0.018\degr$
are displayed in a logarithmic scale for the left panels, whereas the
right panels are shown in a linear scale. The inner ellipse overlaid in these
maps represents a disk of inclination $77\degr$ and radius $2\degr$ (27
kpc), the approximate end of the regular H{\scriptsize I} disk. The
outer ellipse shows a $4\degr$ (55 kpc) radius ellipse which is flattened to
have the axis ratio of $c/a = 0.6$ [i.e., standard guide ellipse firstly drawn by
\citet{Ferguson2002}]. In addition, the cyan concentric dashed circles
show projected radii of 80 kpc. We note that the logarithmic-scaled maps are useful
for elucidating the global spatial variation of the M31 populations in
the inner halo and the large-scale smooth spread of the diffuse outer
halo, while the linear-scale maps are useful for the presentation of some faint
substructures in the outer halo, with signal boosted by the adopted CMD
filter. We summarize the characteristic features in these maps as follows.

\begin{itemize}
\item {\bf South Field\/}:\\
      The southern quadrant of the M31 halo has already been mapped out
      by \citet{Ibata2007}. We have confirmed their finding of some diffuse
      substructures in this part of the halo as given below. 
      \begin{enumerate}
       \item {\it The GSS\/}. This is the most conspicuous
	     overdense region in the M31 halo firstly detected by
	     \citet{Ibata2001}. As discussed in the
	     previous section, the GSS has high mean
	     metallicity of ${\rm [Fe/H]}=-0.7\pm0.5$ and slightly young
	     population with intermediate age of about 8 Gyr. It is interesting
	     to note that in Fig.~\ref{fig:m31map}c for the most metal-poor range,
	     the GSS has almost disappeared. 
       \item {\it Stream C\/} at $(\xi \sim 3.5, \eta \sim -2.5)$, which was
	     discovered by the MegaCam survey of \citet{Ibata2007}. This
	     stream is visible in the more metal-rich range except for
	     Fig.~\ref{fig:m31map}c. 
       \item {\it Stream D\/} at $(\xi \sim 2.5, \eta \sim -1.8)$, which was
	     also found by \citet{Ibata2007} for the first time. This strong,
	     stream-like structure is detected as a clumpy substructure
	     in the intermediate metallicity range 
	     (in particular, Fig.~\ref{fig:m31map}b).  
       \item {\it Major-Axis Diffuse Structure\/} at $(\xi \sim -2.5, \eta
	     \sim -4)$. A subtle overdense structure may be a part of
	     the diffuse structure extending on the major axis between a
	     projected distance of 50 and 100 kpc, as \citet{Ibata2007}
	     have mentioned in their paper. One can more clearly
	     identify the faint substructure in the map over the full
	     metallicity range of Fig.~\ref{fig:m31map}d. 
      \end{enumerate}
\item {\bf North Field}:\\
      Previous works have investigated only the limited northern parts of
      the M31 halo near the center. Here, we first explore the regions far from
      the center, i.e., the outer halo.
      \begin{enumerate}
       \item {\it Western Shelf\/} (W-shelf), which is a faint stream-like feature
             on the western side of M31, located at $(\xi \sim -1.5, \eta
	     \sim 1)$. This structure has already been detected in the map of
	     \citet{Ferguson2002}, having similar color to the GSS.
             There is a sharp cutoff of its stellar
	     distribution in the most metal-rich range of
	     Fig.~\ref{fig:m31map}a.  
       \item {\it A previously unknown stream\/} is seen at $(\xi \sim -3.5,
	     \eta \sim 3)$. This structure is most clearly identified
	     in the map of the intermediate metallicity range, i.e., in
	     Fig.~\ref{fig:m31map}b. We refer to this stream as ``Stream
	     E'' in the discussion below. 
       \item {\it A further faint low-surface-brightness overdense
	     structure\/} is detected toward $(\xi \sim -5.3, \eta \sim
	     4.3)$. The increased sensitivity with the full metallicity
	     range as shown in Fig.~\ref{fig:m31map}d has revealed this
	     structure, which we refer to as ``Stream F''.
       \item {\it The extension of the underlying halo\/} in the metal-poor
             range of the North-West minor axis field is comparable to
             that found in the South-East minor axis. This is not
             the case in the metal-rich range as shown in Fig.~\ref{fig:m31map}a. 
      \end{enumerate}
\end{itemize}

The maps displayed in Figure~\ref{fig:m31map} show the distribution of
the matched filter statistics, whereby the resulting counts are therefore
rather difficult to interpret directly. This is principally because
the matched filter method relies on a model of the stellar population
that one desires to detect and the statistics we measure depend on
a combined CMD of our survey fields, which however do not cover the
entire region of the M31 halo. 
In the next subsection, we discuss in more detail the populations that are
highlighted in Figure~\ref{fig:m31map}, by determining their fundamental
properties such as the metallicity and age based on the methods we have
already described in the previous subsection.

\section{Spatial Substructures}
%%% 5.1 %%%
\subsection{Western Shelf}\label{sec:Wshelf}
This W-shelf substructure, which has low surface brightness compared with even
the core of the GSS, was already mapped out as the overdense region of
the RGB stars by \citet{Ferguson2002} and \citet{Ibata2007}. Some
$N$-body simulations suggest that several other substructures in the M31 halo
including the W-shelf can be explained as the forward
continuation of the GSS
\citep[e.g.,][]{Fardal2007,Mori2008}. However, despite the results obtained
by such simulations, the detailed observational studies of the faint W-shelf
substructure have been little made \citep[e.g.,][]{Richardson2009},
especially to set useful limits on 
its origin. In this subsection, we discuss the fundamental properties of this
shelf by comparing them with those of the GSS based on our
results (\S~\ref{sec:stream}). In the analysis we describe below,
it is possible to remove systematic errors in this comparison between
the W-shelf structure and GSS, since the basic data of both fields,
such as distance, metallicity and age, are extracted based on
the same technique.

\begin{figure}[htpd]
 \begin{center}
  \epsscale{1}
  \plotone{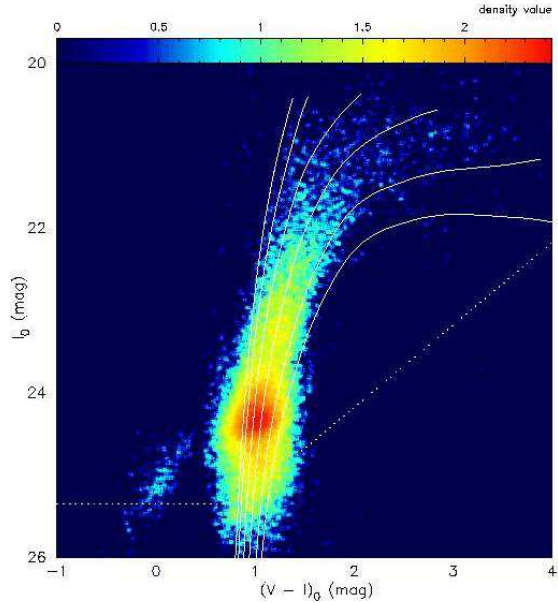}
  \caption[a]{A CMD in the $I_0$ vs. $(V-I)_0$ plane for stellar
  sources detected in the W-shelf field. The dotted lines denote the
  full range of the 50\% completeness level. The superposed solid lines
  in the CMD show theoretical RGB tracks from \citet{VandenBerg2006} for
  an age of 12 Gyr, [$\alpha$/Fe]$=+0.3$, and metallicities ({\it left
  to right\/}) of [Fe/H]$ = -2.31$, $-1.71$, $-1.14$, $-0.71$,
  $-0.30$ and $0.00$. The foreground/background contaminations
  have been subtracted based on the scaled control field of
  CompL103Bm16\_5.
  }
  \label{fig:cmdWS}
 \end{center}
\end{figure}

Figure~\ref{fig:cmdWS} shows a background/foreground subtracted CMD for
the W-shelf, combining the catalogs of the 17 kpc to 24 kpc fields along the
North-West minor axis. This range is chosen to avoid the contamination of
NGC~205 and to be outside of the W-shelf edge of $\sim 1.8\degr$ from
the center. This edge feature was also reported in Figure~1 and Table~1 of
\citet{Fardal2007}. The superposed solid lines in
the CMD show the same theoretical RGB tracks as
Fig.~\ref{fig:cmdall}. This CMD has similar morphology to that of the GSS,
as represented by a broad/metal-rich RGB and a slightly young
population attributable to a RC extending toward brighter magnitude (see also
Fig.~\ref{fig:cmdGS1}). However, it is noteworthy that the population of
the W-shelf with ${\rm [Fe/H]} \sim -0.3$ is somewhat sparser than the
GSS. Furthermore, because of the high quality data of the W-shelf
compared with that of the GSS, an old/metal-poor blue horizontal
branch (HB) extending toward $(V-I)_0 \sim 0$ and $I_0 \sim 25.2$ is clearly
identified, as well as the brightest part of the RGB bump, which indicates
the presence of metal-rich populations.
The slightly diffuse feature of the blue HB is simply
due to the large photometric error of $\pm0.2$ in color. We
cannot detect a prominent AGB bump as was seen in Fig.~\ref{fig:cmdGS1},
probably because the W-shelf field has lower surface brightness than the
GSS field. Note that our
CMD of the W-shelf contains yet an extra contamination of the underlying
halo component, although the effect appears to be not significant
if we consider the approximately 1 mag arcsec$^{-2}$ fainter correction
for the underlying halo than that for the W-shelf (see below Fig.~\ref{fig:m31sb}).

The distance to the W-shelf has been estimated by the sobel edge
detection algorithm as was already described in \S~\ref{sec:GStrgb}. We have
found the distance modulus of 24.51 mag corresponding to 798 ($\pm 40$) kpc,
on the assumption that the absolute $I$-band magnitude of the TRGB is $M_I^{\rm
TRGB} = -4.1\pm0.1$ for metal-poor TRGB stars. It suggests that the
W-shelf region probed by our Suprime-Cam pointing lies about 85 ($\pm
10$) kpc relatively in front of the GSS probed by our observation.
The error in this estimate depends only on a small photometric calibration
error, without considering any systematic errors such as the determination of
a TRGB absolute magnitude. 

\begin{figure}[htpd]
 \begin{center}
  \epsscale{1}
  \plotone{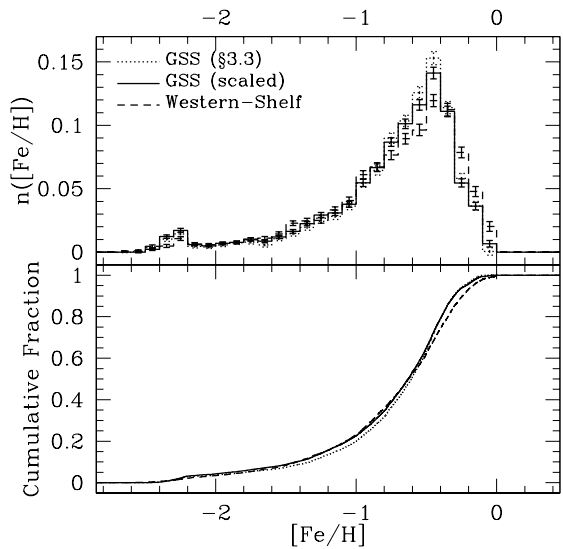}
  \caption[a]{{\it Upper\/}: Metallicity distribution functions (with error bars
  denoting Poisson uncertainties) for the W-shelf sample and the GSS
  sample, as interpolated from the chosen Victoria-Regina
  isochrones. The same photometric limits as Fig.~\ref{fig:mdfGS1} have
  been imposed. The control field, normalized with the gradient shown in
  Fig.~\ref{fig:diskdwarfgrad}, have been used to subtract the expected
  foreground/background counts in each of the metallicity bins. The
  dotted histogram is the same MD as the top-left histogram in the
  right-hand panel of Fig.~\ref{fig:metal_err}. {\it Lower\/}: The cumulative
  MDs for the two substructures. The solid and dashed distributions appear
  very similar but the K-S statistics suggests that they are
  different. 
  }
  \label{fig:mdfWS}
 \end{center}
\end{figure}

Based on our estimated distance to the W-shelf, we now proceed to
construct its MD, assuming stellar population with [$\alpha$/Fe]$ = +0.3$
and age of 10 Gyr based on the Victoria-Regina theoretical isochrones from
\citet{VandenBerg2006}. Figure~\ref{fig:mdfWS} shows the MD and its
cumulative distributions in the W-shelf field plotted as the dashed
histogram. The foreground/background contaminations have been subtracted
based on the scaled control field as described earlier.
In contrast, solid and dotted histograms denote
the MDs of the GSS. The dotted histogram is the same
MD as the top-left histogram in the right panel of
Fig.~\ref{fig:metal_err}, and the solid histogram is the scaled MD of
the GSS; the contamination for the former dotted MD is
estimated based on the SE3 field which is located near our observed 
GSS along the minor axis (see Fig.~\ref{fig:map}), while the
contamination for the latter solid MD is derived from a control field
far from the M31 center, that is, free from the population of the underlying
halo in M31. This consistent correction procedure for the
contamination makes it possible to deduce the nature of the W-shelf itself,
based on the comparison between the dashed and solid histograms. 

As is evident, these two MDs are basically the same in the metal-poor
range of ${\rm [Fe/H]} \la -1$, characterized by a prominent metal-poor
tail. A blue HB feature identified in the CMD of
Fig.~\ref{fig:cmdWS} also suggests that this metal-poor tail is
attributable not to the contamination of the metal-rich AGB stars
but to the existence of the remarkably metal-poor and old population.
This suggests that a progenitor(s) of such overdense substructures
in the inner part of the halo had undergone primordial star formation
before it (they) interacted dynamically with M31 2 Gyr ago,
as simulated by \citet{Fardal2007}. 
In contrast, the properties of these two MDs in the metal-rich range appear
slightly different, although mean/median/peak metallicities of the W-shelf
are in agreement with those of the GSS, ${\rm [Fe/H]_{\rm mean}}
= -0.75$, ${\rm [Fe/H]_{\rm med}} = -0.63$ and ${\rm [Fe/H]_{\rm peak}}
= -0.5$. The two-sided Kolmogorov-Smirnov (K-S) test \citep{Press1992} actually
rejects the null hypothesis that the two data sets are drawn from the same
distribution, with a probability less than $P=0.03$, which is
derived from the maximum value of a vertical deviation
between the two cumulative distributions (as in a bottom panel of
Fig.~\ref{fig:mdfWS}). This indicates that these two MDs are different
in a statistically significant manner.

We proceed to investigate the mean age of the W-shelf population based on
the constructed MD and the RC magnitude, through the same manner as the
analysis of the GSS in the previous section. By least-square
fitting to the LF model with a Gaussian function and a straight line to the
measured luminosity function, we estimate $I_0^{\rm RC} = 24.31 \pm
0.16$ and $M_I^{\rm RC} = -0.20$, whereby we find $8.2_{-3.1}^{+1.6}$ Gyr
for the mean age of the W-shelf stars. 

Based on these analyses, we can conclude that the fundamental properties
of the W-shelf population such as the mean metallicity and the mean age
are excellently consistent with those of the GSS probed by
our Suprime-Cam pointing. Also, there exists a small but statistically
meaningful difference in the fraction of metal-rich stars between
the W-shelf and the GSS.
We emphasize that this difference in the metal-rich population is 
robust because our data are taken with higher
detection efficiency, much smaller photometric errors
and larger statistics than previous studies.
To understand the origin of this small difference in the metal-rich
population, it is useful to consider the simulation results, such as
\citet{Fardal2007}, which show that
the W-shelf structure was formed by the third pericentric passage of the 
same progenitor as that producing the GSS.
In particular, the model by \citet{Fardal2008} suggests the
existence of a population gradient in the progenitor galaxy, i.e.,
spatial gradient in metallicities, as found mainly in
disk galaxies. The observational evidence for the population gradient
was discovered by \citet{Ibata2007}, in their comparison between the core and
envelope population of the GSS. Our finding of the small
difference in the fraction of metal-rich stars between the W-shelf and
the GSS may be caused by such a population gradient within
the progenitor galaxy. Of course, there remains yet another possibility
that the origins of the GSS and the W-shelf are different.

%% 5.2 
\subsection{South-East Streams~C and D}\label{sec:SCD}

\begin{figure*}[htpd]
 \begin{center}
  \epsscale{1}
  \plottwo{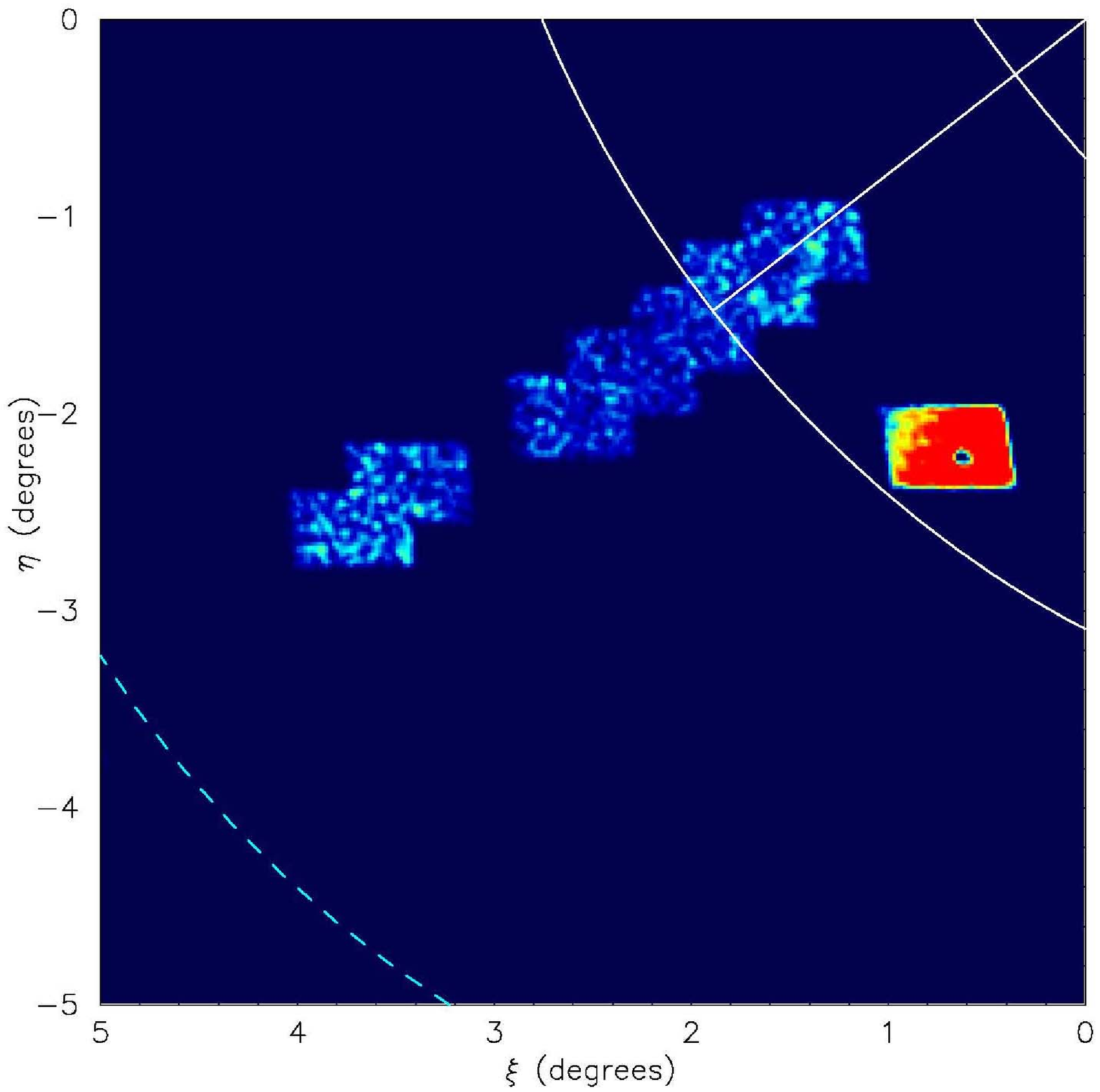}{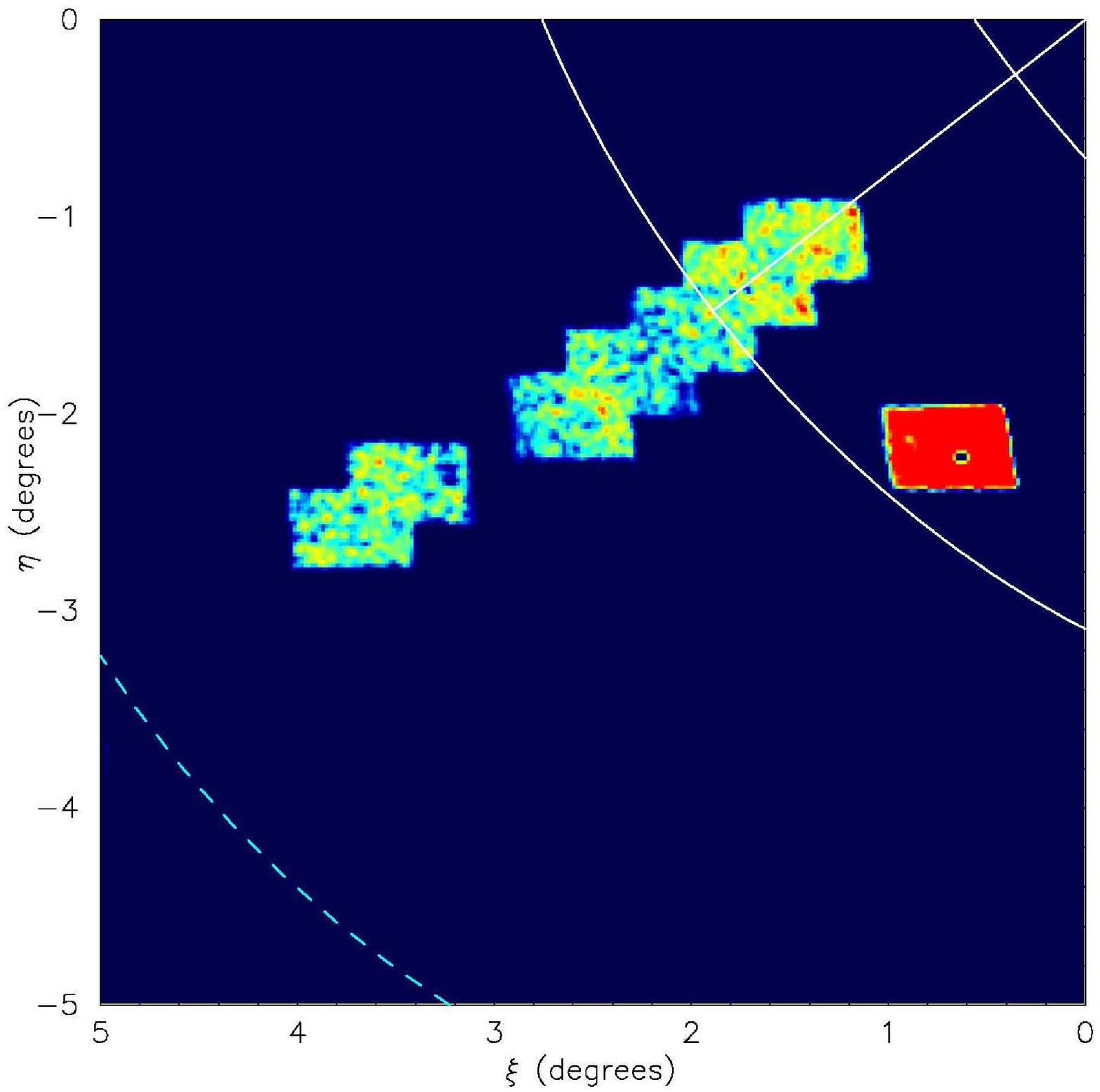}
  \caption[a]{The Stream~C and D regions shown zoomed-in (from
  Fig.~\ref{fig:m31map}) with slices in [Fe/H] ($-0.71 < {\rm [Fe/H]} <
  +0.00$ for left panel and $-1.71 < {\rm [Fe/H]} < -0.71$ for
  right panel).
  }
  \label{fig:m31mapCD}
 \end{center}
\end{figure*}

Figure~\ref{fig:m31mapCD} shows close-up maps of the south-east
minor-axis region in the proximity of our GSS field at $(\xi
\sim 0.5, \eta \sim -2.2)$, divided [Fe/H] into two non-overlapping
ranges ($+0.00$ to $-0.71$ and $-0.71$ to $-1.71$). This division for metallicities
makes it clear that two overdense regions are indeed present, contrary to the
expected smoothly declining stellar density of the halo. These overdense
regions, which were first discovered by \citet{Ibata2007}, are called
Stream `D' (at $\xi \sim 2.5, \eta \sim -1.8$) and `C' (at $\xi \sim 3.5, \eta \sim
-2.5$) in order of increasing declination. Actually, they are aligned
in a direction perpendicular to the minor axis and more extending like a
stream feature on both sides of our observed regions; we see only a part
of the two streams. 

\begin{figure*}[htpd]
 \begin{center}
  \epsscale{1}
  \plottwo{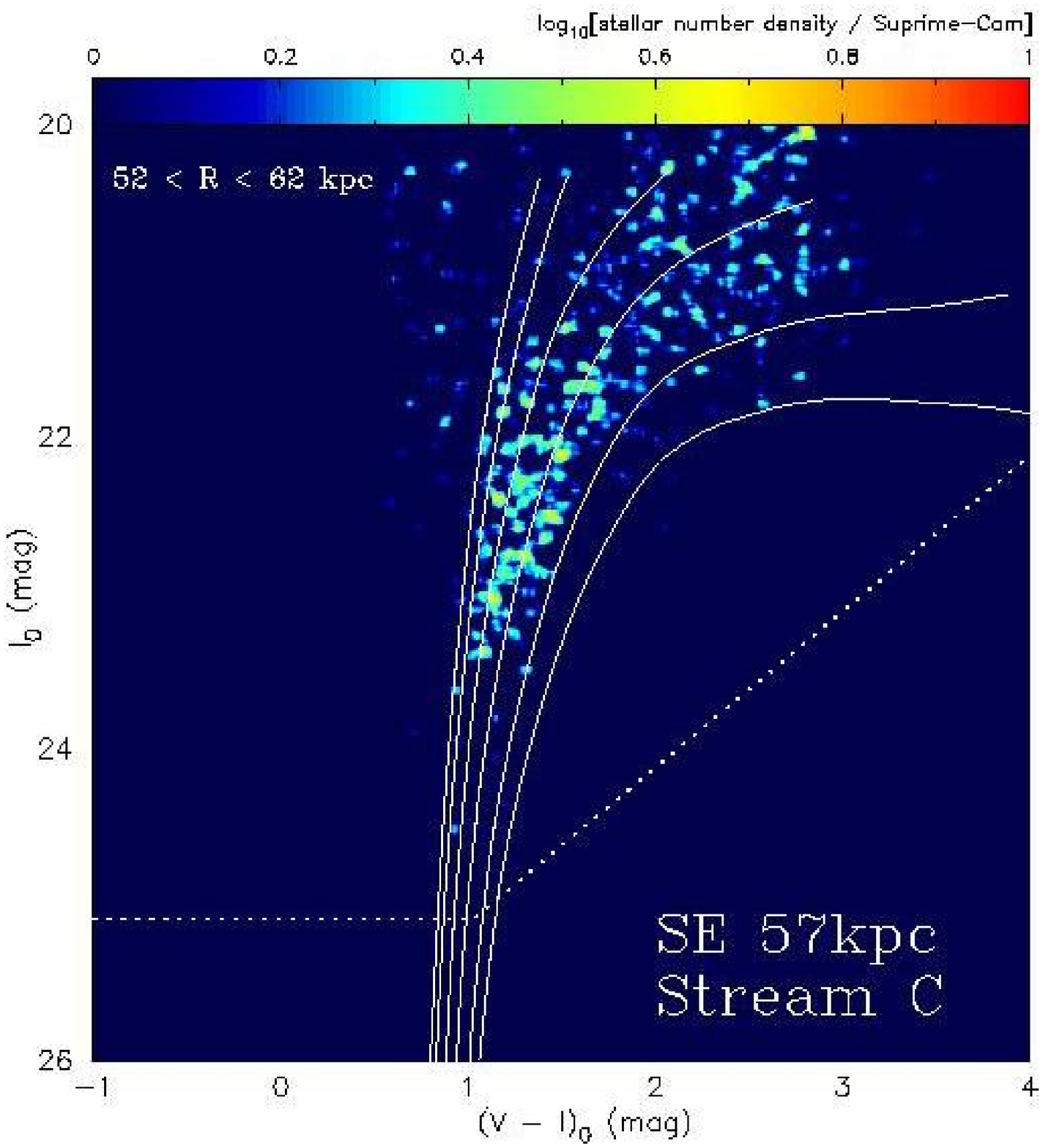}{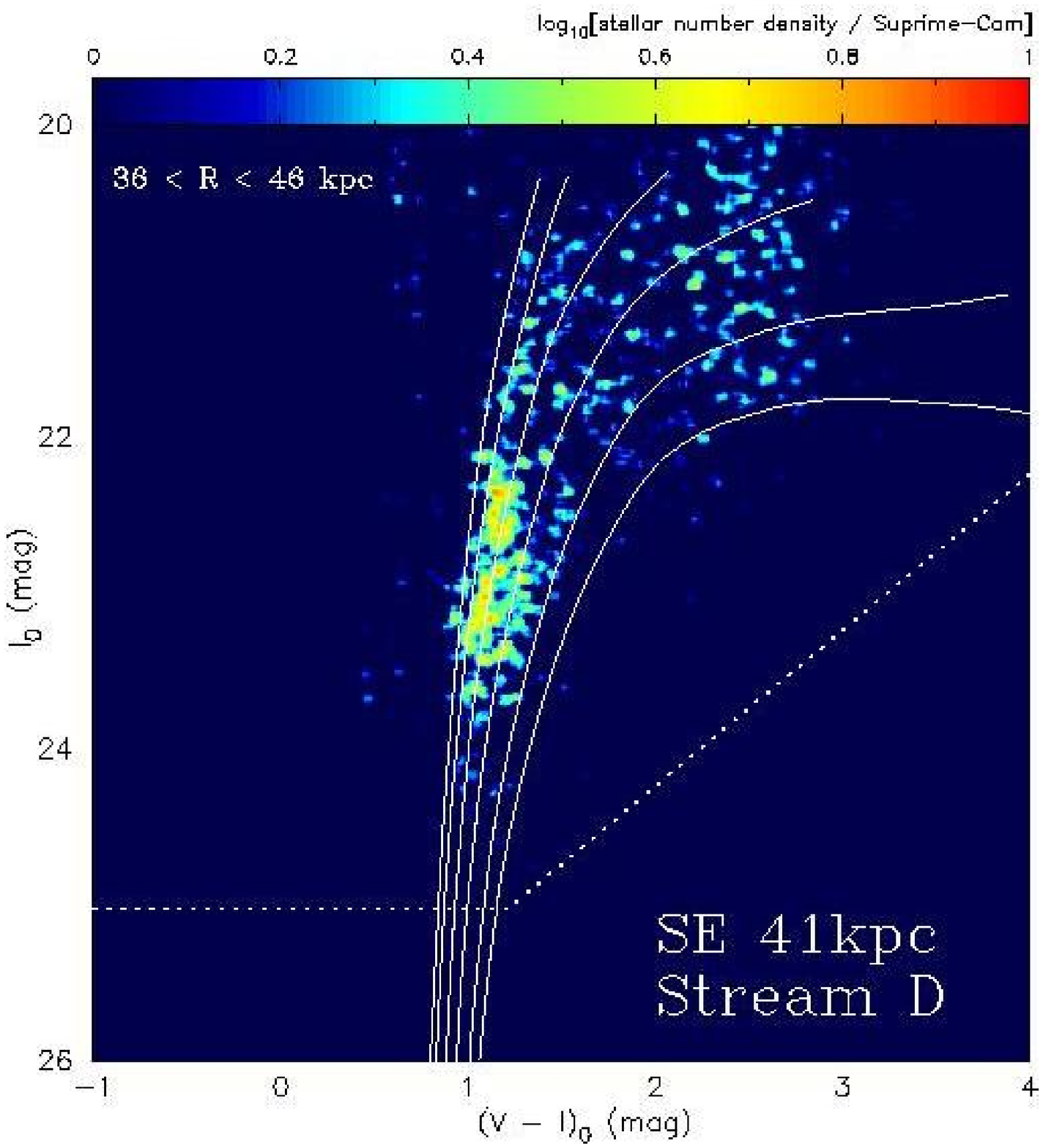}
  \caption[a]{Contamination-uncorrected CMDs for two adjacent Suprime-Cam
  fields along the south-east minor axis, evaluated with stringent criterion
  in SExtractor (see text). The superposed solid lines in the CMD show
  theoretical RGB tracks from \citet{VandenBerg2006} for an age of 12
  Gyr, [$\alpha$/Fe]$=+0.3$, and metallicities ({\it left to right\/}) of
  [Fe/H]$ = -2.31$, $-1.71$, $-1.14$, $-0.71$, $-0.30$ and $0.00$. 
  }
  \label{fig:cmdSCD}
 \end{center}
\end{figure*}

Unfortunately these substructures show about $1/10$ lower surface
brightness than the GSS and the W-shelf, and the detection limits
for these data are somewhat worse than for the control field data.
Thus, the method for subtracting the background contaminations in the CMD
would need caution, when we simply compare with the control field having
much better detection efficiency.
Therefore, in Figure~\ref{fig:cmdSCD} we display the
contamination-uncorrected CMDs of the stream-like structures within the
range of projected $R\sim$ 50--60 (35--45) kpc for Stream~C (Stream~D). 
These CMDs are smoothed with a Gaussian kernel over 3 pixels ($d(V-I)=0.03, dI=0.03$) 
in order to diminish the effect of spiky noises attributable to low statistics. 
Here, as a further step, we have tried to roughly remove background
galaxies in combination with a criterion based on the FWHM parameter
derived from SExtractor \citep{Bertin1996}. The criterion was chosen
along the stellar sequence in the space of magnitude vs. FWHM
\citep[e.g.,][]{Kashik2004}. Since the detection method of sources in
SExtractor suffers from blending objects, the detection limit gets into
somewhat shallower in response to changes in worse seeing
size. Nonetheless, both of the CMDs of Stream~C and D as in
Fig.~\ref{fig:cmdSCD} imposed by the stringent criterion present a clear
RGB sequence along the theoretical isochrones at $I_0 \la 23.5$ and
$(V-I)_0 \sim 1$. 

\begin{figure}[htpd]
 \begin{center}
  \epsscale{1}
  \plotone{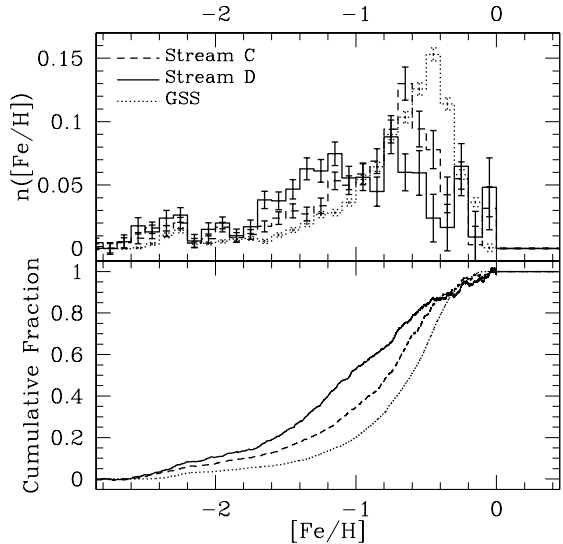}
  \caption[a]{MDs and cumulative MDs of the Stream~C and D,
  using the Victoria-Regina theoretical isochrones for an
  age of 12 Gyr and [$\alpha$/Fe]$ = +0.3$ (as shown in
  Fig.~\ref{fig:cmdSCD}), on the assumption that these substructures are
  as old as a diffuse, classical halo.}
  \label{fig:mdfSCD}
 \end{center}
\end{figure}

A prominent feature of the CMDs is that Stream~C is predominantly
metal-rich, being comprised of the stars in the range $-1.14 <
{\rm [Fe/H]} < -0.30$, while in Stream~D metal-poor stars with the
metallicity of $-1.71 < {\rm [Fe/H]} < -0.71$ are well populated. The
corresponding MDs of the two streams given in Figure~\ref{fig:mdfSCD}
statistically confirm these features, where we have made the subtraction
of contaminations as was done for the GSS and W-shelf fields. The mean
metallicity with the standard deviation of Stream~C (Stream~D) is ${\rm
[Fe/H]} = -0.89 \pm 0.55$ ($-1.03 \pm 0.61$) based on 1684 (1288) stars,
assuming an age of 12 Gyr, [$\alpha$/Fe]$=+0.3$ and distance modulus of
$24.43$ from \cite{Freedman1990}. The characteristics of this population
difference are also suggested by the previous study based on the
CFHT/MegaCam survey \citep{Ibata2007}; we have confirmed the existence
of these faint outer streams based on our Subaru/Suprime-Cam data with
higher S/N.

\begin{figure}[htpd]
 \begin{center}
  \epsscale{1}
  \plotone{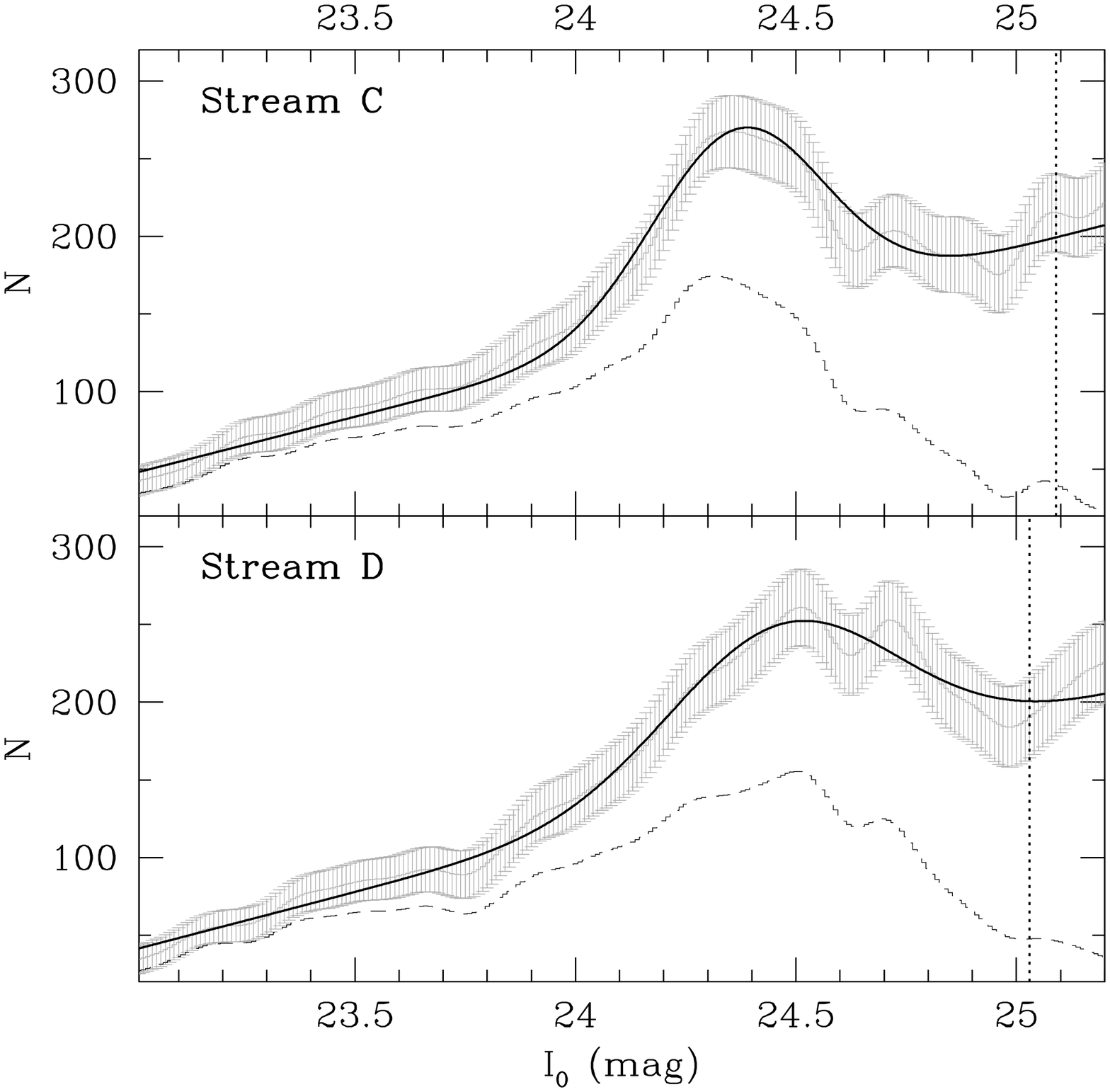}
  \caption[a]{Luminosity functions of the Stream~C and D around the
  RC. Solid line is the completeness-corrected LF with Poisson
  uncertainties, while the dotted line is the completeness-uncorrected LF. Red
  line is the best-fitted model with a simple model consisting of a Gaussian
  plus a straight line. We determine a Gaussian peak as the magnitude of the
  RC. Vertical dotted line denotes the 50\% completeness limit. 
  }
  \label{fig:LFSCD}
 \end{center}
\end{figure}

Now we further determine the mean age of the dominant
populations based on the MD and RC magnitude. We present the luminosity
functions of the Stream~C and D in the expected magnitude range of the
RC when assuming the aforementioned distance in
Figure~\ref{fig:LFSCD}. The RC magnitudes of Stream~C and D are $I_0 =
24.37$ and $24.47$, respectively. By determining a Gaussian peak as RC
magnitude and convolving the MDs of Fig.\ref{fig:mdfSCD} with the
magnitude, we estimate the mean ages of the dominant populations of
these streams as $\sim 9.3^{+0.9}_{-1.4}$ Gyr for Stream~C and $\sim
9.6^{+0.7}_{-1.1}$ Gyr for Stream~D. The errors are based on the
photometric errors, which strongly depend on the relation between RC
magnitude and metallicity distribution at younger part as shown in
Fig.~\ref{fig:rejkuba}. 

%% 5.3 
\subsection{Major-Axis Structure}\label{sec:MA}

\begin{figure}[htpd]
 \begin{center}
  \epsscale{1}
  \plotone{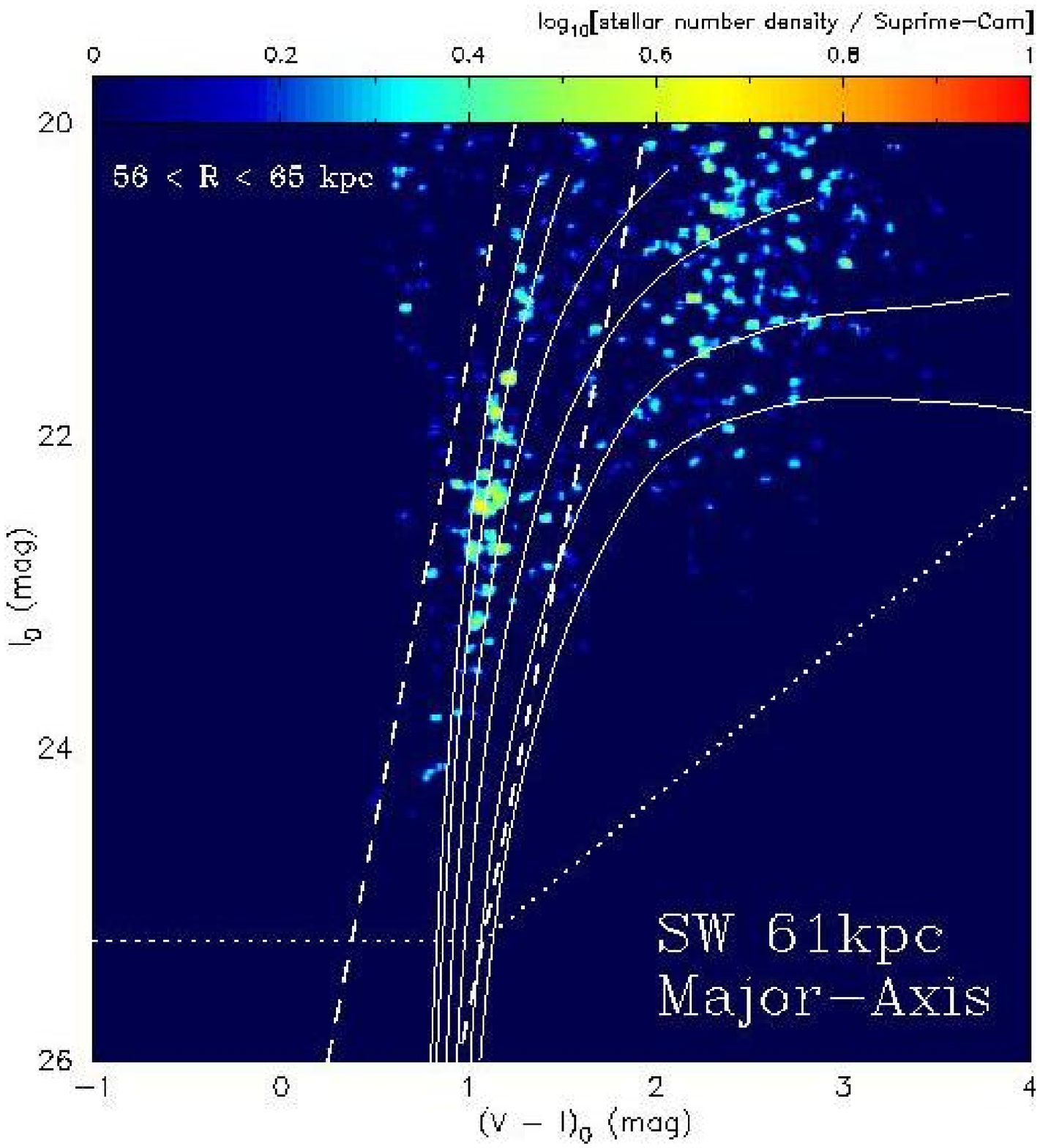}
  \caption[a]{Contamination-uncorrected CMD of the major-axis diffuse
  population, evaluated after stringent criterion in SExtractor (see
  text). 
  }
  \label{fig:cmdMA}
 \end{center}
\end{figure}

\begin{figure}[htpd]
 \begin{center}
  \epsscale{1}
  \plotone{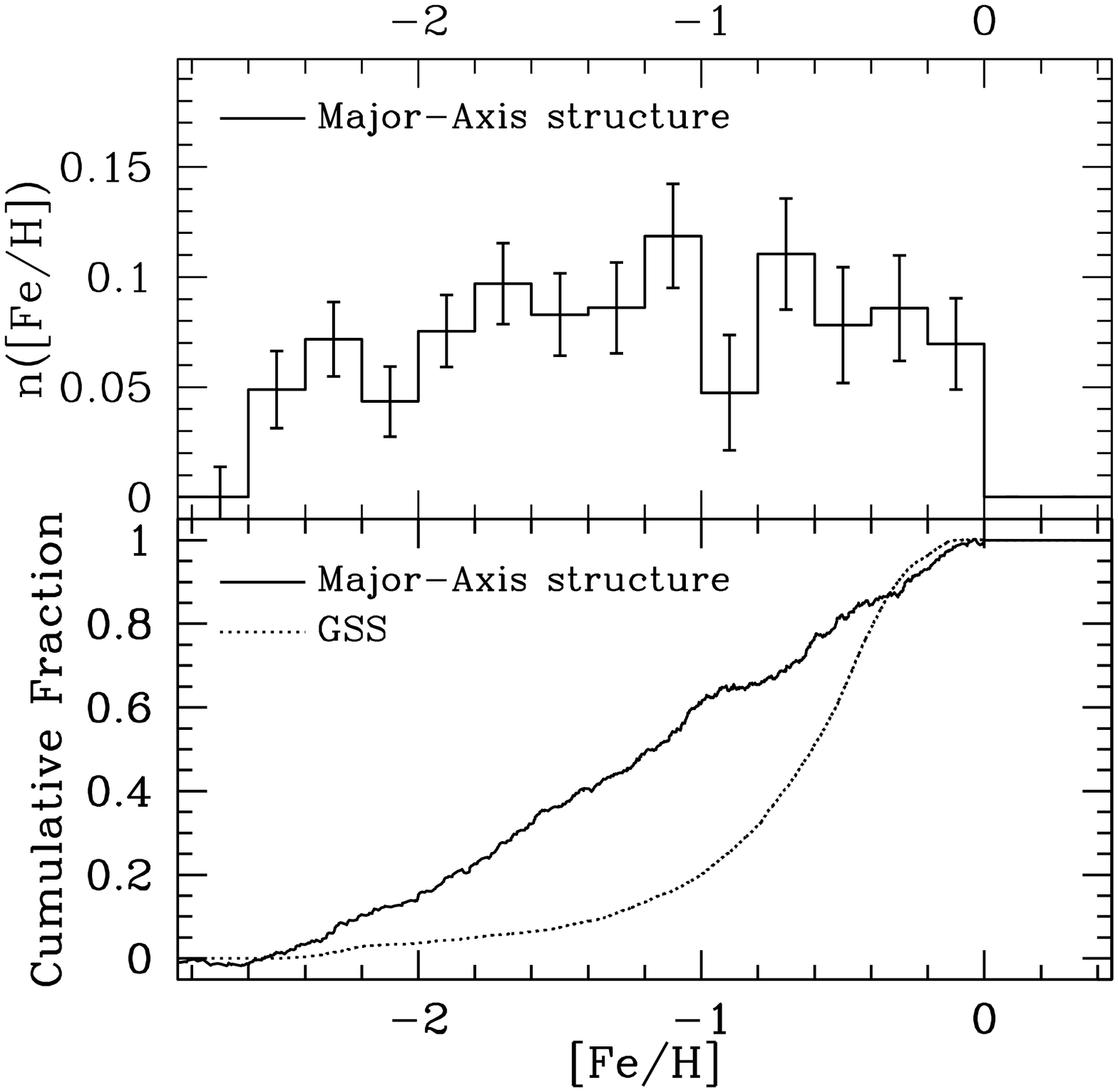}
  \caption[a]{The MD and the cumulative MD of the major-axis diffuse
  structure, using the Victoria-Regina theoretical isochrones for an age
  of 12 Gyr and [$\alpha$/Fe]$ = +0.3$. We also select a bluer sample
  with $(V-I)_0 < 1.8$ to avoid the heavy contamination of disk dwarfs. 
  }
  \label{fig:mdfMA}
 \end{center}
\end{figure}

The faint diffuse structure extended along the south-west major axis
between a projected distance of 50 and 100 kpc was mentioned in
\citet{Ibata2007}. Although our Suprime-Cam survey regions are slightly
different from those covered by CFHT/MegaCam and both of the
two surveys do not encompass the entire region of the diffuse
major-axis structure, the CMD of these areas shows a clearly
well-populated metal-poor RGB sequence with a dominant population in
the metallicity range of $-2.31 < {\rm [Fe/H]} < -1.14$
(Figure~\ref{fig:cmdMA}). The corresponding MD and cumulative MD using
about 420 stars and the Victoria-Regina theoretical isochrones for an
age of 12 Gyr and [$\alpha$/Fe]$ = +0.3$ are displayed in
Figure~\ref{fig:mdfMA}. They are constructed by imposing further color
criterion with $(V-I)_0 < 1.8$ in order to avoid the heavy foreground
disk dwarf contamination, considering the existence of the dominantly
populated metal-poor stars as shown in the CMD. The mean metallicity of
the population is ${\rm [Fe/H]} \sim -1.2$. Our confirmed stellar
population in the diffuse structure on the south-west major-axis is in
nice agreement with that of \citet{Ibata2007}, and there is no
population gradient of the diffuse structure in both of the survey
fields. 

\begin{figure}[htpd]
 \begin{center}
  \epsscale{1}
  \plotone{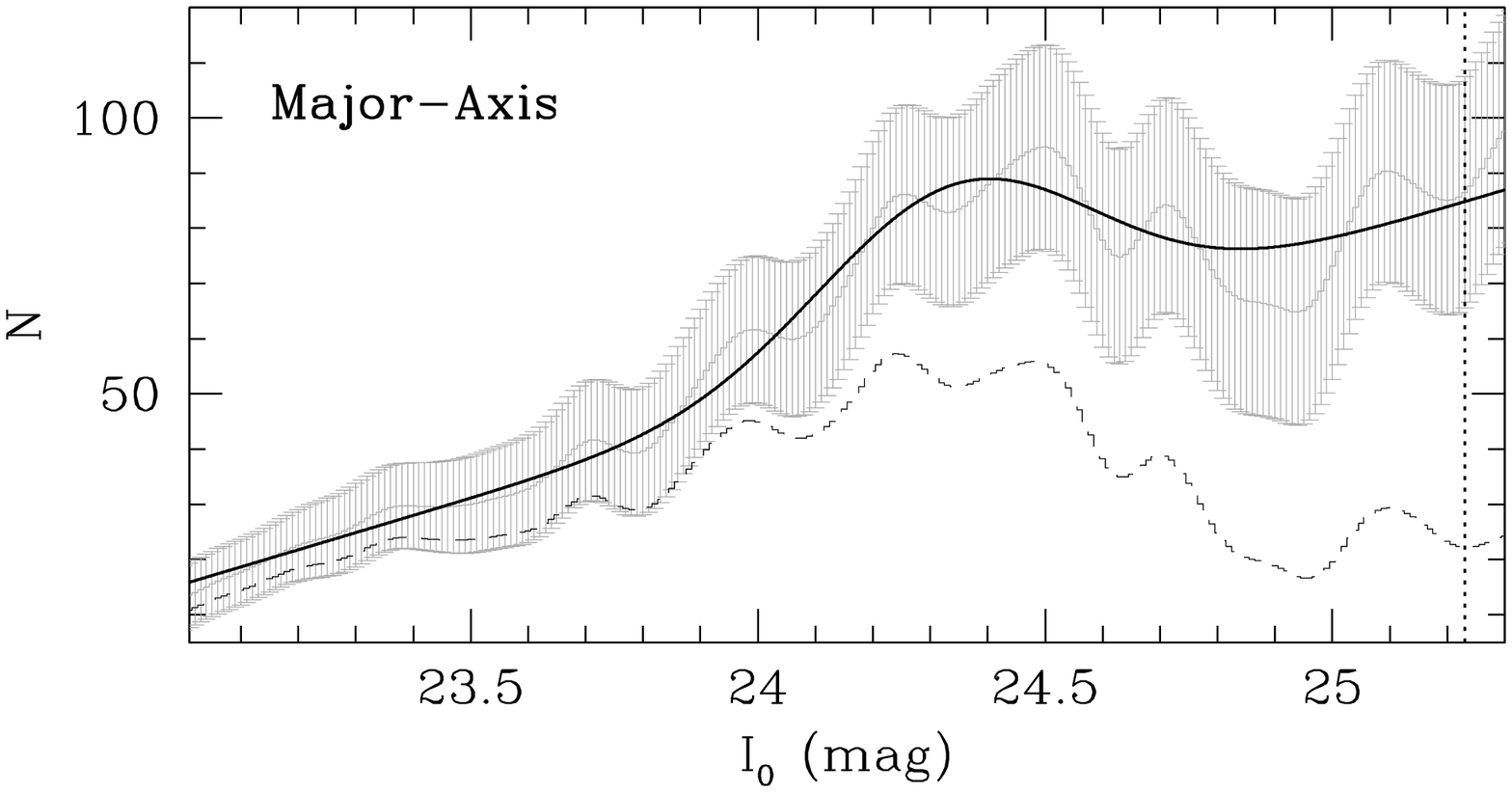}
  \caption[a]{Luminosity functions of the major-axis diffuse structure
  around the RC.
  }
  \label{fig:LFMA}
 \end{center}
\end{figure}

\citet{Ibata2007} concluded that the two stellar populations of the
diffuse structure and the GSS are different and likely
unrelated despite the visual impression that the density map of their
Figure~21 gives indications that the major-axis population merges with the GSS.
However, in the LF shown in Figure~\ref{fig:LFMA}, we detect a subtle peak
similar to the RC over Poisson noise, suggesting that there is an
intermediate-age population in the diffuse halo structure. If the peak
morphology with $I_0 = 24.34$ is actually attributable to the RC, we
can estimate the mean age of the population as $\sim
8.6^{+0.9}_{-1.5}$ Gyr. This estimation provides a comparable age to the GSS.
Based on their numerical simulation, \citet{Fardal2008} actually reproduced
the decline in the mean metallicity from the central core of the GSS
to its cocoon to the southwest, using an accretion model of a disk-like
progenitor having a strong radial metallicity gradient.
Thus, this may suggest that the metal-poor/young population of the diffuse
halo structure originated from the debris of the GSS's
progenitor. So we cannot readily rule out the association of the diffuse
structure with the GSS. A much deeper and wider survey reaching down
to the main-sequence turn-off by a planned thirty meter telescope is
necessary to more accurately determine the age and the distance to the
diffuse structure. Additionally the spectroscopic survey to measure
kinematics will be needed to understand the origin of these halo
structures in detail.

\subsection{New North-West Streams~E and F}\label{sec:SEF}

Here, we report on two new substructures seen in the
north-west portion of the outer halo of M31. In contrast to the south quadrant of M31's halo
which has been studied by a few other groups, the north quadrant has remained
unexplored until now, in part because its low Galactic latitude
($b\sim-10^\circ$) results in a high degree of contamination by foreground
Milky Way dwarf stars.

\begin{figure*}[htpd]
 \begin{center}
  \epsscale{1}
  \plottwo{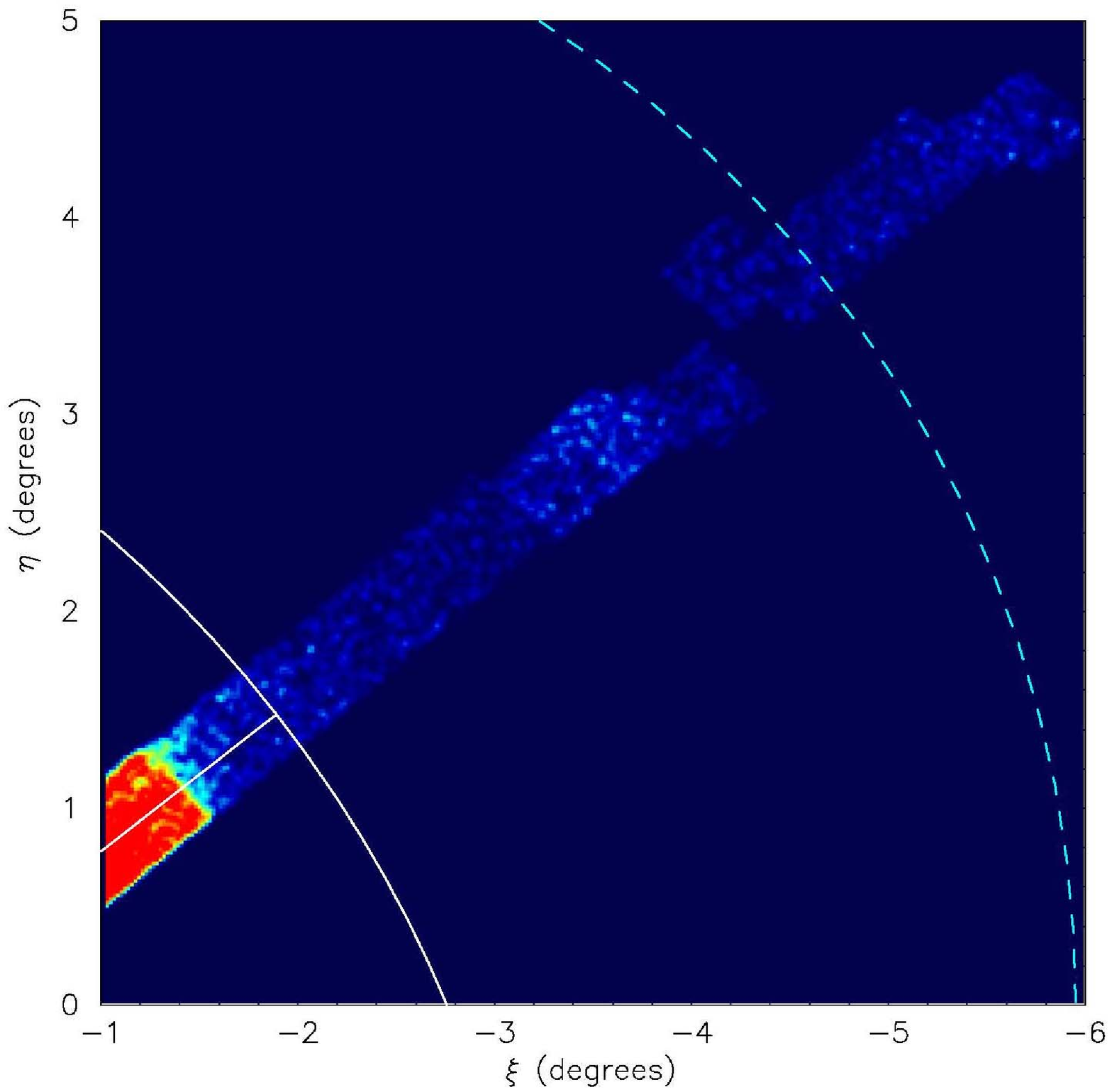}{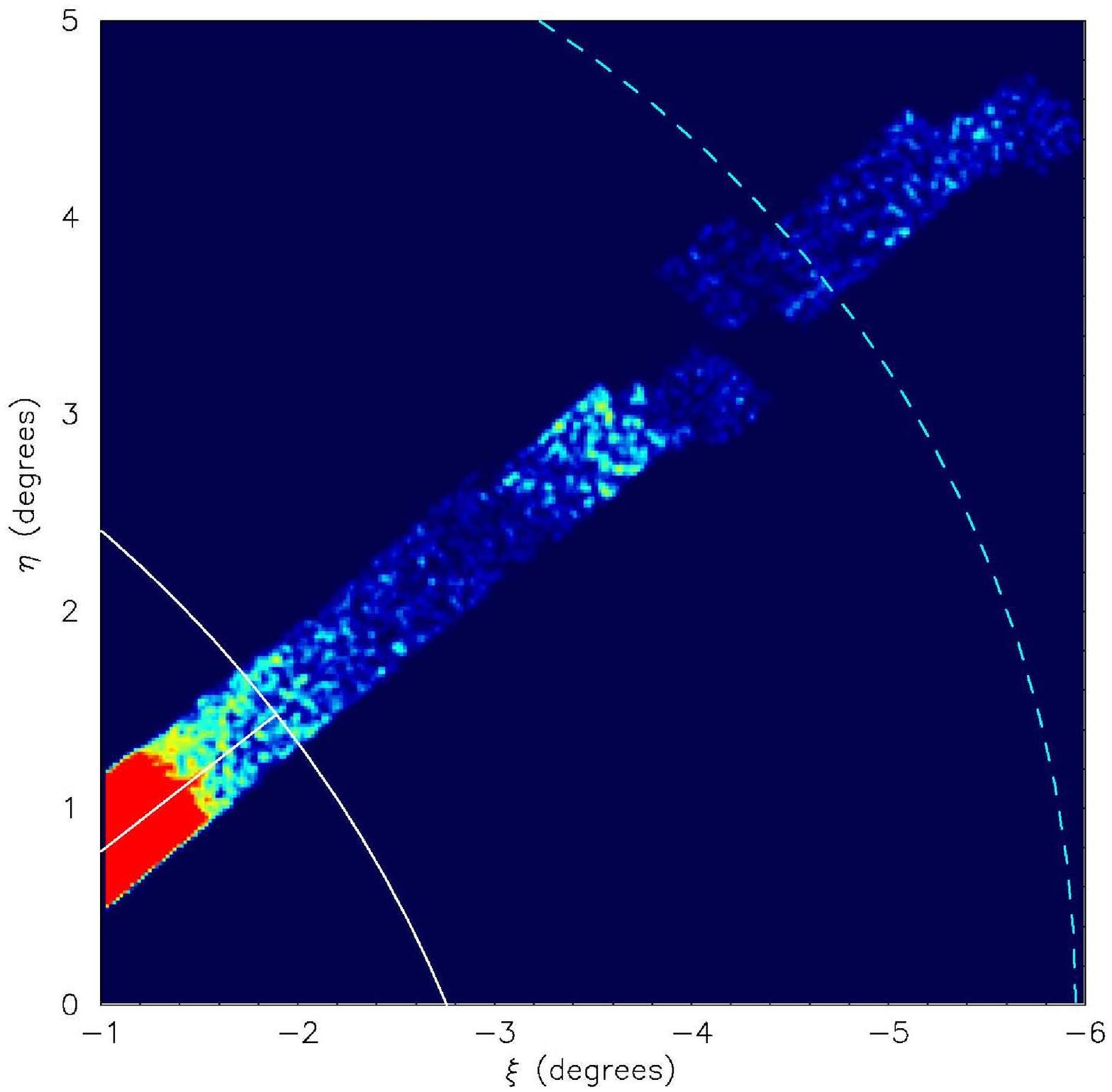}
  \caption[a]{The Stream~E and F region shown zoomed-in (from
  Fig.~\ref{fig:m31map}) with slices in metallicities ($-0.71 < {\rm [Fe/H]} <
  +0.00$ for left panel and $-1.71 < {\rm [Fe/H]} < -0.71$ for
  right panel).
  }
  \label{fig:m31mapEF}
 \end{center}
\end{figure*}

\begin{figure*}[htpd]
 \begin{center}
  \epsscale{1}
  \plottwo{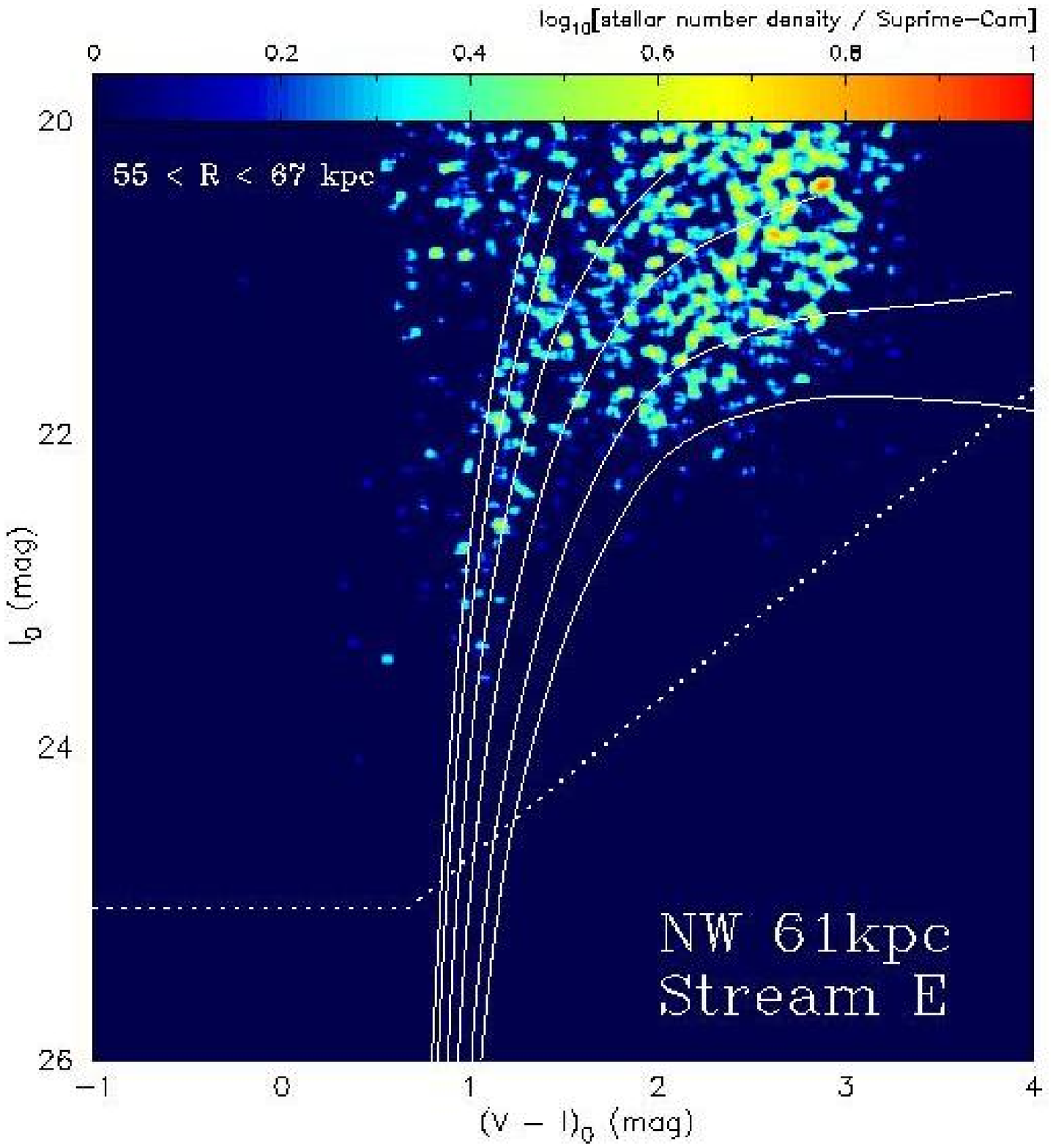}{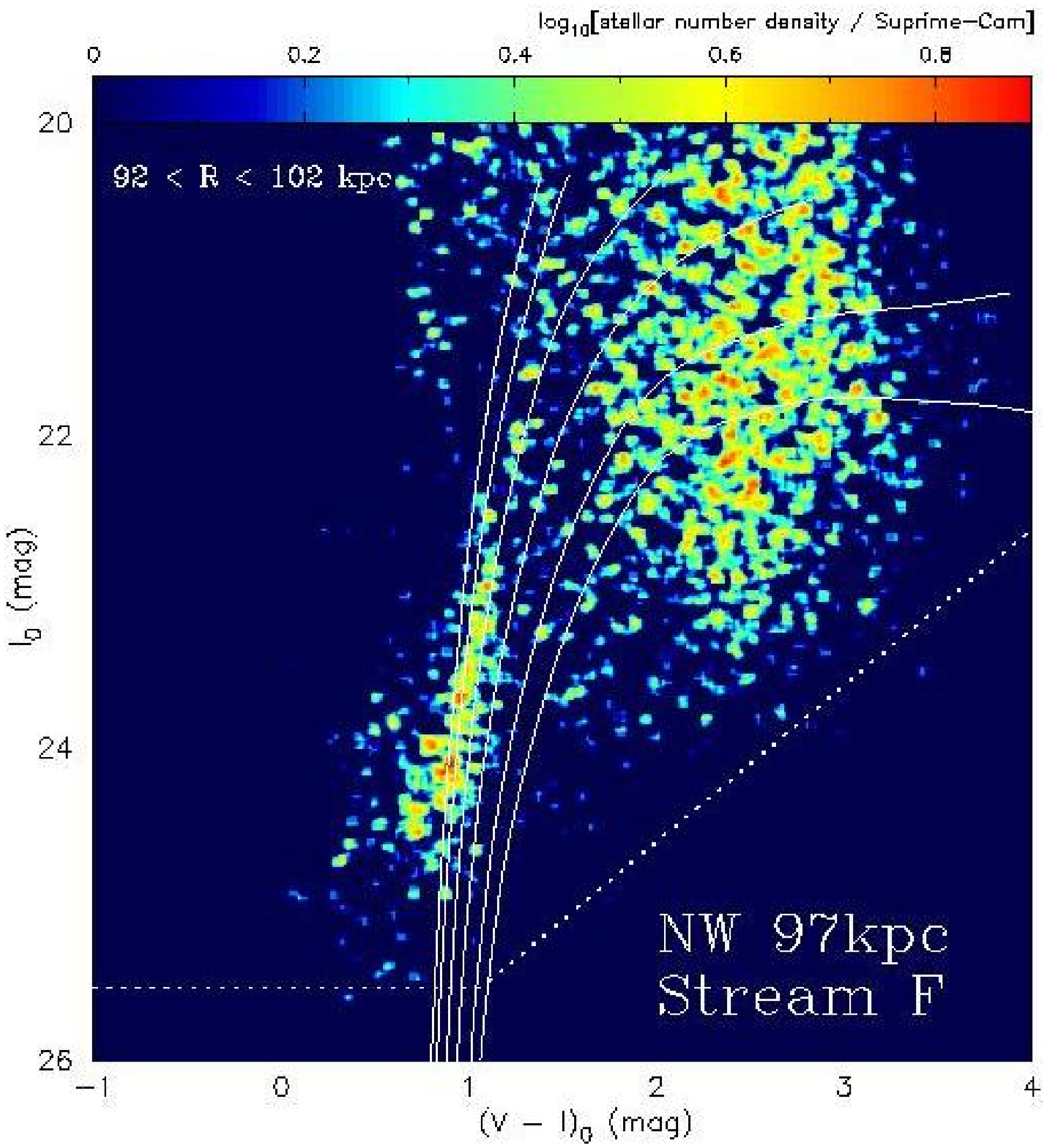}
  \caption[a]{The CMDs of Stream~E and F, as derived in the same manner as
  Fig.~\ref{fig:cmdSCD}. The CMD of Stream~E is made of combined
  catalog from 55 to 67 kpc, while the one of Stream~F is constructed
  by summed-up catalogs from 92 to 105 kpc. The superposed solid lines in
  the CMD show theoretical RGB tracks from \citet{VandenBerg2006} for an
  age of 12 Gyr, [$\alpha$/Fe]$=+0.3$, and metallicities ({\it left to
  right\/}) of [Fe/H]$ = -2.31$, $-1.71$, $-1.14$, $-0.71$, $-0.30$ and 
  $0.00$.  
  }
  \label{fig:cmdSEF}
 \end{center}
\end{figure*}

The two overdense structures on the north-west minor axis at $R \sim 60$
and 100 kpc can be deduced from the visual inspection of the matched
filter maps in Figure~\ref{fig:m31mapEF}. This figure shows close-up maps of the
north-west minor-axis region, divided by [Fe/H] into two non-overlapping
ranges ($+0.00$ to $-0.71$ and $-0.71$ to $-1.71$). 
Here, we refer to these as Stream~E and F, corresponding to the inner 60 kpc and
outer 100 kpc substructures in the discussion below. Their dominant
RGB population has a metallicity range of $-2.31 < {\rm
[Fe/H]} < -0.71$ in the Victoria-Regina isochrones, as can be seen in
Figure~\ref{fig:cmdSEF}. Although the data of Stream~E were obtained
under the worst conditions in our observing run, we can confirm
well-populated metal-poor stars near the bright end of the M31 RGB. This
population has a very similar CMD and surface brightness to those of the
major-axis diffuse structure discussed in \S~\ref{sec:MA}. It also
follows that Stream~F shows a clear metal-poor RGB with a narrow
metallicity range, as shown in the right panel of
Fig.~\ref{fig:cmdSEF}. This population has a very similar CMD to that of
faint Stream~A located at $R \sim 120$ kpc along the south-east minor
axis as found by \citet{Ibata2007}. 

\begin{figure}[htpd]
 \begin{center}
  \epsscale{1}
  \plotone{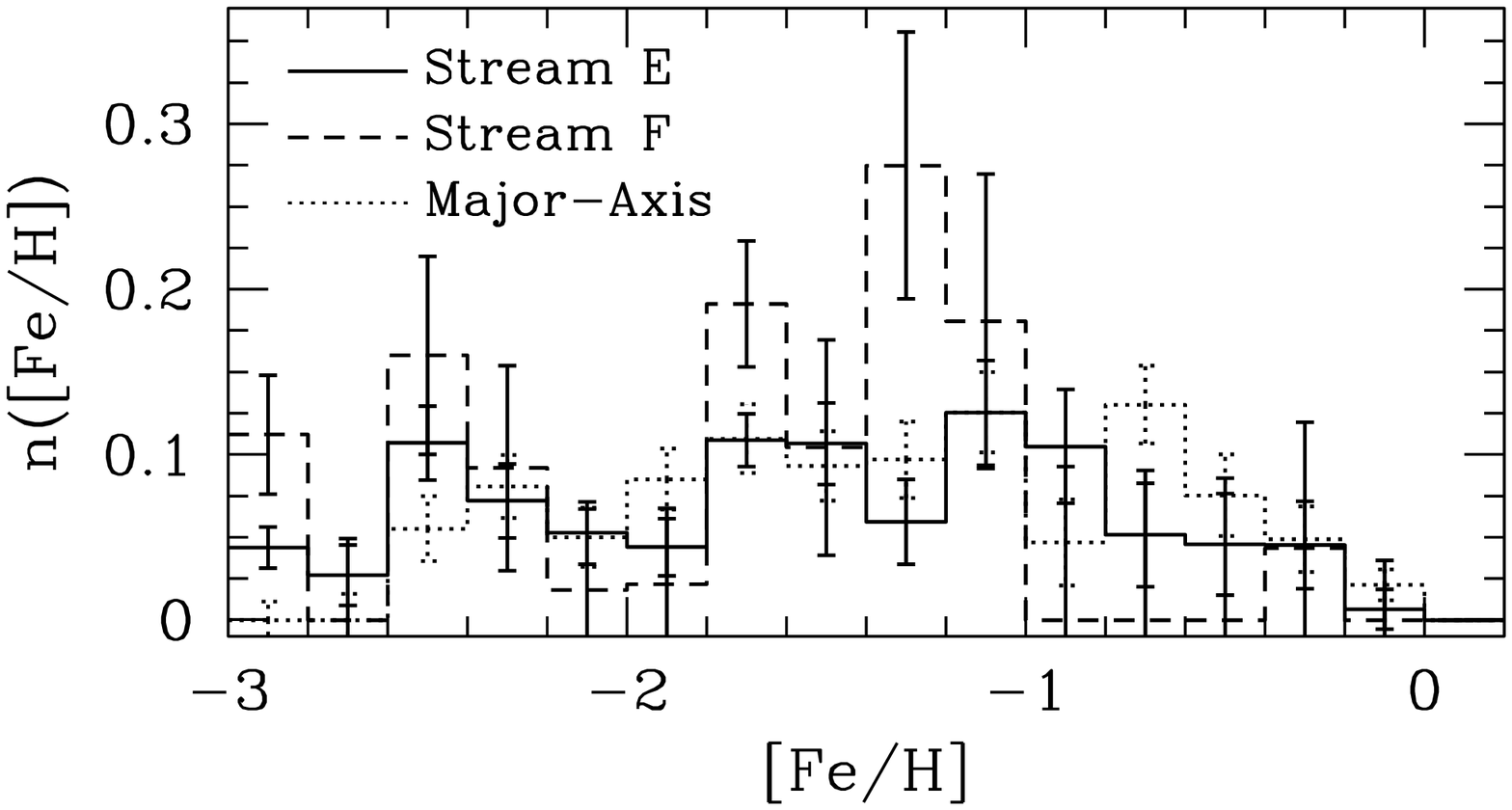}
  \caption[a]{Metallicity distribution function and the cumulative
  distribution of the Stream~E and F, statistically constructed in the
  same manner as the MD of the major-axis diffuse structure. 
  }
  \label{fig:mdfSEF}
 \end{center}
\end{figure}

The MDs corresponding to these two newly-found streams are shown in
Figure~\ref{fig:mdfSEF}, where we use the Victoria-Regina theoretical
isochrones for an age of 12 Gyr and [$\alpha$/Fe]$ = +0.3$. In the north
region of the M31 halo, the number of Galactic dwarf stars exponentially
increases as shown in Fig.~\ref{fig:diskdwarfgrad}, while the stellar
density of the M31 halo exponentially decreases with increasing radius
\citep[e.g.,][]{Gilbert2006}. Therefore, because of the large uncertainty in
correcting the disk contamination and the low statistics in extracting
the intrinsic population of the M31 halo,
we further impose the additional criterion of $(V-I)_0 < 1.8$
to construct the MDs, as was done for the MD of the
major-axis structure (Fig.~\ref{fig:mdfMA}). The properties of the MDs clearly
confirm the visual inspection of the above-discussed CMDs.
Note that in this procedure we may miss some amount of the
metal-rich stars with [Fe/H] $\ga -0.7$ because of this stringent criterion.

%%% 6 %%%
\section{The Global Halo Properties Along the Minor Axis}

%%% 6.1 %%%
\subsection{Population Variation with Increasing Radius}\label{sec:m31cmd}

\begin{figure*}[htpd]
 \begin{center}
  \epsscale{0.28}
  \plotone{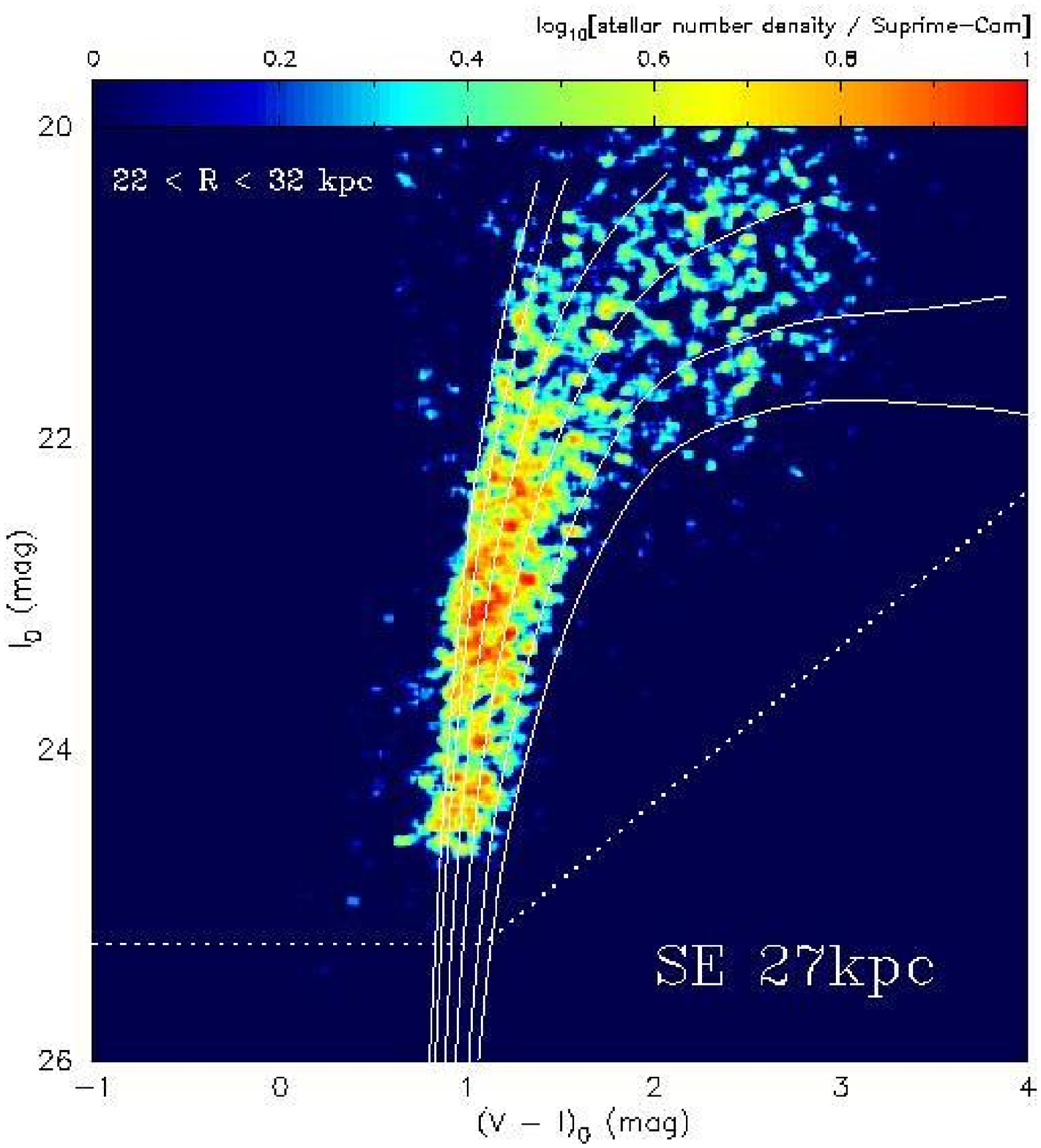}
  \plotone{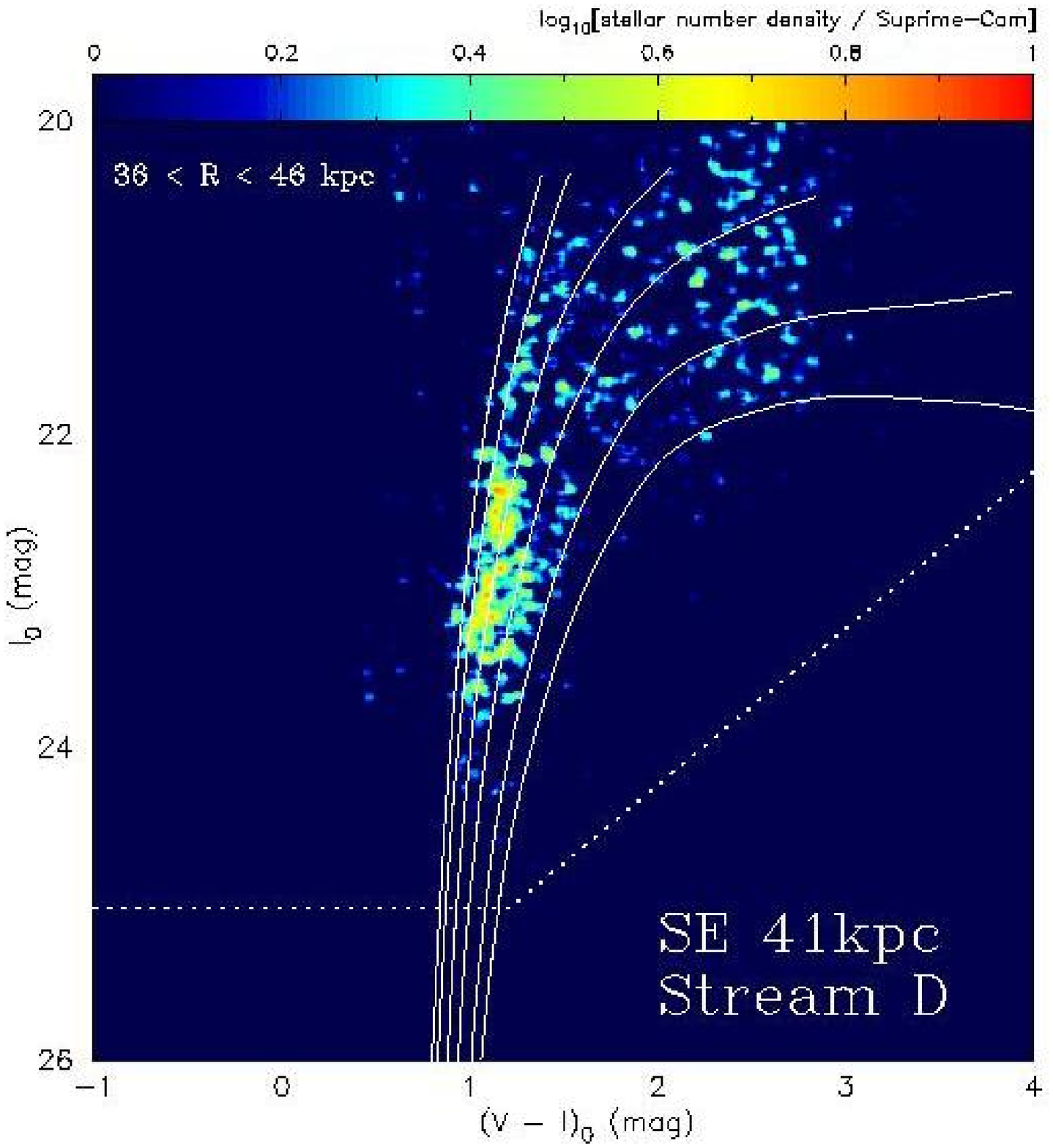}
  \plotone{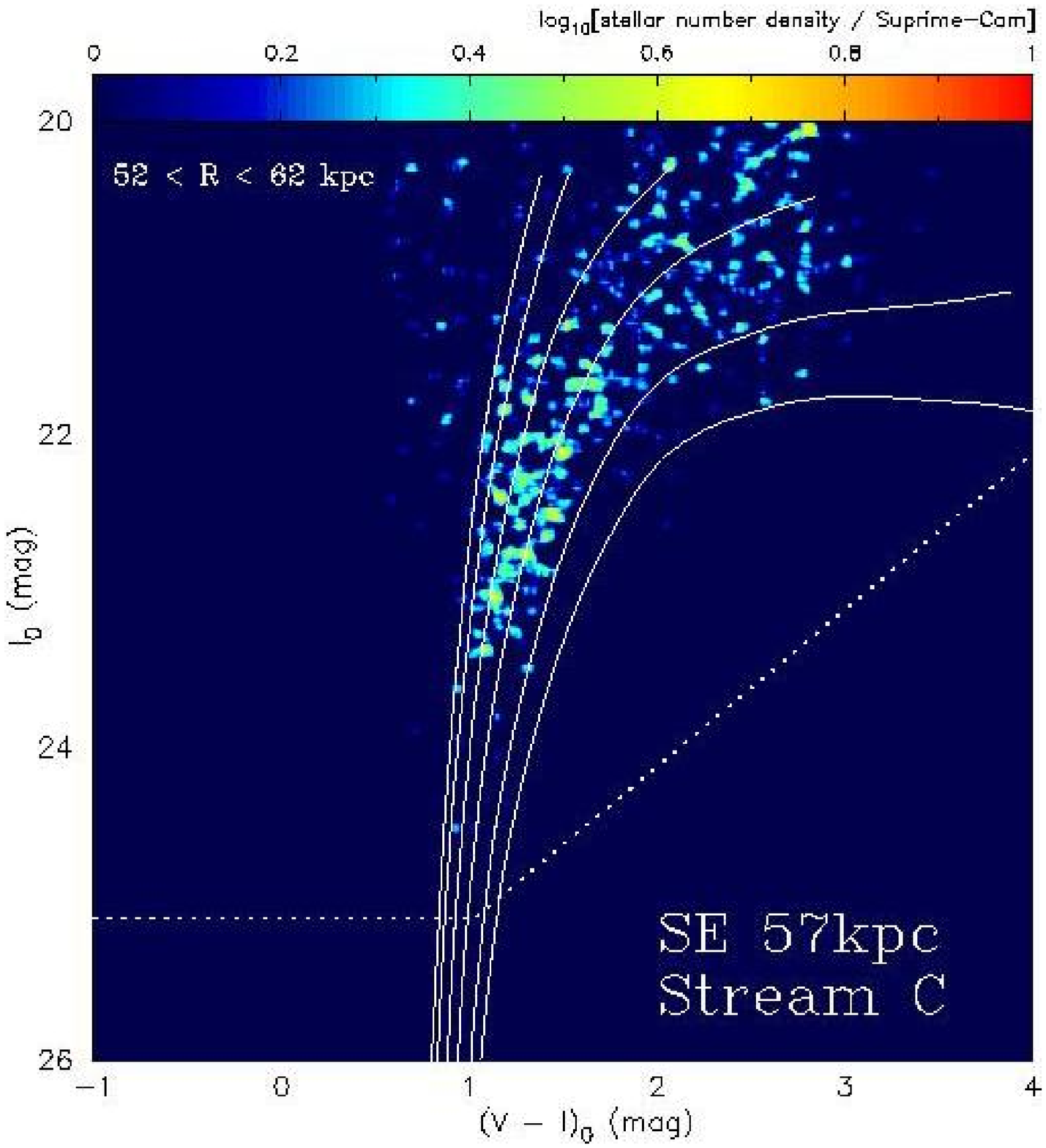}
  \plotone{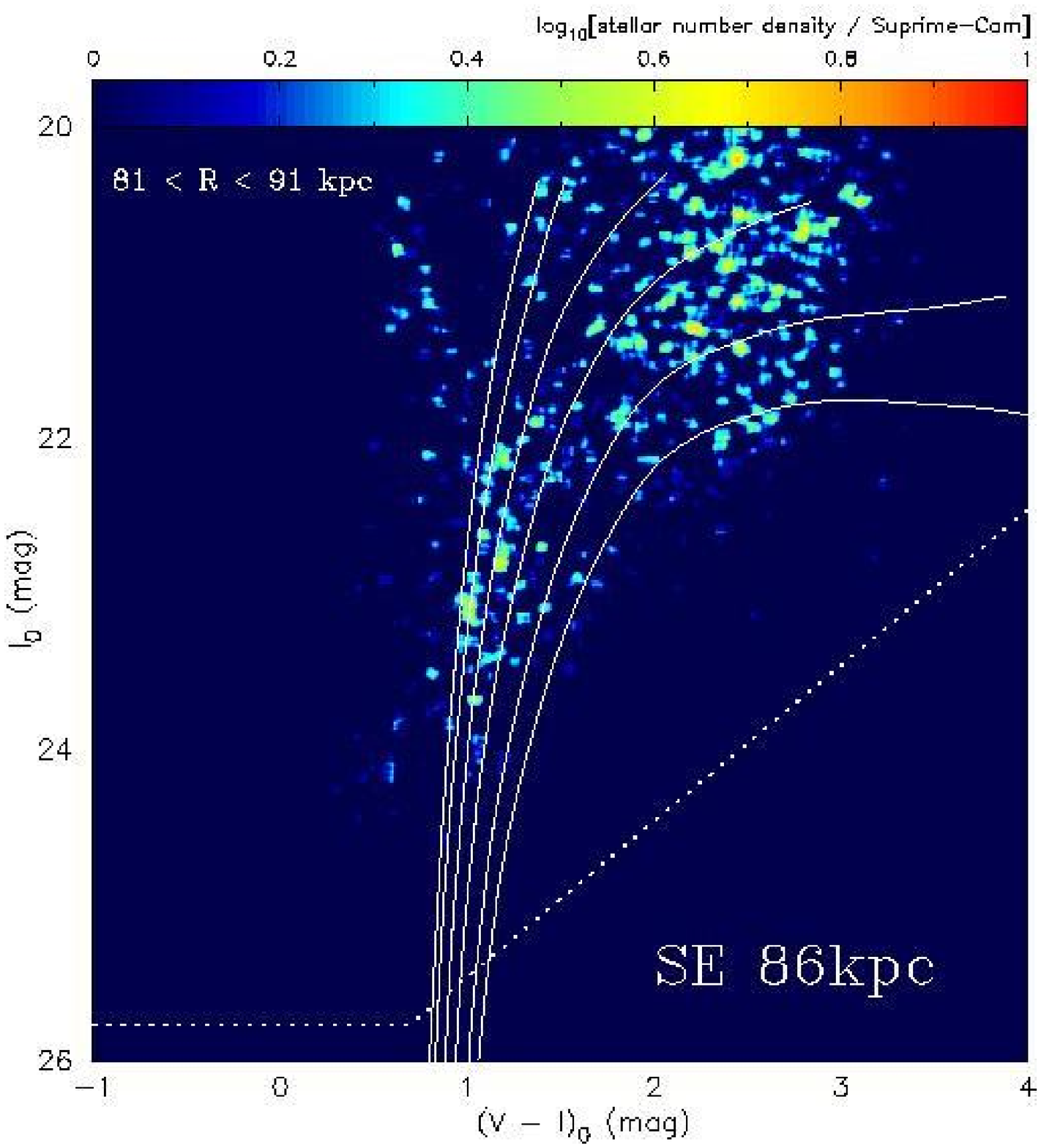}
  \caption[a]{The CMDs along the south-east minor axis of the M31 halo. The
  superposed solid lines in the CMD show theoretical RGB tracks from
  \citet{VandenBerg2006} for an age of 12 Gyr, [$\alpha$/Fe]$=+0.3$, and
  metallicities ({\it left to right\/}) of [Fe/H]$ = -2.31$, $-1.71$,
  $-1.14$, $-0.71$, $-0.30$ and $0.00$. 
  }
  \label{fig:cmdSE}
 \end{center}
\end{figure*}

\begin{figure*}[htpd]
 \begin{center}
  \epsscale{0.28}
  \plotone{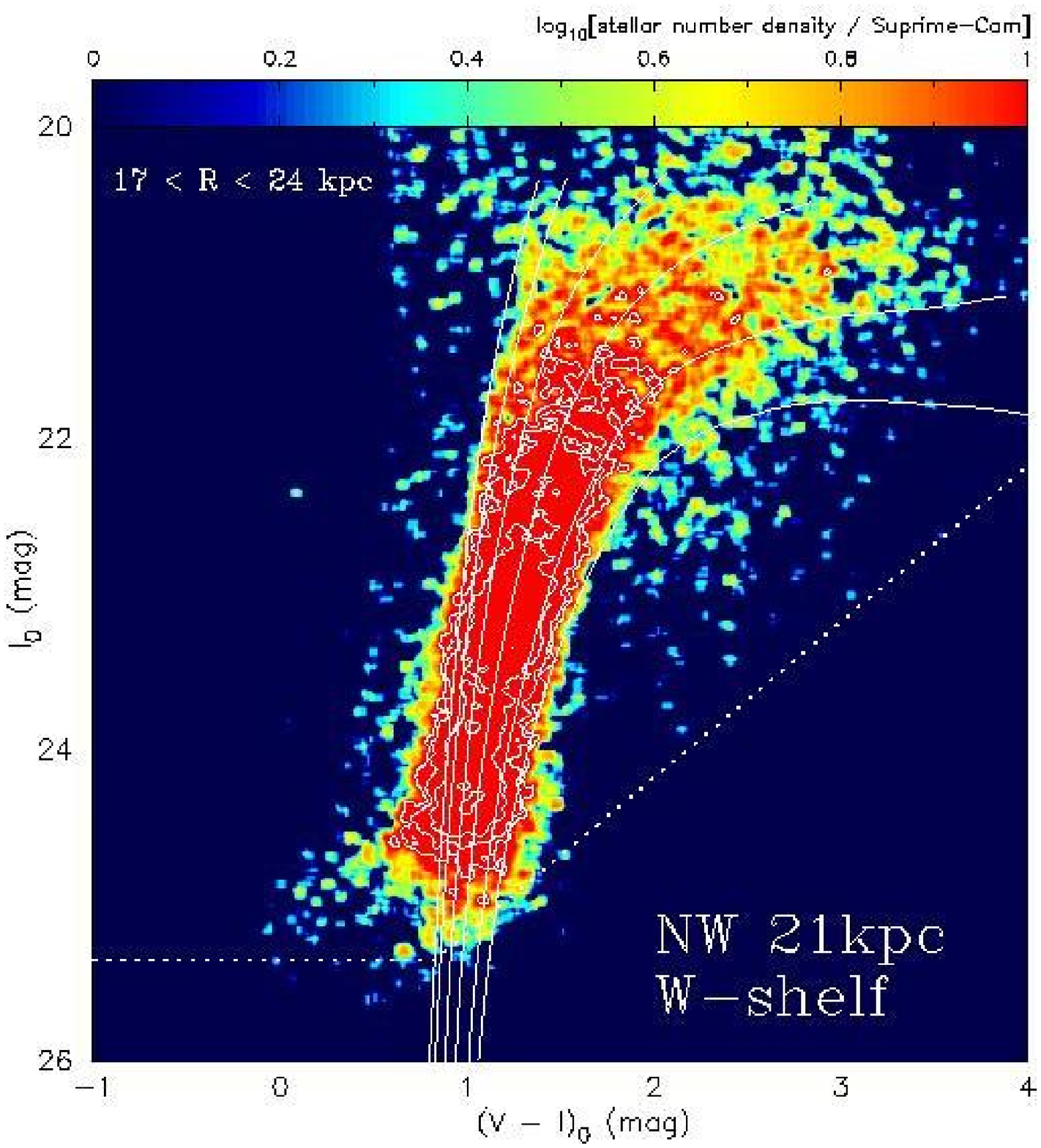}
  \plotone{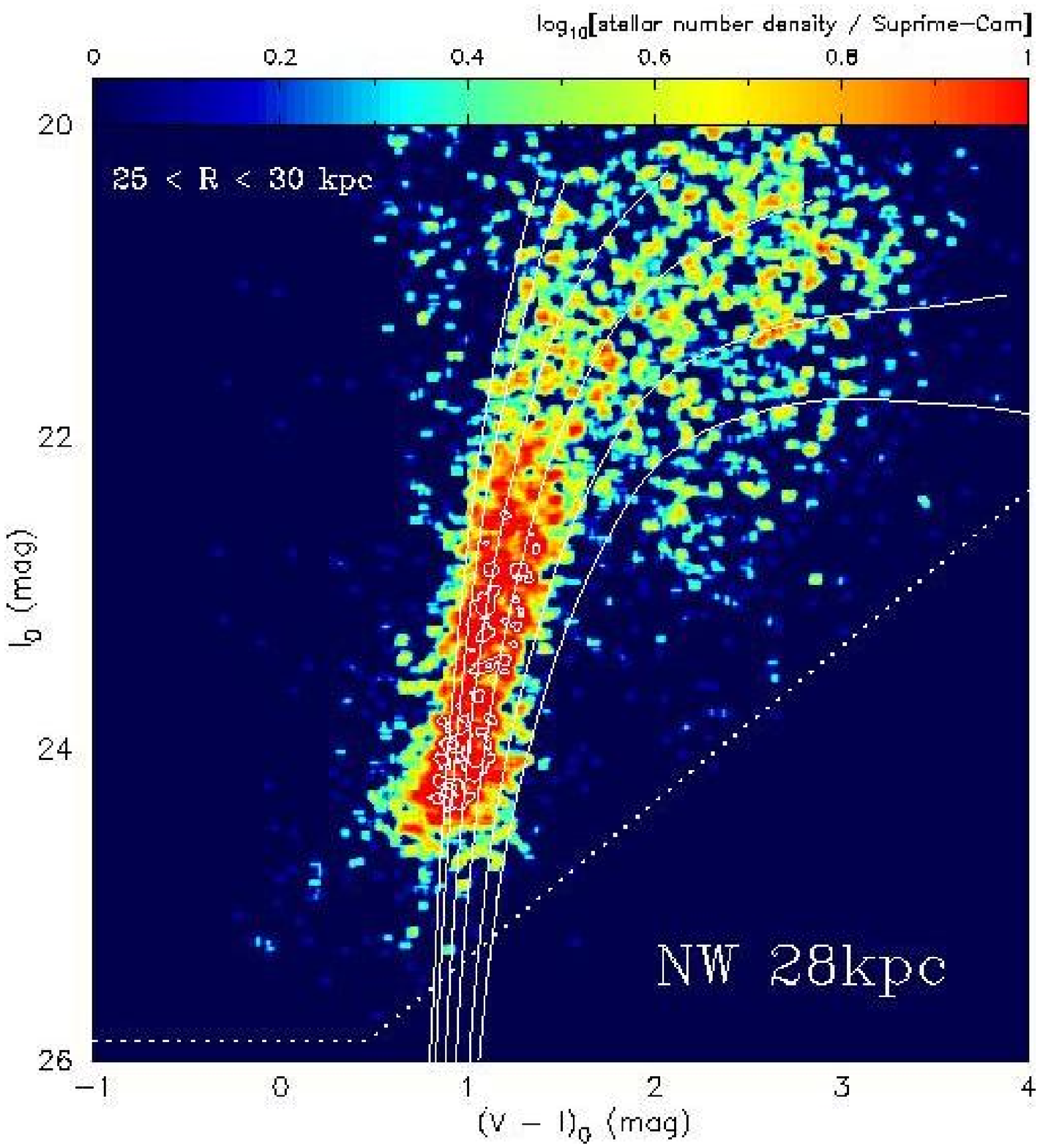}
  \plotone{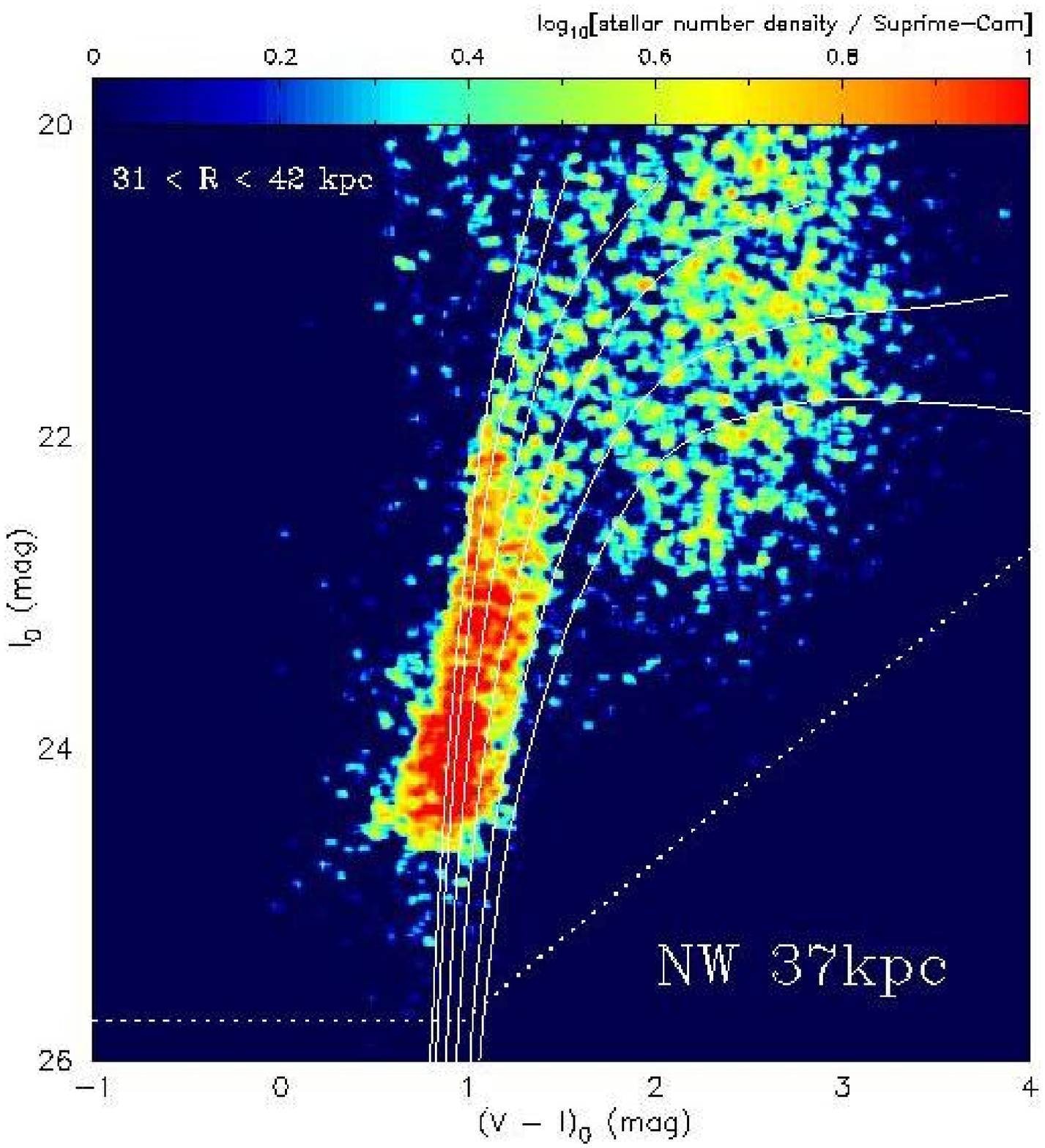}
  \plotone{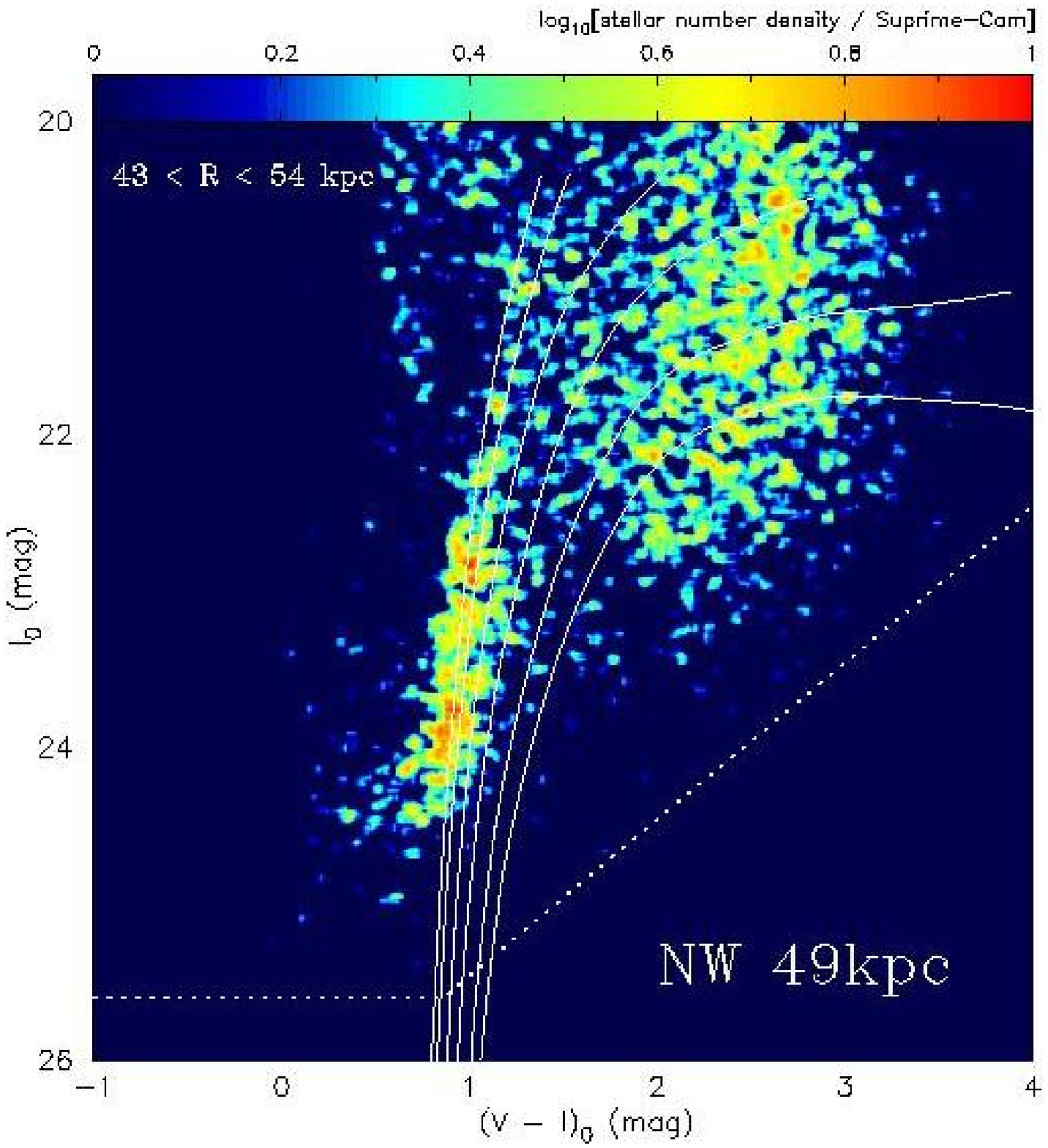}
  \plotone{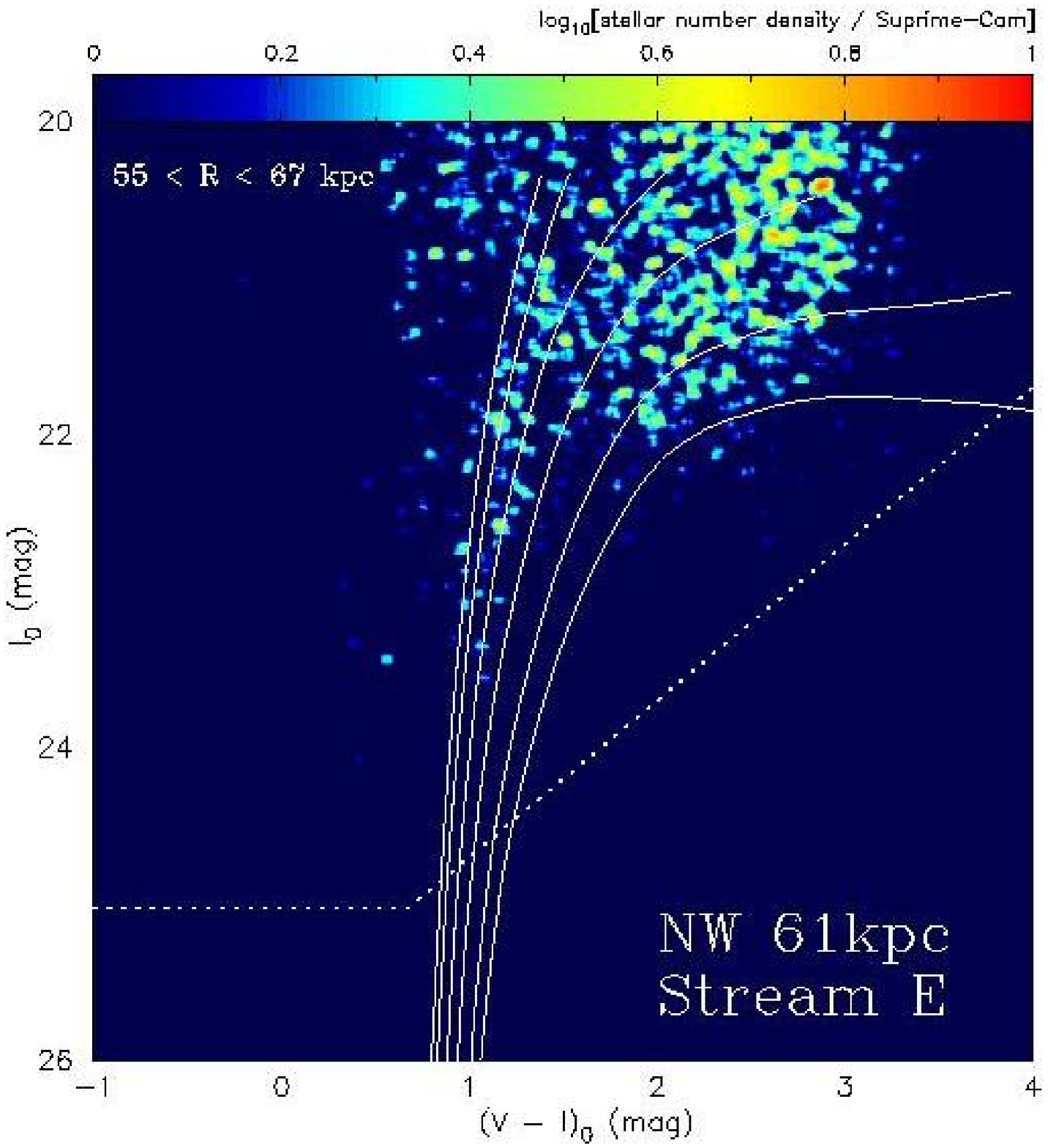}
  \plotone{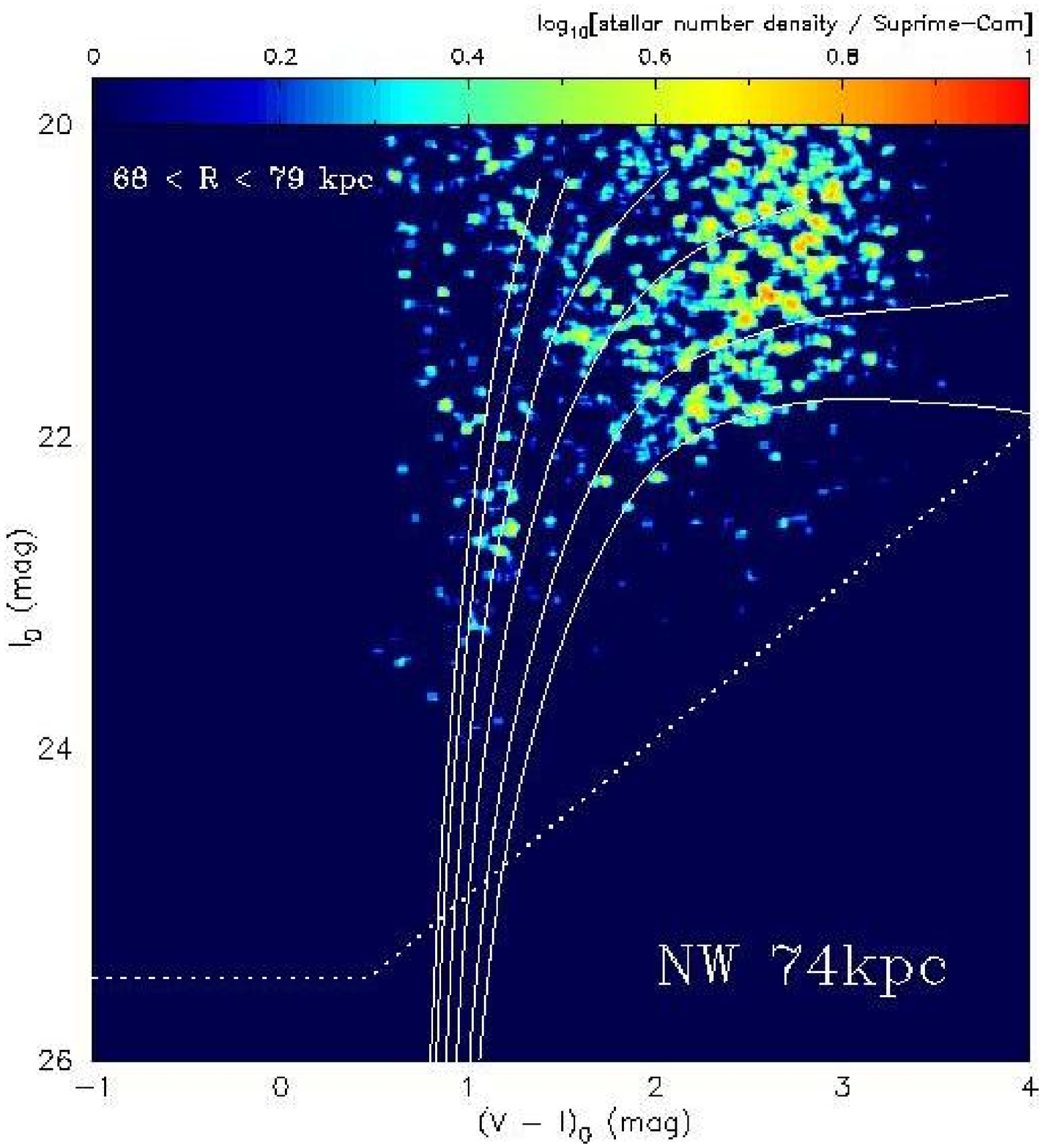}
  \plotone{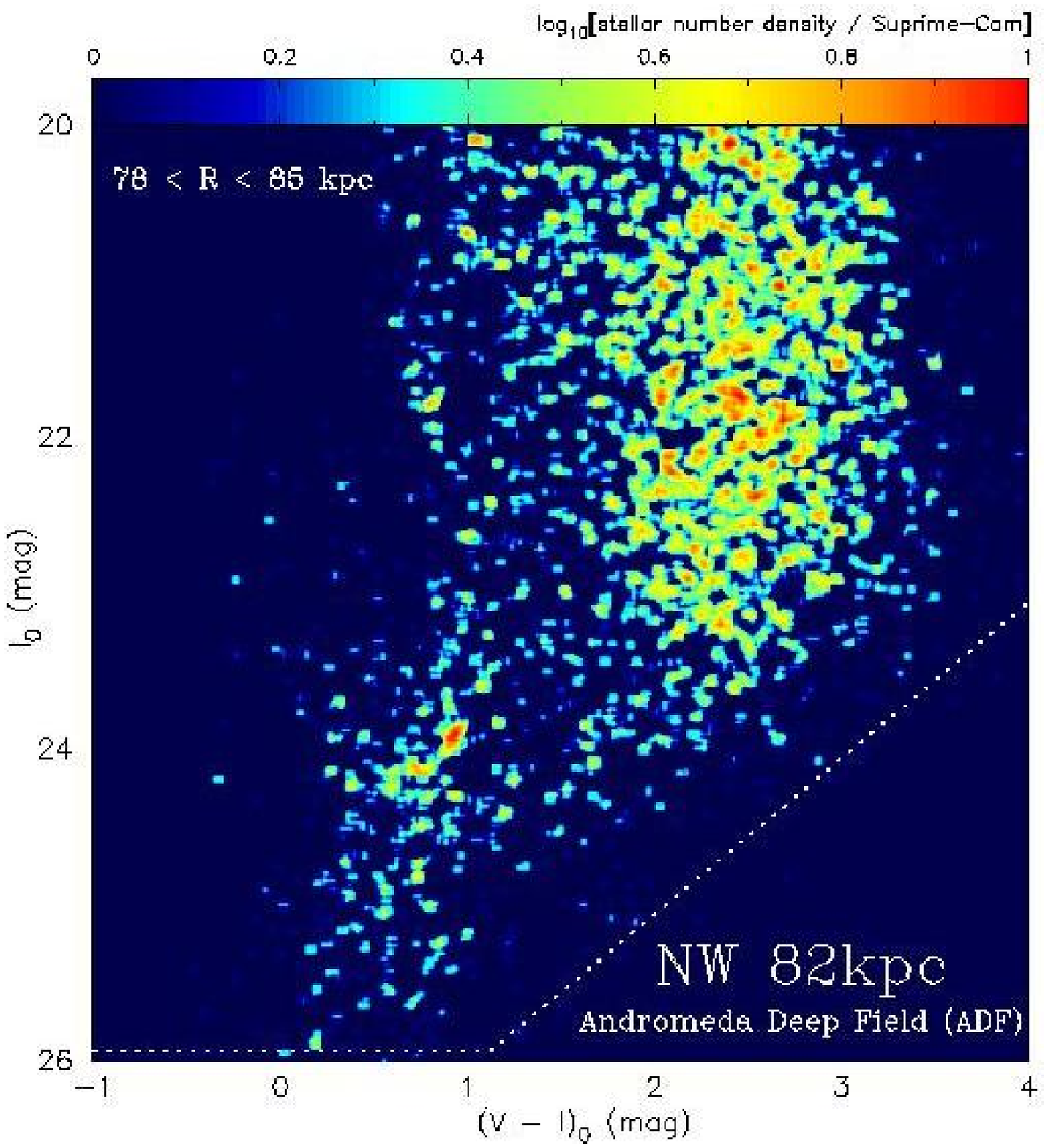}
  \plotone{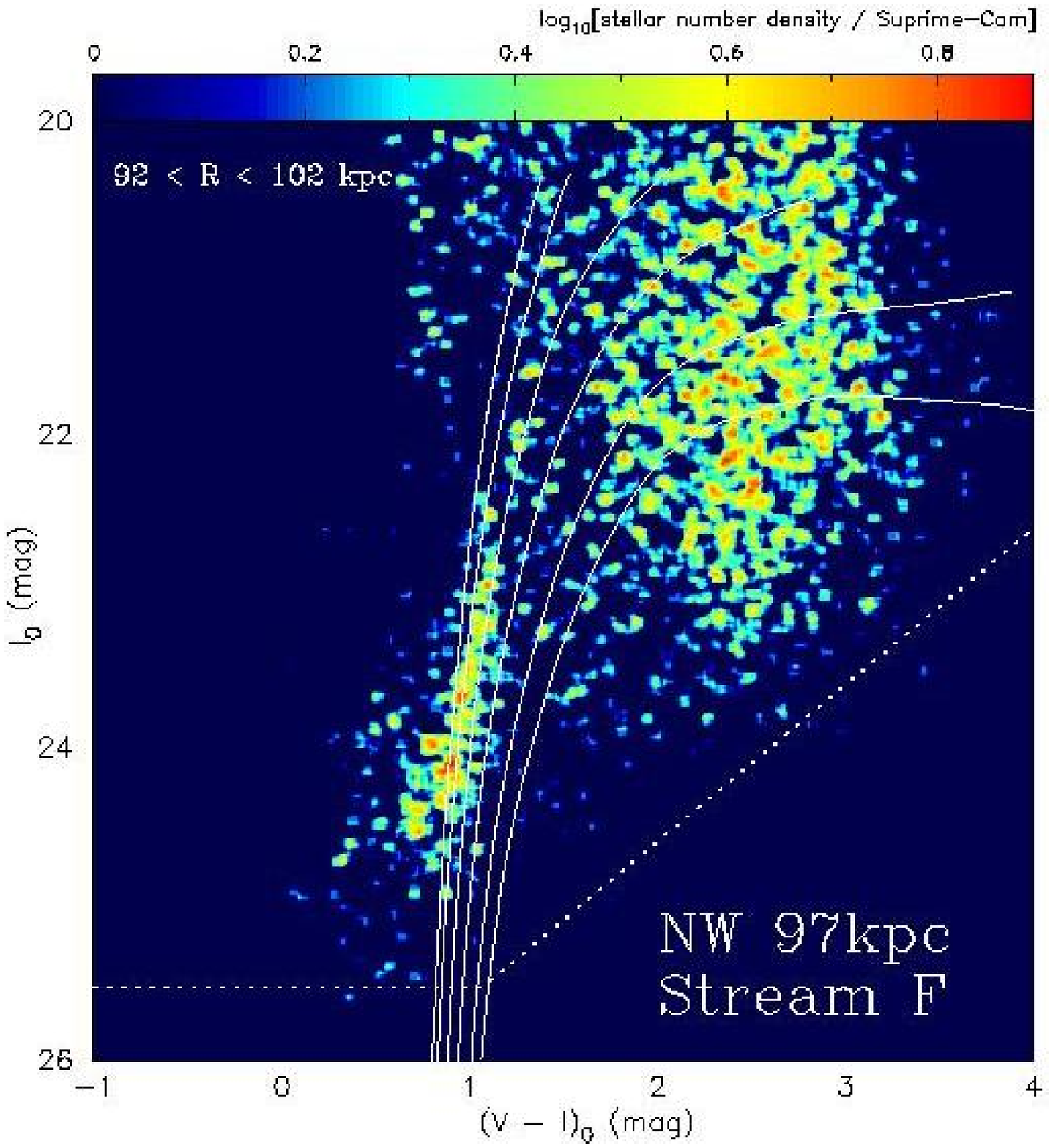}
  \caption[a]{The CMDs along the north-west minor axis of the M31 halo. The
  superposed solid lines in the CMD show theoretical RGB tracks from
  \citet{VandenBerg2006} for an age of 12 Gyr, [$\alpha$/Fe]$=+0.3$, and
  metallicities ({\it left to right\/}) of [Fe/H]$ = -2.31$, $-1.71$,
  $-1.14$, $-0.71$, $-0.30$ and $0.00$. 
  }
  \label{fig:cmdNW}
 \end{center}
\end{figure*}

In Figures~\ref{fig:cmdSE} and \ref{fig:cmdNW} we display the CMDs along
the minor axis in order of projected distance from the M31
center. After the background contaminations in the catalogs are
subtracted by combination of DAOPHOT and SExtractor softwares in the
same manner as our example above, we merge each catalog into final
catalogs covered in 0.8--1.6 times field-of-views of a Suprime-Cam
field and normalize them in the instrumental sky coverage with
$34\arcmin \times 27\arcmin$. It appears that the number density of the
foreground Galactic disk dwarf stars predictably increases with decreasing
galactic latitude, that is, in the northern direction from the M31 center
and the number density of Galactic halo stars is practically constant along
these galactic latitudes within the Poisson noise. This suggests that our
method for the separation of galaxies and stars works successfully
at the brighter magnitude. In contrast, at the fainter
magnitude of, for instance, $I_0\ga23.5$ $(I_0\ga23)$ for the SE 27kpc
(41kpc) CMD, stellar objects seem to be over-subtracted because of the
rapidly declining detection efficiency based on the above-mentioned
stringent criterion. 

The two CMDs among the four south-east CMDs show the stream-like
substructures (Stream~C and D) as shown in \S~\ref{sec:SCD}. The inner
most CMD at SE 27 kpc has wide-spread RGB stars and it is very similar to that
of the north-west counterpart field (NW 28 kpc), suggesting the existence of
complex mix of metal-rich and metal-poor populations with $-2.5\la{\rm
[Fe/H]}\la0.0$ in the inner halo of M31. Both of the populations in the two
symmetrically-located fields have the same mean metallicity of ${\rm
[Fe/H]_{\rm mean}} \sim -1.0$. A KS test of the two cumulative
MDs imposed by the additional criterion with $(V-I)_0 < 1.8$ (mag) in
order to avoid the foreground contamination yields a probability
of 12.3\% that the two distributions are drawn from the same parent
distribution. 

In our Suprime-Cam survey of the M31 halo, we have obtained data of
continuous fields along the north-west minor axis, facilitating the
investigation of global halo structure. Furthermore, most regions in
the north-west halo have yet been unexplored by any other telescopes. The
continuous change of the CMDs along the north-west minor axis of
Figure~\ref{fig:cmdNW} suggests the existence of clear population
gradients with distance from M31's center. The fraction of the metal-rich
population with ${\rm [Fe/H]} \ga -0.7$ strongly decreases out to $\sim 50$
kpc in projection, while the metal-poor stars with ${\rm [Fe/H]} \la -0.7$
are definitely distributed in the outer part of $\sim 50$ kpc from the
M31 center. This population variation identified from the CMDs reflects
a strong metallicity gradient with increasing radius in the M31 halo. 
In the more remote halo beyond $\sim 50$ kpc, the RGB sequence with
any metallicity range is virtually unobservable, except in the two
overdense regions such as Stream~E and F fields. 

%%% 6.2 %%%
\subsection{Surface Brightness Profiles}\label{sec:m31sb}

Since \citet{Pritchet1994} derived the surface brightness (SB) profile of
the inner dense parts of the M31 spheroid, many researchers have studied
the extended, outer parts of the halo along the south-east minor axis
\citep{Durrell2004,Irwin2005,Guhathakurta2005,Ibata2007}.
In contrast to the inner parts
of the spheroid, it is possible to resolve the individual stars in its
outer parts, because of their sparse density and of the sufficiently
close distance of M31 to us. Thus the SB profiles in the spheroid's outer
parts are available by directly counting individual stars. While
previous studies have been limited to relatively bright RGB stars for
the estimation of the SB in an extended halo, we consider fainter stars,
taking advantage of the light-gathering power of the Subaru telescope
with an 8.2-m primary mirror. This suggests that the results obtained
here  using our deep Suprime-Cam data will be most robust against the
effects of any contaminations over the fields. Furthermore, considering
both the north-west and the south-east halo profiles simultaneously
for the first time, we will arrive at more comprehensive halo structures
than have been studied previously.

We estimate the SB profiles using the RGB and AGB stars selected as
$(V_0,~I_0) < (24.8,~24.0)$, i.e., $I$-band magnitude brighter than
the Red Clump (at $I_0 \sim 24.5$) and colors redder than
$(V-I)_0 = 0.8$, with not less than about 90\% completeness limit.
In addition we set $I_0 \ga 20.3$, i.e., fainter than
the tip of RGB stars of M31 \citep[e.g.,][]{Durrell2001},
to exclude contaminations such as foreground stars, when assuming the
distance modulus of 24.43 from \citet{Freedman1990}.

First, we divide a series of our surveyed fields shown in
Fig.~\ref{fig:map} into subregions with about $0.05$ square
degrees. After extracting the secure RGB stars based on the above
magnitude and color selection criteria, we divide them into the
following two groups using the RGB model isochrones of
\citet{VandenBerg2006} for an age of 12 Gyr and [$\alpha$/Fe]$=+0.3$:
{\it Metal-Poor group\/} (MP) defined as $-2.31<{\rm [Fe/H]}<-0.71$ and
{\it Metal-Rich group\/} (MR) defined as $-0.71<{\rm [Fe/H]}<+0.00$. This
division is motivated by the form of the CMDs as obtained above
(\S~\ref{sec:m31cmd}), showing a rather distinct feature below and above
[Fe/H]$\sim -0.7$ (corresponding to the metallicity of one of the
Galactic clusters, \objectname{47~Tuc}).

Second, we convert the summed-up flux counts of selected stars to the SB
in mag~arcsec$^{-2}$, and subtract the SB of the control field, for
which foreground and background contaminants are removed based on the
statistical method outlined in \S~\ref{sec:daopars}. The sources of
noise in this method of estimating the SB arise from the Poisson
statistics from the finite number of observed stars and subtracted
contaminations.

\begin{figure*}[htpd]
 \epsscale{1}
 \plotone{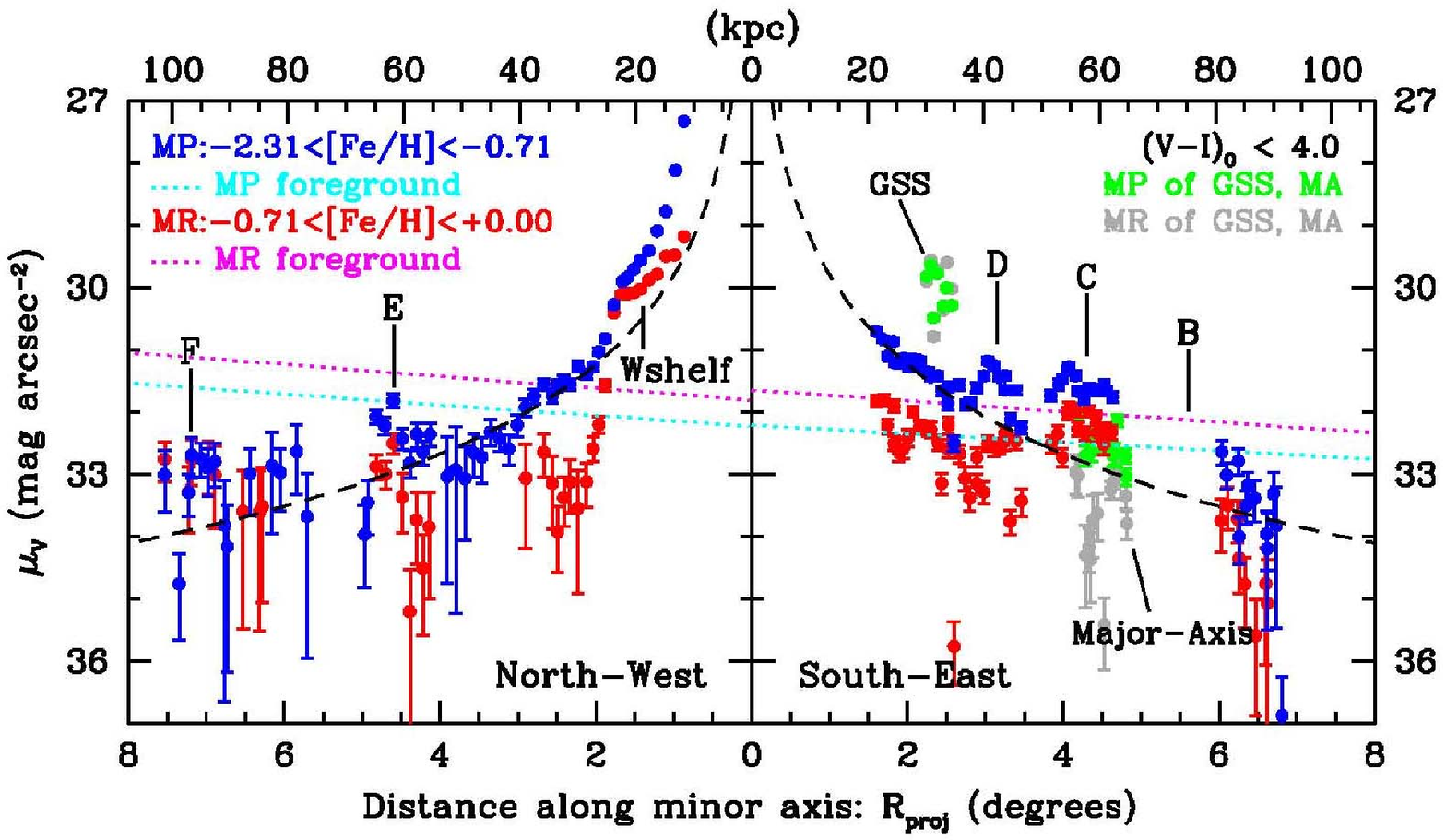}
 \plotone{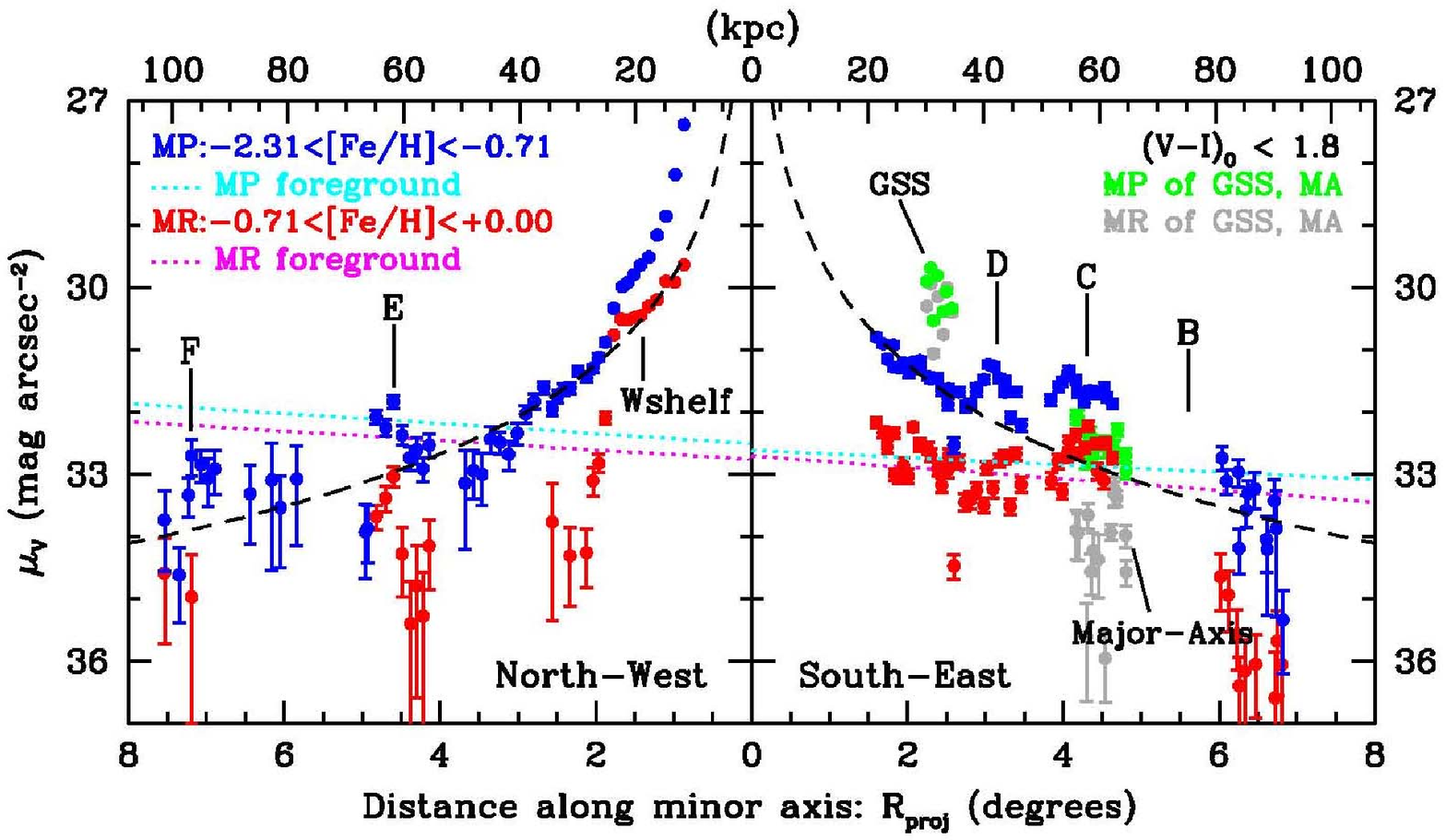}
 \caption[a]{The minor-axis profiles for RGB stars as derived from the MP
 and MR group, which are illustrated with blue and red points,
 respectively. Dotted lines denote the corresponding subtracted
 background levels. The error bars reflect a combination of Poisson
 statistics and background uncertainties. The dashed line shows a
 power-law profile with an exponent of $1.91 \pm 0.12$ for MP group
 derived by \citet{Ibata2007}, for the sake of comparison. The
 distinction between the upper and lower panels reflects different color
 selection (see text). 
 }
 \label{fig:m31sb}
\end{figure*}

Figure~\ref{fig:m31sb} shows the reduced SB profiles of the MP (blue)
and MR (red) groups. In this figure, two colored dotted lines having
flat profiles present the subtracted background levels to obtain the
reduced SB profiles. In addition, we plot the SB in our observed regions
of the GSS and the major-axis diffuse structures (green and
grey). The top panel of the figure denotes the SB profiles selected by
our color cut of $(V-I)_0 < 4.0$ used in the construction of the
MDs in the previous sections, whereas the bottom panel shows the SB
profiles chosen by the bluer color cut of $(V-I)_0 < 1.8$ used in the
construction of the MDs of the above-discussed outer streams. The latter
color cut is motivated to avoid the heavy contamination of the disk dwarf
stars. For guidance, a dashed line corresponds to the SB
profile obtained by \citet{Ibata2007}, based on their survey for south
outer parts of the halo; they identified stream-like substructures in
some parts of our observed fields, whereby they derived the SB profile
for the global halo by avoiding such disturbing substructures. This line
is characterized by a power-law model, $\Sigma(R) \varpropto
R^{-\alpha}$, with an exponent of $\alpha = 1.91\pm0.12$ kpc. We
nominally fit this SB profile to our fields at $R \sim 30$ kpc, which
may lie outside the inner dense spheroid with a $R^{1/4}$ law and trace
an extended halo just beyond this component. It follows that this SB
profile reproduces this inner part of an extended halo. 
At the brighter part of $\mu_V \la 33$ mag arcsec$^{-2}$, the decreasing
trend of the estimated SBs in both the north-east and south-west parts
is reasonably consistent with each other. We are also able to
assess the faint stream-like substructures as
identified in the population maps of Fig.~\ref{fig:m31map}. However, the
absolute SB given in a dashed line is slightly underestimated in the inner
region, which is dominated by
metal-rich stars such as the W-shelf field. In contrast, at the fainter part of
$\mu_V \ga 33$ mag arcsec$^{-2}$, in particular, in the north-west region,
the behavior of the SB profiles are somewhat different. As discussed in
\S~\ref{sec:m31cmd}, it is supposed that we do not detect both metal-rich
and metal-poor stars in the outer region of $R \ga 80$ kpc
because we could not identify the metal-rich RGB with ${\rm [Fe/H]} >
-0.7$ on the CMDs (see Fig.~\ref{fig:cmdNW}). Therefore, in the outer
halo regions beyond $R \sim 80$ kpc we accept the SB profiles based on
the more stringent color criterion of $(V-I)_0 < 1.8$, compared to those
with $(V-I)_0 < 4.0$. Then, a prominent feature of interest is that the number of
metal-rich stars with ${\rm [Fe/H]} \ga -0.7$ in the
north-west halo of M31 steeply decreases with increasing radius, compared
to the south-east halo, because the south-east region within
$R\sim70$ kpc may be polluted by substructures with metal-rich
populations such as Stream~C, D and the diffuse debris of the GSS
as predicted by \citet{Fardal2007}. This picture is slightly
inconsistent with our assumption that the region along the south-east
minor axis at $R \sim 35$ kpc is free from any substructures, so we used
the region for the control field through the analysis of the GSS in
\S~\ref{sec:stream}. However, we note that the effect of such substructures does not
significantly change the final result of our analysis for the GSS
because the number of the contaminated metal-rich stars at the $R
\sim 35$ kpc field is only a small percent of those in our observed field of
the GSS as shown in the right panel of Fig.~\ref{fig:m31sb}.

\begin{figure}[htpd]
 \epsscale{1}
 \plotone{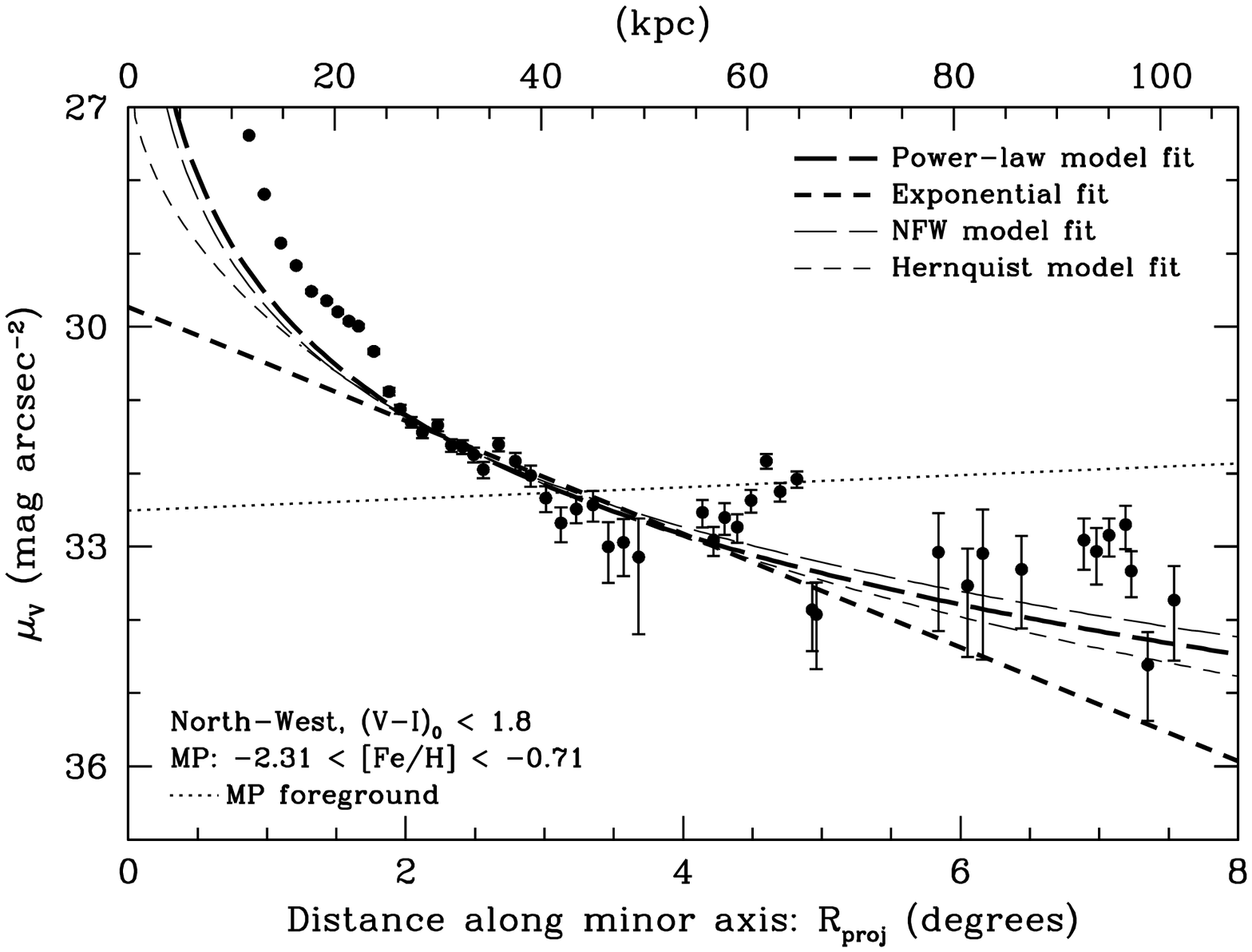}
 \plotone{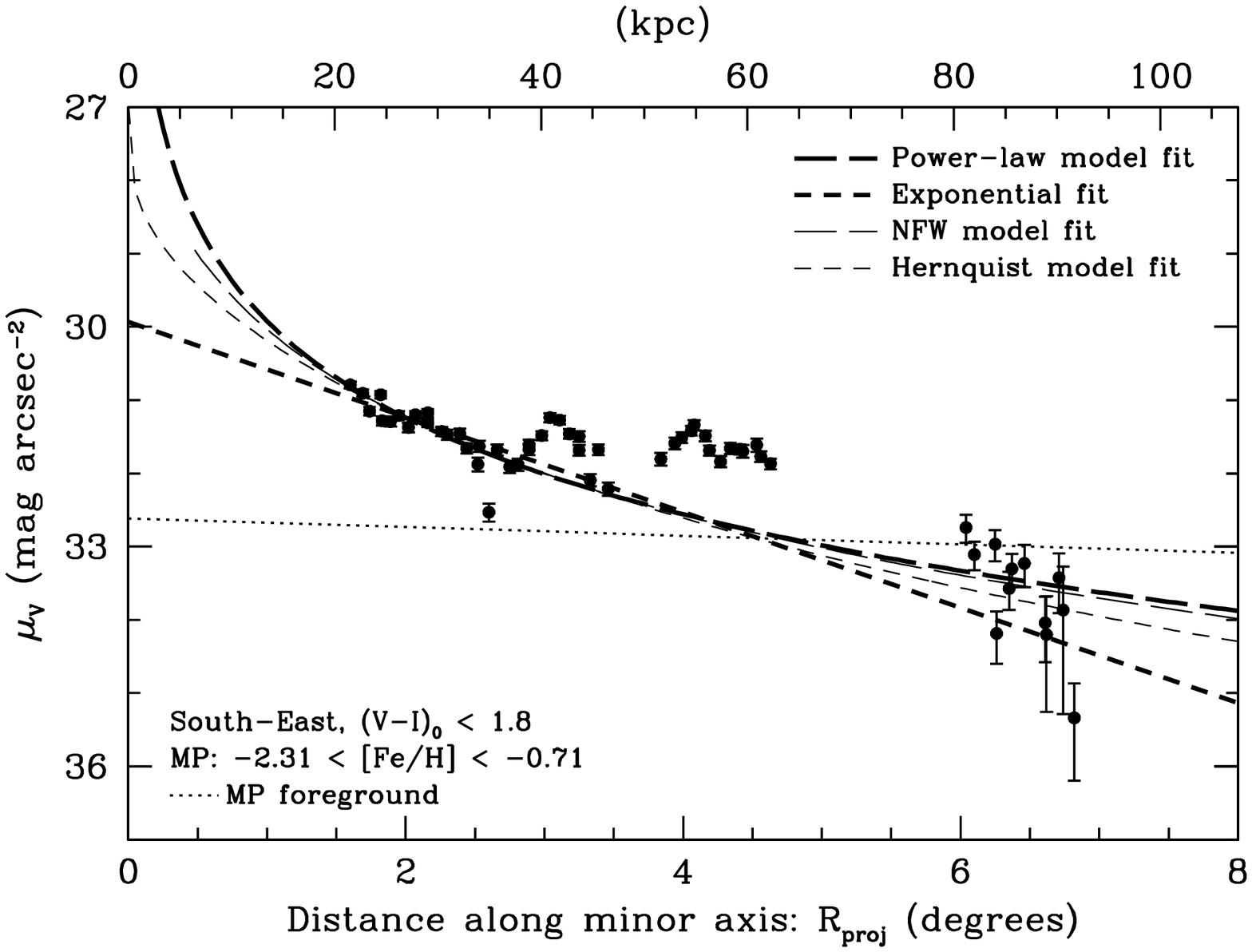}
 \caption[a]{The north-west (south-east) minor-axis radial profile from the
 MP group is shown in upper (lower) panel. We reject the data containing
 the overdense regions as we
 have shown in the text. The thick long-dashed line is a power-law fit to these
 data with $\alpha = 2.17 \pm 0.15$ $(1.75 \pm 0.13)$, while the thick
 short-dashed line is an exponential profile with scale length of $R_h = 18.8
 \pm 1.8$ $(22.4 \pm 2.3)$ kpc. The thin short-dashed line shows the best-fit
 Hernquist model, which has a scale length of $R_h = 17.1 \pm 4.7$ $(31.7
 \pm 6.7)$ kpc. In addition, with the thin long-dashed line, we show the NFW model for the halo mass profile with a scale length of $R_h = 0.2 \pm 3.5$ $(6.3 \pm
 3.9)$ kpc. 
 }
 \label{fig:m31sbfit}
\end{figure}

To investigate the spatial distribution of the M31 halo in more detail,
we attempt to fit the data on the radial profile (by weighting the
Poisson errors in the star counts) of four theoretical models of the
surface density profile. The
power-law and exponential models have been used to reproduce the surface
brightness of some stellar systems \citep[e.g.,][]{BT1987}. The NFW and
Hernquist models representing the (dark matter) density profile are
derived from \citet{NFW1996} (converted into the projected formula by
\citealt{Bartelmann1996}) and \citet{Hernquist1990}, respectively;
this model choice is motivated by the simulations of \citet{BJ2005}. The
dashed lines in Figure~\ref{fig:m31sbfit} show four different model fits
to the selected minor axis regions, which are selected to probe a
smooth, dynamically relaxed halo, where the contamination by
substructures appears small. The inner halo, which may hold complex
structures mixed by the debris of the GSS and/or the edge of
the W-shelf structure, provides the upper limit to the radial SB profile
of the minor-axis halo.

Indeed, the outer halo profile appears remarkably flat in log-linear
representation, thereby indicating essentially an exponential function.
The black dashed
line in Fig.~\ref{fig:m31sbfit} shows an exponential model fit to the
data of MP group; we find a large exponential scale length of $R_h = 18.8
\pm 1.8$ ($22.4 \pm 2.3$) kpc for north-west (south-east) minor axis
halo. We also show a projected Hernquist model fit ({\it thin short-dashed
line\/}) to these data; the best model gives a scale radius of $R_h = 17.1
\pm 4.7$ ($31.7 \pm 6.7$) kpc, suggesting that the scale radius for
north-west halo is comparable to that of \citet{BJ2005} while that for
the south-east halo is about a factor of 2 larger than predicted by
\citet{BJ2005}. \citet{Ibata2007} also indicates the flat density
distribution as well as our south-east result, based on more outer and
wider regions in the south quadrant halo out to 140 kpc, because our
south-east fields from 20 to 90 kpc are polluted by many substructures
such as Stream~B, C and D.

\begin{figure}[htpd]
 \epsscale{1}
 \plotone{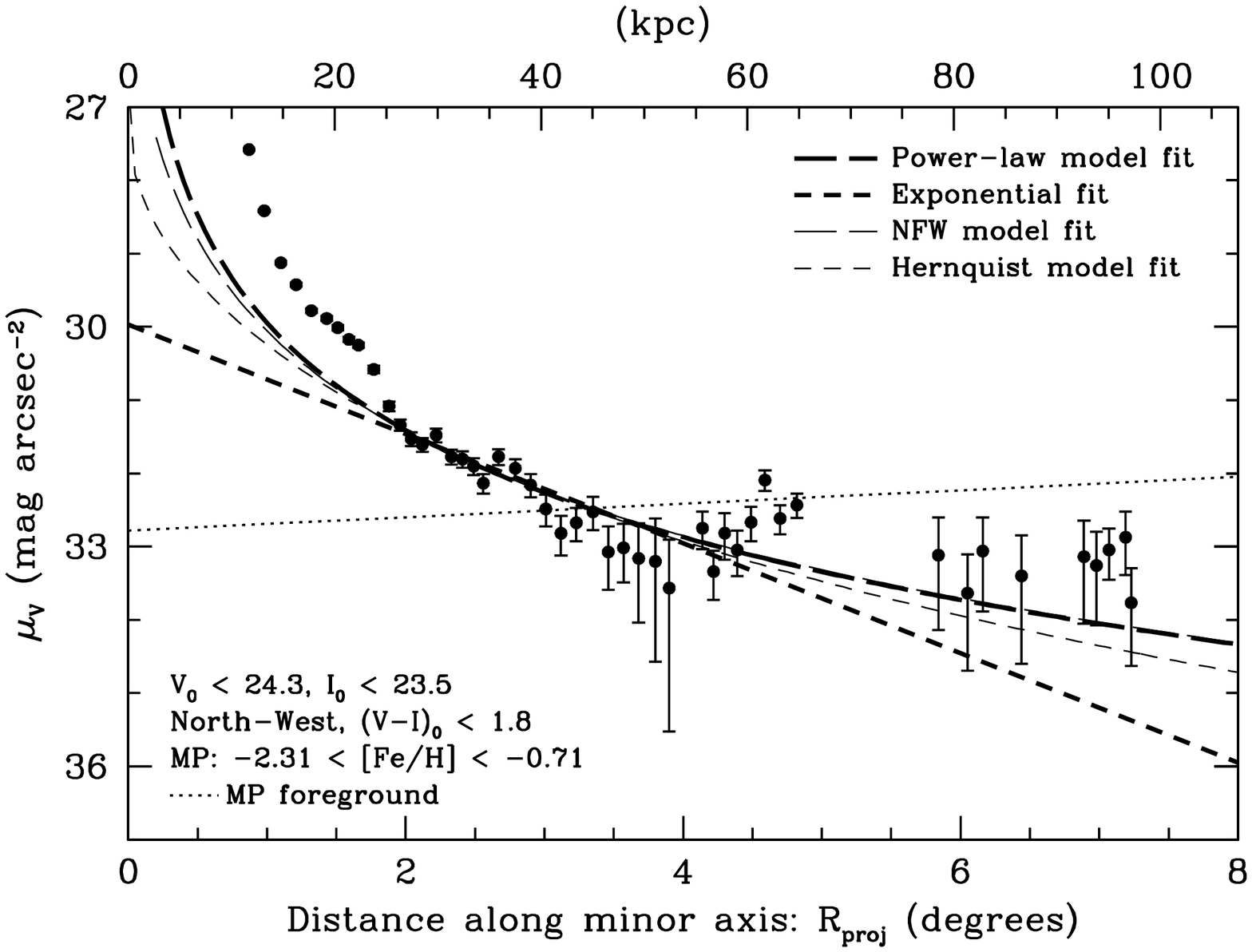}
 \caption[a]{Same as the upper panel of Fig.~\ref{fig:m31sbfit}, but for
 stars restricted in the small color-magnitude region $V_0 < 24.3$ and $I_0
 < 23.5$, in order to more stringently remove the heavy background
 contamination. Thus, these plots are only for the purpose of verifying the
 effects of such contamination on the derived surface brightness profile.
 The thick long-dashed line is a power-law fit to these data
 with $\alpha = 1.95 \pm 0.18$, while the thick short-dashed line is an
 exponential profile with scale length of $R_h = 19.5 \pm 2.3$ kpc. The
 thin short-dashed line shows the best-fit Hernquist model, which has a
 scale length of $R_h = 24.1 \pm 7.0$ kpc. In addition, with the thin
 long-dashed line, we show the NFW model halo mass profile with a scale length of
 $R_h = 2.6 \pm 4.5$ kpc.
 }
 \label{fig:m31sbsha}
\end{figure}

These results, in particular for the north-west profile, are
of great importance for understanding the nature of M31's halo, so these
are to be assessed from a more careful analysis. We know that the number
counts of background galaxies rapidly increases in the fainter magnitude
range as shown in \citet{Kashik2004}. Therefore, there is concern about
the misclassification of background galaxies on the above calculated
SB profiles. To test this effect, we re-calculate
the SB profile along the north-west regions as before for the selected
metallicity range, but this time we adopt an additional selection of
stars: we further limit the halo sample with higher S/N, by imposing
more stringent color-magnitude criteria of $(V_0,~I_0) <
(24.3,~23.5)$. The results are shown in Figure~\ref{fig:m31sbsha}. It is
evident that these plots are qualitatively and quantitatively identical to
those based on the previous selection using deeper data.
Also shown is the predicted
behavior of the Galactic foreground contamination (with the same
color-magnitude selection) as a dotted line in the figure. This
profile of the contamination is nearly flat in this log-linear
representation, so contamination cannot account for the observed
profile. Thus, this smooth decline for the outer halo population of M31,
which is comparable to the density profile predicted by the \citet{BJ2005}
model, is a robust result of this survey. We note that the density profile
of the north-west minor axis halo derived here is steeper than that of the
south-east minor axis halo derived by \citet{Ibata2007}.
This is because their sample is shallower than ours and they have
underestimated the surface brightness in the denser, inner regions of the halo.

%%% 6.3 %%%
\subsection{Metallicity Distributions}\label{sec:m31mdf}

We derive the spatial variation of mean/median metallicities along the
minor-axis of M31's halo, by comparing the RGB stars on the CMD of each 
halo field to theoretical isochrones from \citet{VandenBerg2006} in the
same manner as previous constructions of the MDs. For adopted
theoretical isochrones we assume the old and $\alpha$-enhanced
populations with an age of 12 Gyr and [$\alpha$/Fe]$ = +0.3$ like the
Milky Way halo. Regarding our current method for the estimation of
metallicities using these old isochrones, it is noted that recent
studies by \citet{Brown2003,Brown2006b} suggest an intermediate age,
i.e., younger than 10 Gyr, for about one third of the inner halo stars, in
contrast to the Galactic halo population. More recent studies with {\it
HST\/}/ACS \citep{Brown2007,Brown2008} have revealed the existence of truly
ancient and metal-poor populations as well as younger populations in the inner
halo of M31. While the age of these halo populations is yet to be
fully understood, our method using the old theoretical isochrones
provides appropriate lower limits on the metallicities of
stars. However, as discussed in \S~\ref{sec:metal_err}, for
low-mass RGB stars with ages of 6-14 Gyr the metallicity derived from
interpolation/extrapolation of bright RGB stars hardly depends on
such an age difference \citep{Kalirai2006}.

\begin{figure*}[htpd]
 \begin{center}
  \epsscale{1}
  \plottwo{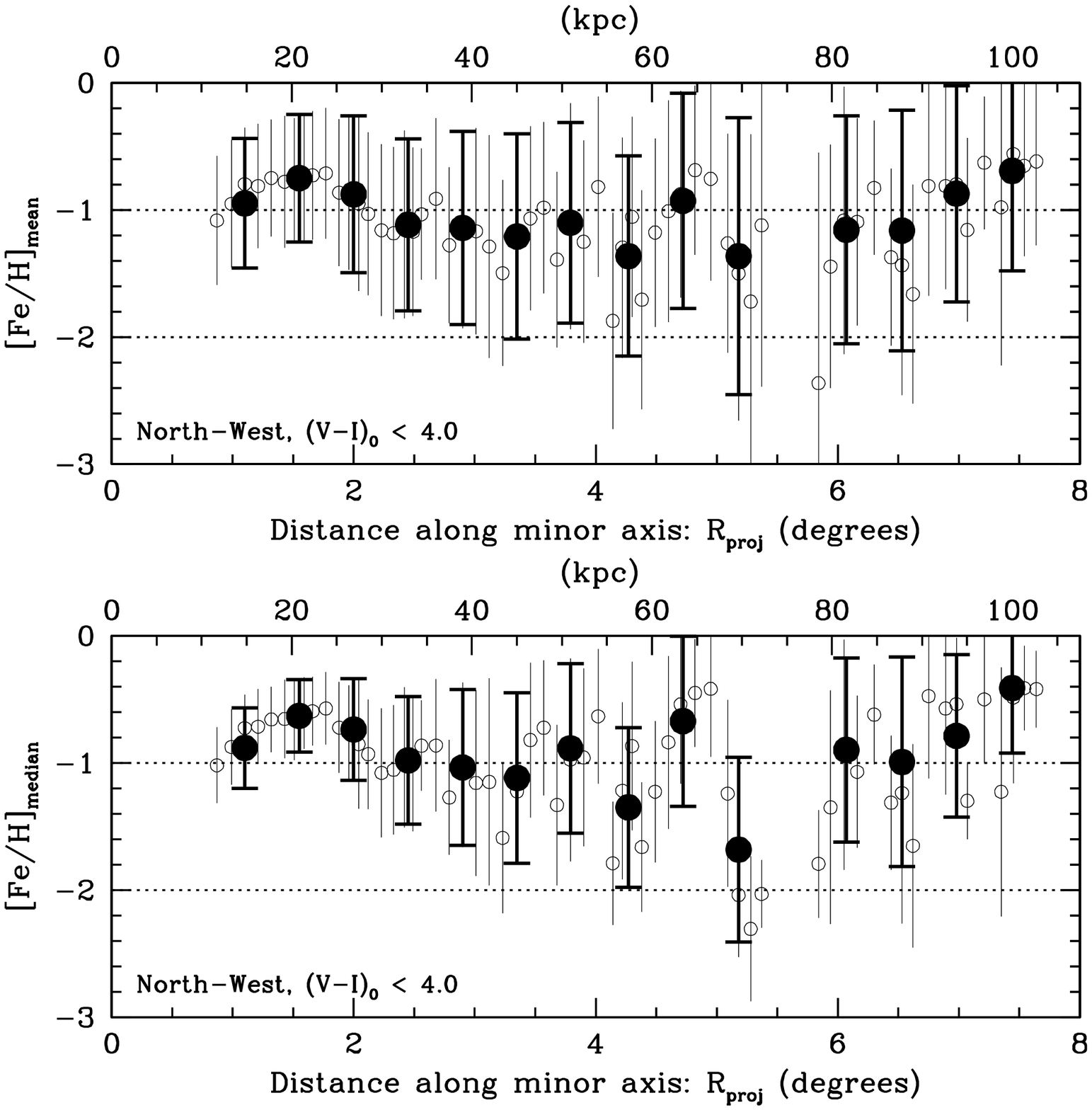}{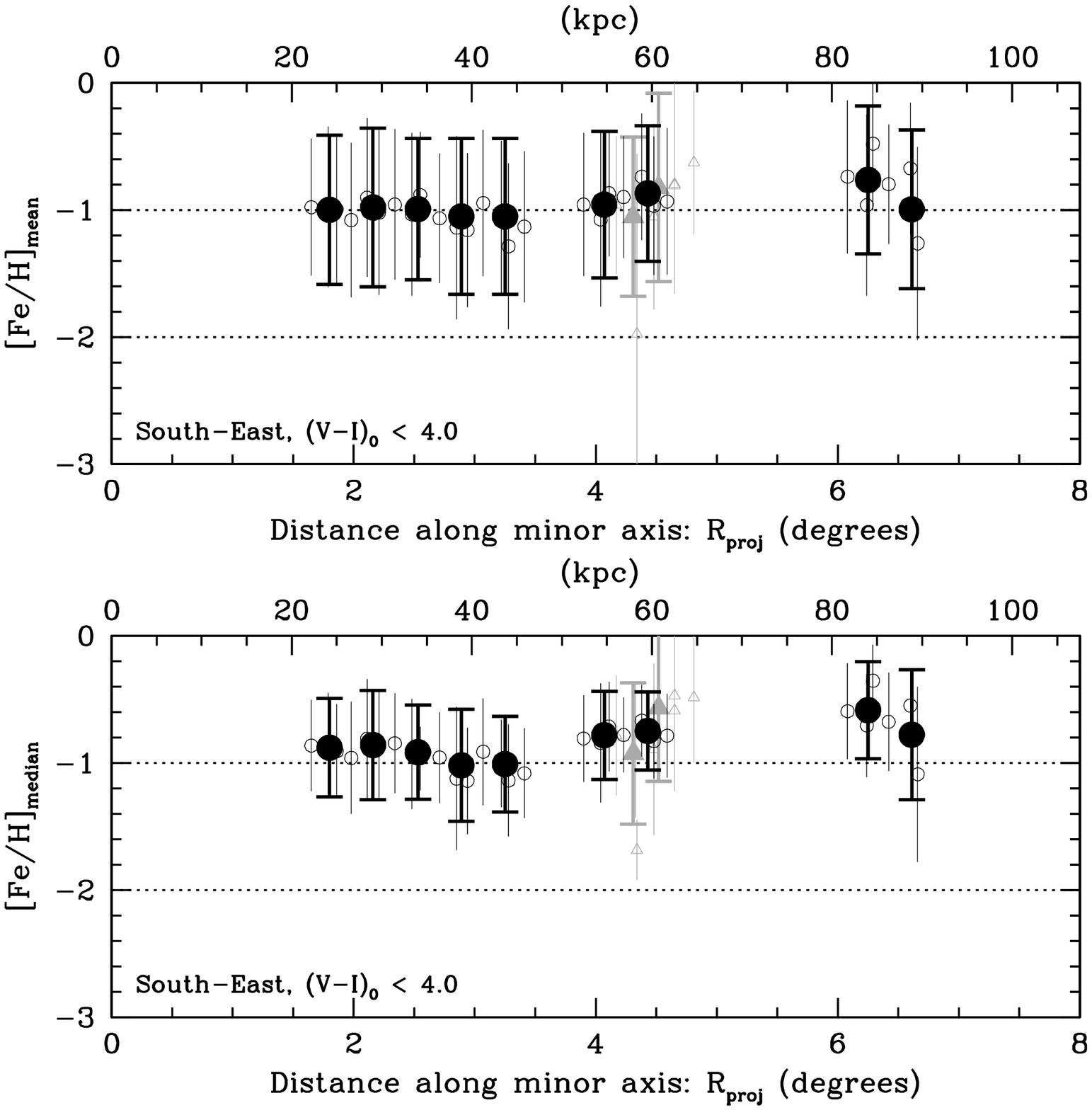}
  \caption[a]{Upper panel shows the mean metallicity of north-west (left)
  and south-east (right) halo regions together with the associated
  standard deviation, plotted as a function of distance from the M31
  center (in projection), while lower panel presents the median one
  together with the associated interquartile interval. Filled black
  circles present the average values of mean/median metallicity in the
  targeted fields, while black open circles denote those in each
  divided subfield (see text). Gray filled and open triangles present
  the mean/median metallicities of summed-up and individual fields of the
  major-axis diffuse structure, respectively. 
  }
  \label{fig:m31md4}
 \end{center}
\end{figure*}

We show their mean (median) metallicities as a function of distance
from the M31 center in the upper (lower) panel of
Figure~\ref{fig:m31md4}. Several black filled circles present the average
values of mean/median metallicities in our targeted fields, while numerous
black open circles are those in the subfields, which are obtained by 
dividing a continuous field into the subfields with about $0.05$ square
degrees. These subfields help us to examine a fine spatial variation
inside a single Suprime-Cam field. Conversely, each filled black filled circle
reflects the mean value of metallicities for the subfields. Vertical
error bars of the upper (lower) panel show the associated standard
deviation (interquartile interval) of the mean (median) metallicities.

\begin{figure*}[htpd]
 \begin{center}
  \epsscale{1}
  \plottwo{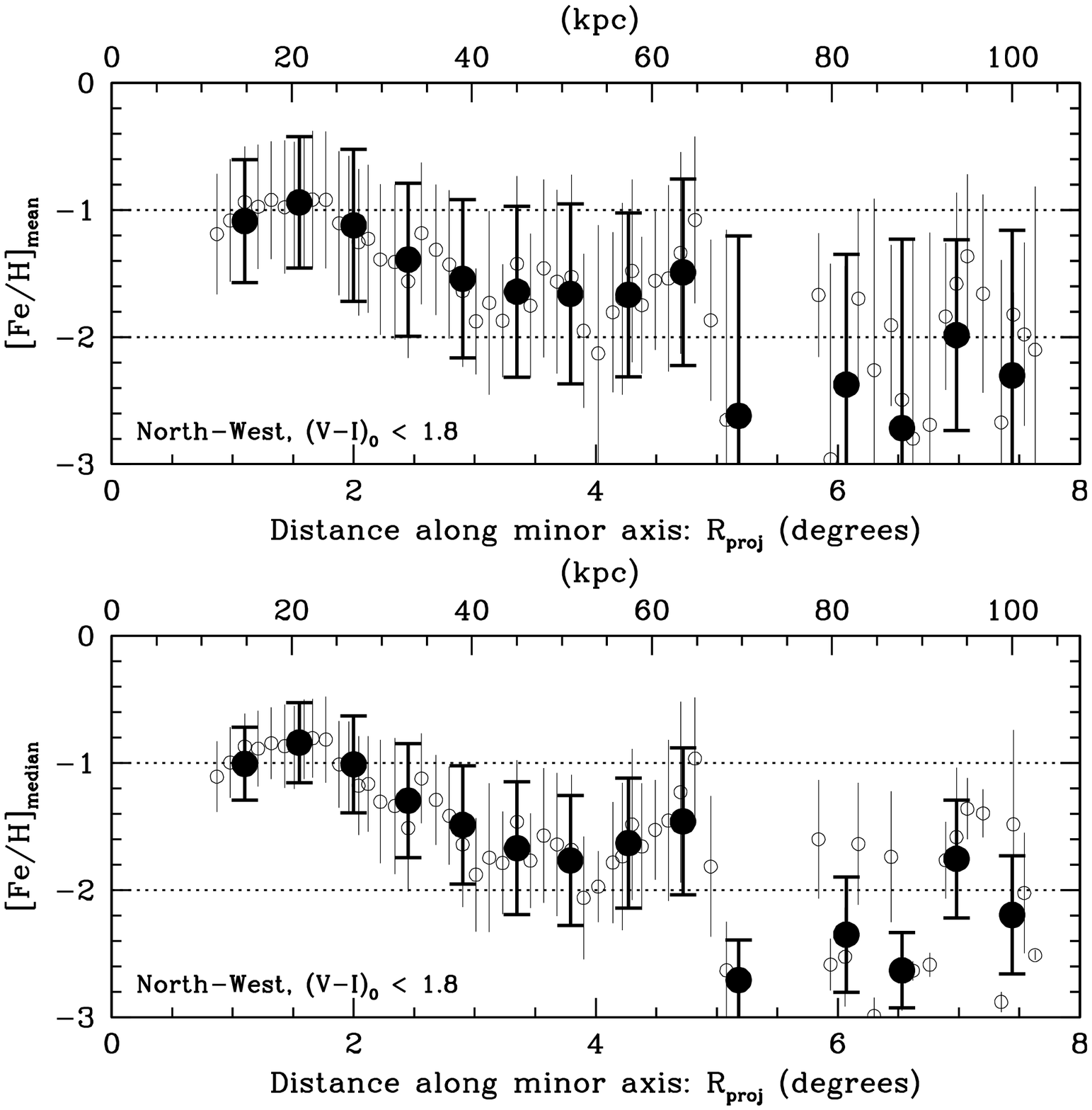}{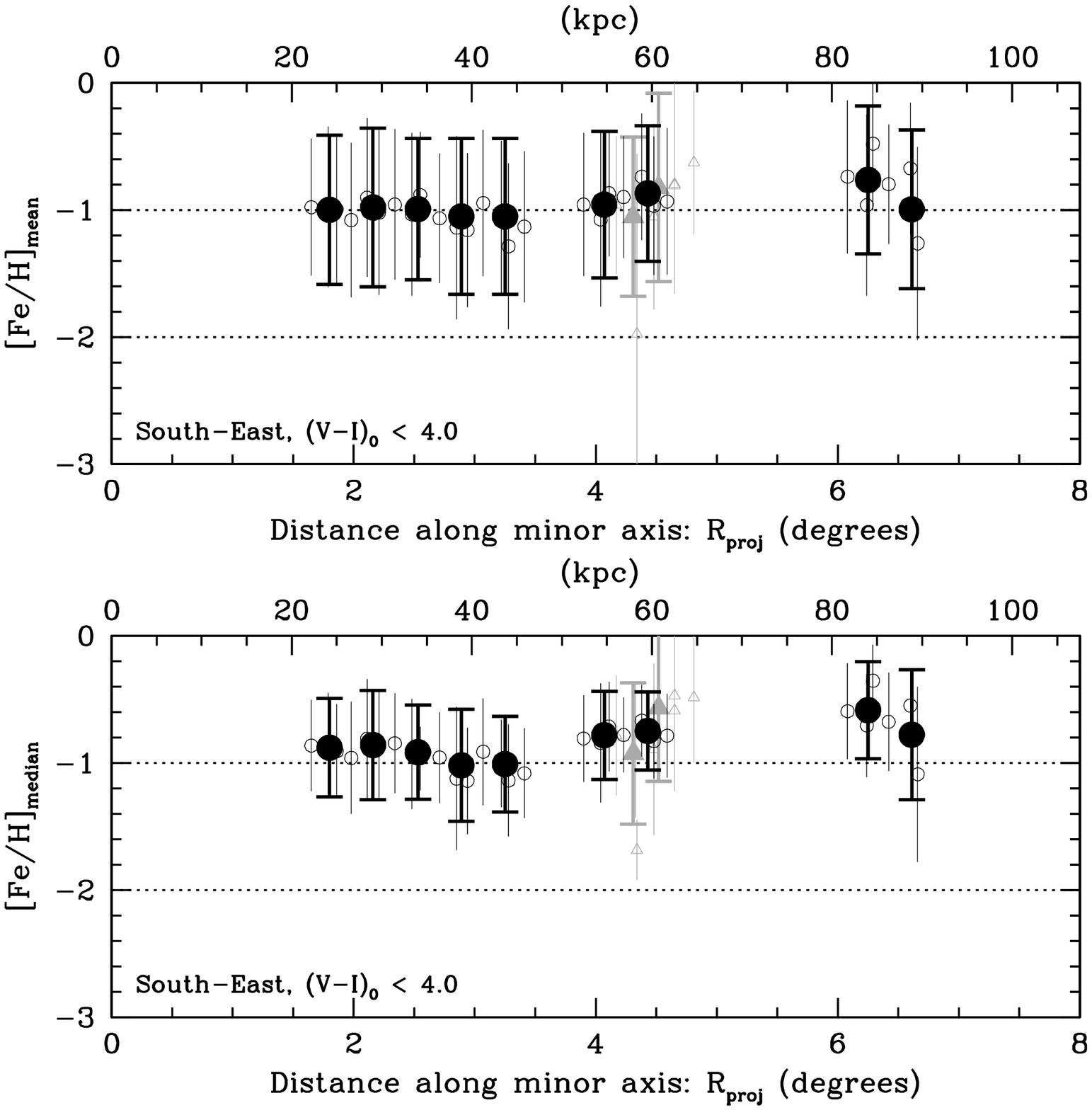}
  \caption[a]{Same as Fig.~\ref{fig:m31md4}, but for stars restricted to
  the small color region $0.9 < (V-I)_0 < 1.8$, to ensure a minimal
  contamination from the Galactic disk. 
  }
  \label{fig:m31md2}
 \end{center}
\end{figure*}

These plots suggest that the stellar content of the M31 halo has no
metallicity gradient within our observed range of fields. However, 
the spatial variation from 20 to 50 kpc along the north-west minor axis
appears to have a slight metallicity gradient. Our analysis of the CMDs discussed
in \S~\ref{sec:m31cmd} robustly confirms the presence of the gradient.
However, while the CMD analysis also predicts the lack of metal-rich stars in
the more remote outer region, the figure shows the unexpected high
fraction of metal-rich stars at $R \ga 80$ kpc. This is because the
derived mean/median metallicities in the outer halo are somewhat
overestimated due to the heavy contamination of the Galactic disk dwarf
stars as discussed in \S~\ref{sec:m31sb}. Therefore, we
re-calculate the mean/median metallicities by imposing a more stringent color
cut of $(V-I)_0 < 1.8$, to ensure a minimal contamination from the
Galactic disk. The results are shown in Figure~\ref{fig:m31md2}. The
mean/median metallicities in the south-east region
show a systematic decrease of about 0.2 dex while keeping no spatial gradient.
In contrast, the variation in the north-west region based on the stringent
color criterion shows now a strong metallicity gradient. 

\begin{figure}[htpd]
 \begin{center}
  \epsscale{1}
  \plotone{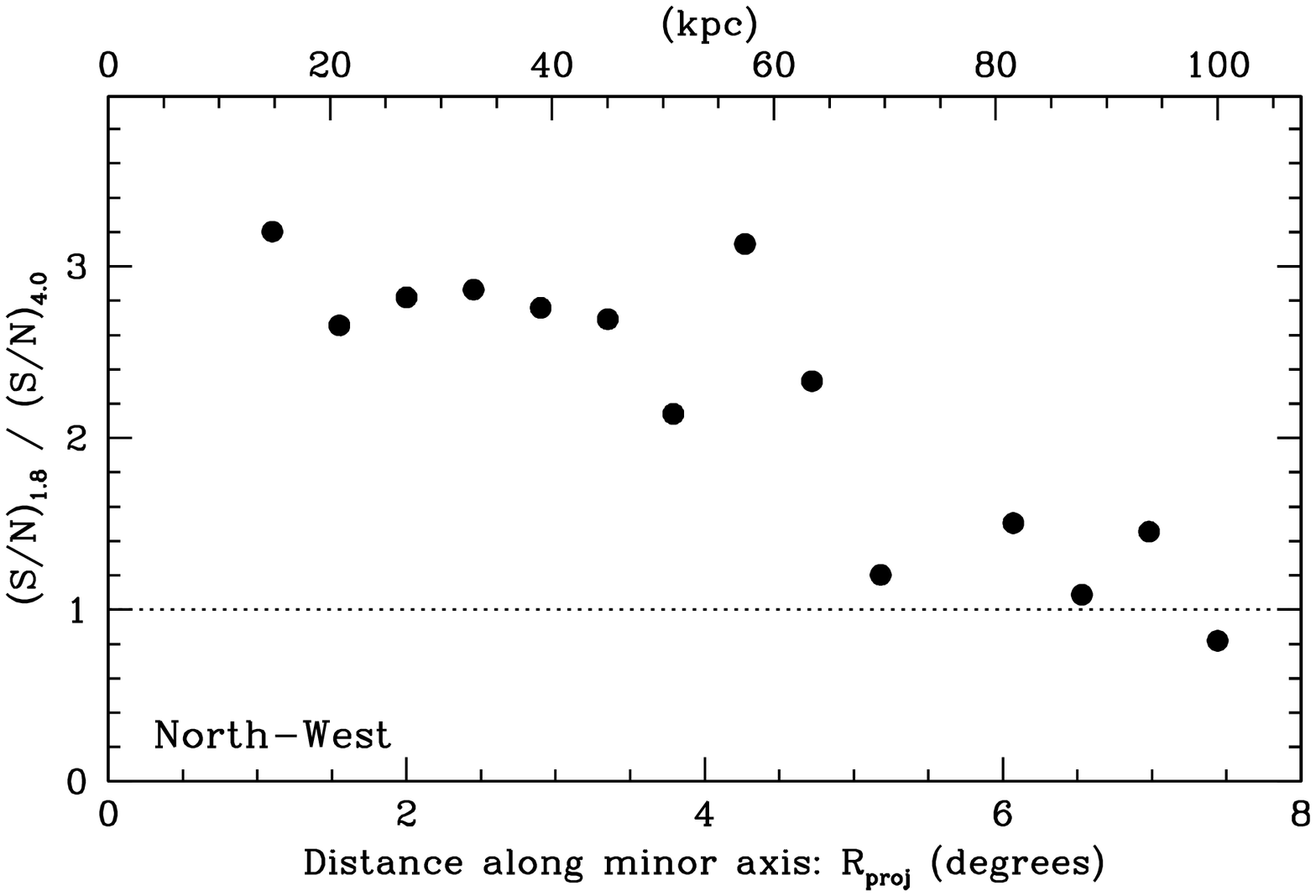}
  \caption[a]{The comparison of the two S/N ratios, (S/N)$_{1.8}$
  [with a color cut of $(V-I)_0 < 1.8$] and (S/N)$_{4.0}$ 
  [with $(V-I)_0 < 4.0$], plotted as a function of
  north-west minor-axis distance from the M31 center (in projection),
  where the S/N ratio means the ratio of the number of the residual
  objects used to calculate the mean/median metallicity in each
  summed-up field to the number of the foreground sources in the scaled
  control field.
  }
  \label{fig:m31mdsn}
 \end{center}
\end{figure}

We further test whether this strong metallicity gradient is a real
feature or not, in addition to the above-confirmed morphological variation
in the CMD along the north-west minor-axis. Figure~\ref{fig:m31mdsn} denotes the
ratio of the signal-to-noise (S/N) ratio, plotted as a function of
north-west minor-axis distance from the M31 center (in projection),
where the S/N ratio means the ratio of the number of 
objects used to calculate the mean/median metallicity in each summed-up
field to the number of the foreground sources in the scaled control
field. The vertical axis of Fig.~\ref{fig:m31mdsn} means the ratio of
the S/N ratio of $(V-I)_0 < 1.8$ selection sample to the signal-to-noise
ratio of $(V-I)_0 < 4.0$ selection. Basically, because the
vertical value is more than 1.0 in most regions of the north-west halo,
the spatial variation of mean/median metallicities based on the
stringent color criterion of $(V-I)_0 < 1.8$ is statistically more
reliable than that based on a $(V-I)_0 < 4.0$ color cut. However, note
that the absolute mean/median metallicities are clearly somewhat
underestimated in the inner fields with a lot of metal-rich stars at $R
\la 40$ kpc along the north-west minor axis. Eventually, we conclude
that this test provides the lower limit to the radial metallicity
gradient photometrically detected in the north-west minor axis halo. 

%%% 7 %%%
\section{Discussion}\label{sec:HALOdisc}

\subsection{The Distribution of the Underlying Halo}

In this study, one of our major purposes is to clarify the general properties
of the underlying, smooth halo component of M31. As mentioned in \S~\ref{sec:intro},
recent large-scale photometric and spectroscopic surveys of the M31 halo have 
revealed the existence of the metal-poor and old populations (like a
classical smooth halo in the Milky Way), which are buried in the inner complex
metal-rich and young substructures \citep[e.g.,][]{Kalirai2006,Ibata2007}. By
the term ``smooth'', what we mean here is not necessarily that the component is
perfectly spatially smooth, but instead that any substructures that may
be present in the halo are below detectability with the current survey. 

The surface brightness profile as shown in \S~\ref{sec:m31sb} is a good
indicator to describe the radial extent of the halo. Since many studies
of the density distributions of the Milky Way halo have already been made,
it is useful to compare such results with our current
studies of the M31 halo. The Milky Way halo has been characterized by
a steep power-law density profile $\rho(r) \propto r^{-\gamma}$:
its power-law index, $\gamma$, is in the range of $2.5$ to $3.6$.
Some of the variance in the value of $\gamma$ appear among different
research groups, which can be attributed to the presence of substructures
in the halo; previous studies have commonly been limited in sky
coverage, so some local structures in the halo can affect a final result.
Surveys covering significant portions of the sky have been
feasible only recently. For instance, the most recent SDSS study by
\citet{Juric2008}, which made use of good distance estimates to halo
stars near main-sequence turnoff covering 6500 deg$^2$ of sky, found
$\rho(r) \propto r^{-2.8}$ corresponding to $\alpha = 1.8$ as a
power-law index of the surface brightness profile. Similarly,
\citet{Bell2008} have found the stellar density distribution with
$\rho(r) \propto r^{-3}$, based on about $4\times 10^6$ color-selected
main-sequence turnoff stars contained in the SDSS DR5 catalog. Although
in these SDSS surveys accurate investigations of the density
profile are limited only within $R \sim 20$ kpc (or at most 40 kpc, relied on
the QUEST RR Lyrae sample of \citealt{Vivas2006}) due to the uncertainty in
the determination of the distance to remote stars, these results for
the Milky Way halo are completely consistent with the present result of
the M31 halo, $\Sigma(R) \propto R^{-2.17 \pm 0.15}$ (i.e., $\gamma
\sim 3.2$), as derived for the north-west minor-axis halo out to $R \sim
100$ kpc.  

It is worthwhile to compare our current result of the radial profile of
the M31 halo with other observational studies dedicated to the outer halo.
Our photometrically-estimated north-west profile is in reasonably agreement
with the south-east profile of $\Sigma(R)
\propto R^{-1.91 \pm 0.12}$ based on the large south quadrant area out
to 140 kpc (in projection) from the M31 center by the photometric
CFHT/MegaCam survey of \citet{Ibata2007}. In contrast, our result is slightly
flatter than \citet{Guhathakurta2005} and \cite{Gilbert2006} who have
found an extended stellar halo in M31 following $\Sigma(R) \propto
R^{-2.6}$ (or $R^{-2.3}$) out to $R \sim 165$ kpc, based on the
kinematically-selected RGB sample by the Keck/DEIMOS
observation. However, we note that the former authors have not
removed unresolved background galaxies which significantly affect fields
under the worse seeing condition as discussed in \S~\ref{sec:daopars}
and their survey is more than one magnitude shallower than our
Suprime-Cam survey. The latter authors have estimated the
radial profile of the M31 halo derived from secure spectroscopic RGB
sample in the remote outer halo, but it is based on only $\sim 40$ red
giants to determine the outer profile beyond $R = 60$ kpc. 

Recent numerical simulations predict that most of the mass in the
stellar halo originates in satellites that were accreted more than 8 Gyr
ago (corresponding to at $z \simeq 1$). The halo is smoothly built from
the inside out within the finite dynamical time scale, and becomes the
dominant contributor at 30--60 kpc and within
\citep{BJ2005,Abadi2006}. Our derived surface density profile of the M31
halo in the external view is in excellently agreement with the density
profile of the accreted stellar halo; it is described as
a Hernquist profile with $R_h \simeq 14$ kpc within the radial range
dominated by the ancient halo, as expected by the \citet{BJ2005} model. 
In the meantime, our NFW fitting based on the observed SB profile
returns several times higher concentration value ($c \equiv R_{\rm
vir}/R_h$) than that derived from the simulations based on dark matter
distributions \citep[e.g.,][]{BJ2005}, assuming the typical virial
radius (e.g., $R_{\rm vir} = 200$ kpc) of Milky Way-type galaxies. This
fact suggests that the visible baryonic matter like stars may well converge
in the galactic center compared to the invisible dark matter because of
baryon dissipation during accretion processes. 

\subsection{Implications for the Origin of Substructures}\label{sec:HALOdisc2}

Another major purpose of the current study is to probe the origin of various
substructures as found in both the inner and outer parts of the M31 halo.
Their metallicity distributions as well as the spatial distributions of
substructures are useful information to untangle
the formation history of the halo because each substructure preserves
such fossil records associated with star formation and merging/accretion events.

\subsubsection{The South-East Minor Axis Substructures}

As shown in the previous subsection, the spatial distribution of
mean/median metallicity along the south-east minor axis of the M31 halo
remains approximately constant with increasing radius from the
center. This is because our observing fields with Suprime-Cam
coincidentally overlap some overdense stream-like substructures,
characterized by Stream~D, C and B of \citet{Ibata2007}, which dominate  
the metallicity distribution of the stars in these fields. Stream~D and
C have similar high mean metallicity but clearly different CMD
morphology and metallicity distributions as shown in \S~\ref{sec:SCD},
suggesting that these two populations are different. Stream~B brushes
our outermost south-east field at $R \sim 90$ kpc, as shown in the south
quadrant map of \citet{Ibata2007}. This constancy in the mean
metallicity variation is consistent with the prediction by many previous
studies, proposing that the halo is built-up through the dissipationless
chaotic assembly from small subgalactic fragments
\citep[e.g.,][]{SZ1978}. 

However, the absence of metallicity gradient conflicts with the
result of \citet{Kalirai2006} and \citet{Koch2007}, who found a strong  
metallicity gradient in the south-east minor axis discrete fields out to 
$\sim 165$ kpc, based on the kinematically-selected RGB sample with the
Keck/DEIMOS spectrograph. The mean metallicities in the Subaru fields
(SE2 and SE9) which overlap with the two fields studied in
\citet{Kalirai2006} (their a0[30kpc] and m6[87kpc]) are actually almost
consistent with those given in their paper ($\langle$[Fe/H]$\rangle \sim
1.1$), although their work showed a metallicity gradient over larger
radii out to $R \sim 165$ kpc. On the other hand, some discrepancies
with the \citet{Koch2007} result are puzzling. These authors re-analyzed
the same data as \citet{Kalirai2006} but using a different method for
metallicity determination and found a strong metallicity gradient, where
the metallicity at $R \sim 90$ kpc is [Fe/H]$ = -2.4$, i.e., more
metal-poor than our result of [Fe/H]$ \sim -1$. Although comprehensive
understanding of this descrepancy is difficult only from our photometric
studies, Stream~B which distributes over our fields at $R \sim 90$ kpc
may affect the determination of a global metallicity distribution, as
mentioned above. To settle this issue, it is necessary to carry out a
wider-field photometric/spectroscopic survey over the large outer part
of the M31 halo. 

In addition, we compare with the main-sequence fitting results based on
the {\it HST\/}/ACS deep observations of \citet{Brown2008} who found the
trend that the mean age becomes simply older beyond 30 kpc in the
south-east inner region. In fact, we have found that at 27 kpc, 41 kpc (corresponding 
to Stream~D), and 57 kpc (corresponding to Stream~C), the mean ages are
$\sim 10$, $\sim 9.6$, and $\sim 9.3$ Gyr, respectively, as derived from
their MDs and RC magnitudes using the theoretically-computed relation of
\citet{Rejkuba2005} (see Fig.~\ref{fig:rejkuba}). These results suggest
that there exist some inhomogeneities of the halo, which is yet on the way
to the equilibrium structure. 

\subsubsection{The North-West Minor Axis Substructures}

The mean metallicity variation along the north-west minor axis, based on
continuous Suprime-Cam pointings out to $\sim 100$ kpc, shows a strong
spatial gradient with increasing radius. This is consistent with the
result of \citet{Kalirai2006} and \citet{Koch2007}. This spatial
dependence of abundance distribution may conflict with the scenario of
galactic halo formation via hierarchical merging, where successive
merging events can erase any spatial gradients in metallicities as
confirmed by numerical simulations. For instance, the study of
\citet{Font2006b} predicts a gradient in [Fe/H] averaged over each full
simulated halo of at most 0.5 dex over a few tens of kpc.

It is yet difficult to clarify the physical origin of the detected
metallicity gradient in the halo because there exist few observations
and simulations for the presence of a strong abundance gradient with
increasing radius in the halo. A possible explanation for the current
result is that the observed spatial distribution of metallicities may
reflect just a snapshot of the dynamical evolution, which is on the way
to the equilibrium structure characterized by a flatter metallicity
gradient. In particular, the halo structure is dominated by transient
substructures, which may largely affect the spatial distribution of
metallicities instantaneously. Indeed, modern galaxy formation
simulations within the context of the standard $\Lambda$CDM cosmology
predict that the outer halo beyond $R \sim 50$ kpc is dominated by
numerous faint arcs, shells and streams, as the relaxation time of these
subsystems as found at large distances from the galaxy center is quite
long at such locations, where even the dynamical time is of order of Gyr
\citep[e.g.,][]{BJ2005,Johnston2008}. Furthermore, dynamical friction is
at work, which acts more strongly on more massive subhalos, making them
fall rapidly into the potential well, where they become disrupted and
their contents are mixed into the evolving galaxy. We then observe a
rather steep metallicity gradient just after this event occurs.

\subsubsection{Dispersed Faint Substructure?}

Satellites with sufficiently long accretion times of more than several Gyr
would contribute to the wide range of the stellar halo, so it is
difficult to find any density substructures in such well-mixed systems at
the present day. It is noteworthy that previous studies found
kinematically cold substructures based on the detailed Keck/DEIMOS
spectroscopic observations in the south regions of the M31 halo, where 
no evidence for the presence of substructures has been deduced from
photometric observations
\citep{Kalirai2006a,Gilbert2007,Chapman2008}. These results suggest the 
possibility that there are also as-yet unrelaxed debris from disrupted
dwarf satellites underneath the smooth halo at the north-west 30-40 kpc
fields. Some $N$-body simulations \citep[e.g.,][]{Bekki2001} predicted
that these optically faint substructures below our detected surface
brightness levels may still exist in phase space, kinematics.  
In this regard, we discuss about the possibility of the existence of
these dispersed faint substructures in this subsection. 

As long as the surface brightness profiles (Fig.~\ref{fig:m31sb} and
\ref{fig:m31sbfit}) are concerned, there is no evidence for 
overdense substructures in the north-west and south-east regions at 30 kpc.
However, we find the age difference between both fields; the mean age of the
stellar population in the north-west part ($\sim 7.5 \pm 1.5$ Gyr) is
about 2.5 Gyr younger than that in the south-east counterpart field
($\sim 10 \pm 1$ Gyr), as derived from their MDs and RC magnitudes using
the theoretically-computed relation of \citet{Rejkuba2005} (see
Fig.~\ref{fig:rejkuba}). In fact, we cannot insist that the two
populations originated from the same stellar population because the KS
probability for this hypothesis based on their two cumulative MDs is
rather low. Therefore, we expect that there might be a dispersed faint 
substructure in the seemingly smooth regions at $R =$ 30--50 kpc in the
north-west minor axis halo, which preserves the information of ancient
accretion events. In addition, note that the age difference could be
occurred by the difference in the distance of the same population
substructures. To distinguish these much fainter substructures from 
the smooth inner halo we need additional kinematic studies as well as
some abundance information (such as $\alpha$-element abundance) as
predicted in \citet{Johnston2008}. 

Suppose that in the north-west region of M31 halo there exists such a
well-mixed substructure with sufficient long accretion time (e.g., 9
Gyr), it may be possible to trace this accretion event from our
photometric data if the event provides some specific stellar populations.
If so, we expect a specific magnitude of the RC/AGBb feature, where a clustered
distribution of stars in color-magnitude diagrams allows us to easily identify
the presence of such stellar populations, compared to attempts to detect faint
spatial substructures from already smeared out surface density distribution.
Also we can estimate the mass of an accreted satellite as $\sim 1 \times 10^9 
M_{\sun}$, which corresponds to ${\rm [Fe/H]_{\rm mean} \sim -
1.0}$. Then, if we suppose that this progenitor galaxy genuinely has an
intermediate-age population, it is likely that such a galaxy possibly
having interstellar gas has undergone a star formation driven by, for
instance, the second or third collision or supernova wind about 7.5 Gyr ago. 

\subsubsection{Comparison with Recent Numerical Simulations}

\begin{figure}[htpd]
 \begin{center}
  \epsscale{1}
  \plotone{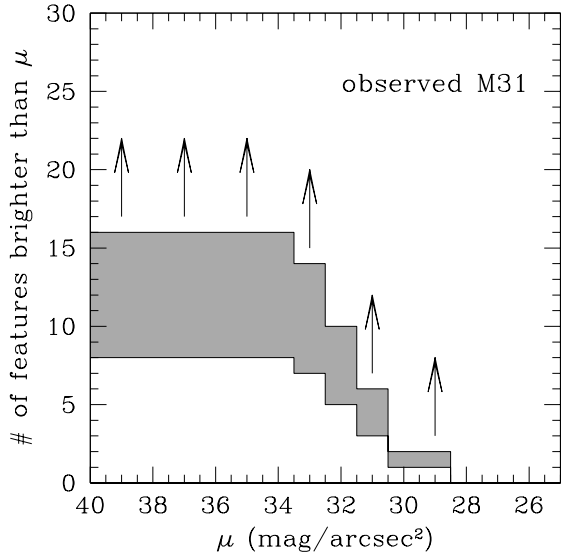}
  \caption[a]{The cumulative number of distinct
  substructures observed in the M31 halo, based on our work and
  \citet{Ibata2007}. The gray zone indicates plausible upper and lower
  limits, if simply extrapolated to the entire survey region. 
  }
  \label{fig:KVJf8}
 \end{center}
\end{figure}

\begin{figure}[htpd]
 \begin{center}
  \epsscale{1}
  \plotone{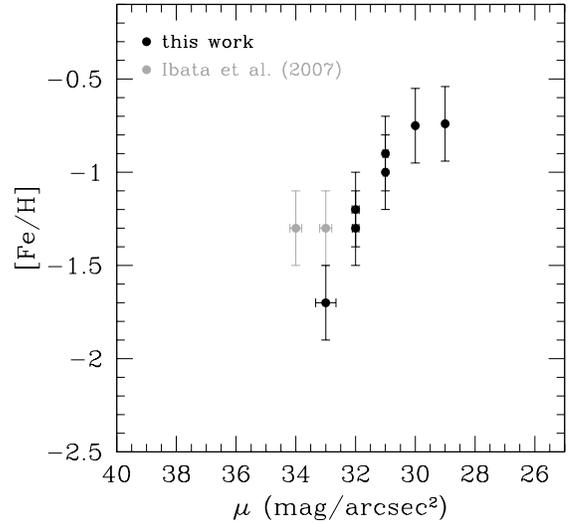}
  \caption[a]{The average metallicity against SB for the
  substructures combining our M31 observations and \citet{Ibata2007}
  (see text). Metallicity uncertainties ($\pm 0.2$ dex) are roughly
  plotted based on the error analysis in \S~\ref{sec:metal_err}. Higher
  SB tidal debris tend to be more metal-rich. 
  }
  \label{fig:font08}
 \end{center}
\end{figure}

The recent models also predict that substructures that we can discover 
in the remote outer halo more likely have low surface brightness
compared to the inner substructures with high metallicity and
high surface brightness \citep[e.g.,][]{BJ2005,Johnston2008}. In
particular, \citet{Johnston2008} predict that in a stellar halo of Milky
Way type galaxies being built entirely from accretion within a
$\Lambda$CDM context, the fraction of the system (of the order of 10\%
of its stars) are comprised of distinct substructures, i.e., a few to a
dozen streams which are brighter than 35 mag arcsec$^{-2}$ at projected
distances greater than 30 kpc from the host galaxy. 
Figure~\ref{fig:KVJf8} shows the same cumulative
distribution derived from our M31 observations and \citet{Ibata2007}:
the GSS, Stream~A to F and the major-axis diffuse
structure. The gray zone indicates plausible upper and lower limits, if
simply extrapolated to the entire survey region. It is evident that both
luminosity functions derived from the models and the observations are
reasonably similar, suggesting that the outer halo of M31 as well as the
Milky Way halo has been built up by recent accretion events of satellite
galaxies. Furthermore, Figure~\ref{fig:font08} plots the average
metallicity against SB for these substructures including our W-shelf
data. Gray circles show the Stream~A and B substructures of
\citet{Ibata2007}. Since metallicity uncertainties are just roughly
plotted based on the error analysis in \S~\ref{sec:metal_err}, in
practice they may be slightly larger. Despite this, we find that higher
SB tidal debris tend to be more metal-rich, which is confirmed by
some recent numerical simulations \citep[e.g.,][]{Font2008,Johnston2008}. This
result implies that higher SB debris may tend to come from the more
luminous, metal-rich dwarf satellites and/or recent encounters. 
However, while more detailed theoretical approaches to the issue are
required, it is also of great importance to explore more and fainter
substructures in the larger region of the M31 halo using the
next-generation wide-field imager such as Hyper Suprime-Cam, to set
tighter limits on the halo formation history. 

It is worth noting that recent high-resolution $N$-body simulation made by
\citet{Mori2008} provides an interesting scenario for the M31 halo:
a merging event of a dwarf satellite into the halo of M31 gives rise to
an elongated substructure similar to the GSS, the giant stellar stream found
in the south-east part of M31's halo. Also, their model suggests
that further stream-like substructures emerge in the form of
concentric shells around a galaxy after collisions with a disk component
and subsequent relaxation process, and that a multiple large-scale shell
system is finally formed in the outer region of the halo; the outermost
large-scale shells have a radius of $>50$ kpc and survive for at least
4~Gyr. If this is the case, then widely-separated shell structures would 
appear in the form of concentric shells around M31's halo. Although with this
anticipation it is difficult to understand the existence of the clearly
different radial metallicity gradients found along the north and south
parts of M31, we actually have been able to identify
two overdense substructures at the north-west counterpart
fields. Considering their CMD morphology, it appears that the
counterpart of Stream~E is Stream~B and that of Stream~F is
Stream~A, although there is some positional shift of about
15--20 kpc. Since it is not yet clear that these 
counterpart streams indeed have the same populations, we need to carry out
further observations of the M31 halo with a wider and deeper imaging
survey, in order to clarify the detailed stellar population and
the true spatial extension of these substructures.

\section{Conclusions}\label{sec:summary}

We have investigated the structure and stellar populations of the stellar halo
in the Andromeda galaxy, using deep and wide-field $V$ and $I$-band images
taken with Subaru/Suprime-Cam.  From the analysis of the CMDs
through comparing isochrones and globular cluster fiducials,
the fundamental properties of each halo
structure identified here have been derived, such as distance, surface brightness
profiles, metallicities and ages of the stellar populations. Our results for
each of the GSS and other halo structures (substructures and smooth halo)
are summarized as follows.

\subsection{The Andromeda Giant Stream}

We have identified an intermediate-age and metal-rich RC population as well as
a broad RGB sequence in the Andromeda GSS.
From the $VI$ CMD and the LF, we have measured the distance to our observed stream
field of $(m-M)_0 = 24.73 \pm 0.11$ ($883 \pm 45$ kpc) based on the $I$-magnitude
of the TRGB and reddening $E(B-V)=0.05$. This result supports the
previous prediction of \citet{McConnachie2003} that their Field~5 ($840
\pm 20$ kpc) which falls near GSS1 is at about 100 kpc further away
along the line of sight relative to M31's main body.

From the photometric comparison of more than 7000 bright RGB stars in
our stream field with the RGB templates of Galactic globular clusters,
we have obtained the metallicity ditributions (MDs) of the GSS for
a statistically significant sample. It shows the characteristics
summarized as the peak metallicity of [Fe/H]$_{\rm peak} \gsim -0.5$,
the average metallicity of [Fe/H]$_{\rm mean} = -0.6$ and the median
metallicity of [Fe/H]$_{\rm med} = -0.5$. This 
result is in excellent agreement with that of kinematically selected RGB
stars in the same stream field \citep{Guhathakurta2006}. 
Furthermore, based on the stellar
mass/luminosity-metallicity relation observed for the Local Group dwarf
galaxies shown in \citet{Cote2000} and \citet{Dekel2003}, the mass of the
plausible progenitor is in the range of $10^7$ to $10^9 M_{\sun}$, which 
is consistent with the work of \citet{Mori2008}.

The feature near RC, showing an extended brighter part relative to its peak,
is an unambiguous signature
of an intermediate age population, which is also supported by the
existence of an AGB bump. We have estimated the
age of the stellar population in the GSS, by convolving
the MDs and the RC magnitudes with the theoretical models from
\citet{Rejkuba2005}. As a result, we have found that the dominated population
of the GSS has the mean age of $\sim 8$ Gyr, consistent with the
main-sequence fitting results with the {\it HST\/}/ACS observations by
\citet{Brown2006b}. 

The GSS possibly originated from the accretion or minor merger of a
massive dwarf satellite within the mass range of $10^7$ to $10^9
M_{\sun}$, like M32, but previous studies suggests that M32 itself is
probably not a progenitor of the GSS, as we already mentioned in
Section~1. Considering the finite dynamical time of such a mass
accretion, the progenitor has to be bound to M31 several billion years
ago as judged from the current residual substructure. The first
pericentric merging may well induce the third star forming event from
the remaining gas, whereas 30\% of the GSS stars were provided in this
event.

\subsection{The Andromeda Stellar Halo}

Our method for analyzing the CMDs of the GSS has also been
applied to other large survey regions of the M31 halo.
Main results obtained from the outer halo survey are summarized as follows:

\begin{itemize}
\item
In the outer halo beyond $R \sim 30$ kpc,
     M31 contains a multitude of overdense substructures
     compared to the smooth radial profile, such as streams, arc and
     shells. We confirm the two substructures, namely Stream~C and D,
     discovered by \citet{Ibata2007}. Furthermore, from our constructed
     stellar density maps we have found two as-yet unknown overdense
     substructures in the north-west halo, which we call Stream~E
     and F in this work. It is yet unclear to what extent
     these substructures are spatially distributed. 
\item
We have derived the fundamental properties of these four stream-like
     structures in addition to the major-axis diffuse structure implied
     by \citet{Ibata2007} and the Western shelf seen in the star count
     map of \citet{Ferguson2002}. The derived properties are summarized
     in Table~\ref{tab:Sub}. We have found that higher SB tidal
     debris tend to be more metal-rich, which is confirmed by some
     recent numerical simulations
     \citep[e.g.,][]{Font2008,Johnston2008}. This result implies that
     higher SB debris may tend to come from the more luminous, metal-rich
     dwarf satellites and/or recent encounters. 
\item
By separating regions showing obvious substructures, we have found that the
     remaining area of the survey exhibits a smooth metal-poor stellar
     halo component in the north-west minor axis halo. This structure
     needs not to be perfectly spatially smooth, but the intrinsic
     inhomogeneities are below the sensitivity of this study. The smooth
     halo is vast, extending out to the radial limit of the survey, at
     $R = 100$ kpc. The SB profile of this component can be modeled with a
     Hernquist profile as suggested by simulations, and the resulting
     scale radius of $\sim 17$ kpc is comparable to what modern halo
     formation simulations predict. A power-law profile with $\Sigma(R)
     \propto R^{-2.17 \pm 0.15}$ [$\rho(r) \propto r^{-3.17 \pm 0.15}$]
     can also be fitted to the data. Our NFW fitting based on the
     observed SB profile returns a several times higher concentration
     value than that derived from the simulations based on dark matter
     distributions \citep[e.g.,][]{BJ2005}, assuming the typical virial
     radius (e.g., $R_{\rm vir} = 200$ kpc) of Milky Way-type
     galaxies. This fact suggests that the visible baryonic matter
     may well converge into the galactic center compared to the
     invisible dark matter because of energy dissipation in the course of
     accretion process. 
\item
We have detected metallicity inhomogeneities in different radial
     directions. The spatial distribution of mean metallicity along the
     south-east minor axis of the M31 halo remains approximately
     constant, while that along the north-west minor axis shows a strong
     spatial gradient with $R$ out to 100 kpc. This fact suggests that
     the halo may be yet in the course of the dynamical evolution,
     leading subsequently to a flatter metallicity gradient. 
\end{itemize}

In conclusion, the outer halo of the Andromeda galaxy as well as that of the
Milky Way can be formed from distinct stellar substructures
deposited from over $\sim 15$ independent accretion events of
subhalos with masses similar to typical dwarf galaxies, i.e., of order $10^7$--$10^9
M_{\sun}$. The Andromeda galaxy has quietly undergone the accretion
of these intermediate-mass subhalos without the destruction of the disk
\citep[e.g.,][]{Kazantzidis2008} since $z\sim1$ corresponding to $\sim
8$ Gyr ago. However, in the stellar halo of M31 there are probably much
fainter substructures below detectability with the current survey,
as suggested by the existence of further relaxed substructures (e.g., 
slightly younger substructures discussed in \S~\ref{sec:HALOdisc2}) and
kinematic substructures recently discovered in some spectroscopic
studies \citep{Kalirai2006a,Gilbert2007,Chapman2008}. Therefore, we will
need further observations with higher sensitivity to arrive at a better 
understanding of the formation of M31's stellar halo. 

\begin{deluxetable*}{lccccc}
 \tablewidth{0pt}
 \tablecaption{Fundamental properties of the substructures\label{tab:Sub}}
 \tablehead{
 \colhead{Name} & \colhead{Distance} & \colhead{$\mu_V$(typical)} &
 \colhead{$\langle$[Fe/H]$\rangle$} & 
 \colhead{$\langle$age$\rangle$} & \colhead{Progenitor mass\tablenotemark{$\star$}}\\
 \colhead{(Projected Distance from M31)} & kpc & mag arcsec$^{-2}$ & dex & Gyr & $M_{\sun}$
 }
 \startdata
 Giant Stream ($\sim 30$ kpc) & $883 \pm 45$ & $\lsim 29$ & $-0.74$ &
 $\sim 8$ & $\sim 10^9$\\
 W-shelf ($\lsim 25$ kpc) & $798 \pm 40$ & $\sim 30$ & $-0.75$ & $\sim 8$ & $\sim 10^9$\\
 Stream~D ($\sim 35$-45 kpc) & -- & $\sim 31$ & $\sim -1.0$ & $\sim 10$ & $\sim 10^9$\\
 Stream~C ($\sim 50$-60 kpc) & -- & $\sim 31$ & $\sim -0.9$ & $\sim 9$ & $\sim 10^9$\\
 MA structure ($\sim 60$ kpc) & -- & $\sim 32$ & $\sim -1.2$ & $\sim 9$? & $\sim 10^8$\\
 Stream~E ($\sim 60$ kpc) & -- & $\sim 32$ & $\sim -1.3$ & -- & $\sim 10^8$\\
 Stream~F ($\sim 100$ kpc) & -- & $\sim 33$ & $\sim -1.7$ & old? & $\sim 10^7$\\
 \enddata
 \tablenotetext{$\star$}{Based on the stellar mass-metallicity relation of \citet{Dekel2003}.}
\end{deluxetable*}

\acknowledgments

We are grateful to the anonymous referee for useful comments that helped
improve the manuscript.
We thank the staff at the Subaru telescope for their excellent support during
our observing runs. We wish to recognize and acknowledge the very significant
cultural role and reverence that the summit of Mauna Kea has always had
within the indigenous Hawaiian community. We are most fortunate to have
the opportunity to conduct observations from this mountain.
We are grateful to Steve Majewski and Rachael Beaton who provided the KPNO 
photometric catalog of the M31 halo so that we could compare our photometry 
with his group's photometry. We would like to thank Evan Kirby who helped 
generate the metallicity distribution from the theoretical isochrones. 
We are also grateful to Marina Rejkuba for sending us
the table for $I$ magnitudes of the AGB bump and RC features vs. stellar
ages based on her calibration of the theoretical stellar evolutionary tracks.
Data reduction and analysis were carried out on general common use computer
system at ADAC (Astronomical Data Analysis Center) of the National
Astronomical Observatory of Japan.
This work has been supported in part by a Grant-in-Aid for
Scientific Research (20340039) of the Ministry of Education, Culture,
Sports, Science and Technology in Japan.

\end{document}